\documentclass{ptephy_v1}

\usepackage{url} 

\begin{document}
\title{Overview of KAGRA : KAGRA science}



\author[1,2]{T.~Akutsu}
\author[3,4,1]{M.~Ando}
\author[5]{K.~Arai}
\author[5]{Y.~Arai}
\author[6]{S.~Araki}
\author[7]{A.~Araya}
\author[3]{N.~Aritomi}
\author[8]{H.~Asada}
\author[9,10]{Y.~Aso}
\author[11]{S.~Bae}
% ...
\affil[1]{Gravitational Wave Science Project, National Astronomical Observatory of Japan (NAOJ), Mitaka City, Tokyo 181-8588, Japan}
\affil[2]{Advanced Technology Center, National Astronomical Observatory of Japan (NAOJ), Mitaka City, Tokyo 181-8588, Japan}
\affil[3]{Department of Physics, The University of Tokyo, Bunkyo-ku, Tokyo 113-0033, Japan}
\affil[4]{Research Center for the Early Universe (RESCEU), The University of Tokyo, Bunkyo-ku, Tokyo 113-0033, Japan}
\affil[5]{Institute for Cosmic Ray Research (ICRR), KAGRA Observatory, The University of Tokyo, Kashiwa City, Chiba 277-8582, Japan}
\affil[6]{Accelerator Laboratory, High Energy Accelerator Research Organization (KEK), Tsukuba City, Ibaraki 305-0801, Japan}
\affil[7]{Earthquake Research Institute, The University of Tokyo, Bunkyo-ku, Tokyo 113-0032, Japan}
\affil[8]{Department of Mathematics and Physics, Hirosaki University, Hirosaki City, Aomori 036-8561, Japan}
\affil[9]{Kamioka Branch, National Astronomical Observatory of Japan (NAOJ), Kamioka-cho, Hida City, Gifu 506-1205, Japan}
\affil[10]{The Graduate University for Advanced Studies (SOKENDAI), Mitaka City, Tokyo 181-8588, Japan}
\affil[11]{Korea Institute of Science and Technology Information (KISTI), Yuseong-gu, Daejeon 34141, Korea}
% ...

\collaborator{KAGRA Collaboration}


\date{\today}
\begin{abstract}%
KAGRA is a newly build GW observatory, a laser interferometer with 3~km arm length, located in Kamioka, Gifu, Japan. 
In this paper in the series of KAGRA-featured articles, we discuss the science targets of KAGRA projects, considering not only the baseline KAGRA (current design) but also its future upgrade candidates (KAGRA+) for the near to middle term~($\sim$5 years).
\end{abstract}

\subjectindex{xxxx, xxx}

\maketitle

\tableofcontents



\section{Introduction}\label{ptep06_sec1}

Advanced LIGO (aLIGO)~\cite{aligo} and advanced Virgo (AdV)~\cite{AdV} have detected gravitational waves (GWs) from ten mergers of binary black holes (BBH) \cite{GWTC1} and one merger of binary neutron stars (BNS) \cite{GW170817PRL} in O1 and O2 observing runs\footnote{More GWs have been detected in O3 observing run, but not been published yet as of July 27th, 2020.}. The observations of these GW events have broadened many science opportunities: the astrophysical formation scenarios of BBH and BNS, the emission mechanisms of short gamma-ray bursts (sGRB) and following electromagnetic (EM) counterparts associated with BNS mergers, equation of state (EOS) of a neutron star (NS), the measurement of Hubble constant, the tests of gravity, the no-hair theorem of a black hole (BH) and so on. In addition to these topics studied based on the observational data, there are still several sources of GWs to be observed in the future with ground-based detectors such as those from supernovae, isolated NSs, intermediate-mass BHs, and the early Universe.

\begin{figure}[t]
\centering
\includegraphics[width=0.80\linewidth]{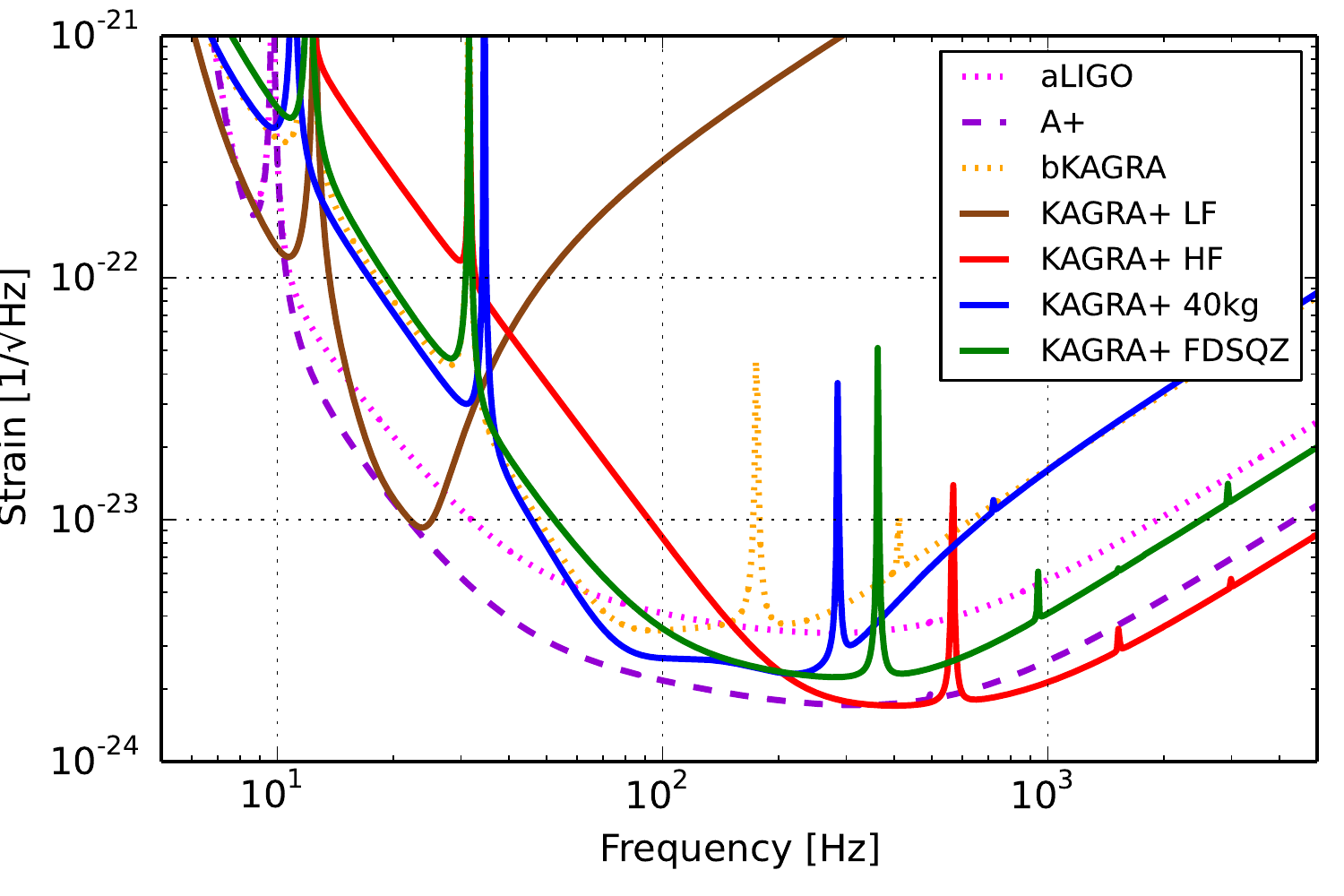}
\caption{Sensitivity curves for bKAGRA and upgrade candidates for KAGRA+. Sensitivity curves for aLIGO, A+, and bKAGRA are shown for comparison~\cite{kagraplus}.}
\label{fig:kagraplus_cmp}
\end{figure}

The current ground-based detectors, aLIGO and AdV, are planned to start upgrading their instruments around 2024 right after O4 observation ends and improve their sensitivities (called A+~\cite{A+} and AdV+~\cite{AdV+}, respectively)~\cite{LVKScenario2018}. KAGRA \cite{KAGRA:NatureAstro} will also be upgraded in the same term as aLIGO and AdV, though the concrete upgrade plan has not been decided officially yet. In this paper in a series of KAGRA-featured articles, we review scientific cases available with the second-generation ground-based detectors including aLIGO, AdV, the baseline KAGRA (bKAGRA) and their future upgrades, A+, AdV+, and KAGRA+, respectively, and discuss KAGRA's scientific contributions to the global detector networks composed of the detectors above. Since the official design of KAGRA+ has not been decided yet, we consider four possible upgrade options for the near to middle term~($\sim$5 years): one focused on low frequencies~($\sim 10$ - $50\,{\rm Hz}$)~[LF], one focused on high frequencies~($\sim 300\,{\rm Hz}$ - $5\,{\rm kHz}$)~[HF], to use heavier mirrors to improve middle frequencies~($\sim 50$ - $300\,{\rm Hz}$)~[40kg], and to inject frequency dependent squeezing for the broadband improvement~[FDSQZ]\footnote{We divide frequencies into three ranges: low, middle, and high, depending on the shape of the noise curves in Fig.~\ref{fig:kagraplus_cmp}.}. In addition to these, we also consider a slightly optimistic case in which technologies for four KAGRA+ upgrades are combined (called Combined). The noise curves of bKAGRA and its upgrade candidates are shown in Fig.~\ref{fig:kagraplus_cmp}. More details on the concrete plans and technological aspects are discussed in a companion paper in the series \cite{KAGRA:PTEP07}.\\

\section{Stellar-mass binary black holes}
\subsection{Formation scenarios} 

\noindent {\it - Scientific objective}


The existence of massive stellar-mass BBH provides us a new insight about 
formation pathways of such heavy compact binaries \cite{GW150914PRL,GW151226PRL}.
Since massive stars likely loose their masses by stellar winds driven by metal lines, dust, and pulsations 
of the stellar surface, heavy BHs with masses of $\gtrsim 20~M_\odot$ are not expected to be left as 
remnants at the end of their lifetime of massive stars with metallicity of the solar value \cite{LIGO_ApJ}.
In fact, the masses of BHs we have ever observed in EM waves (e.g., X-ray binaries) are 
significantly lower than those detected in GWs \cite{GWTC1}. 
This fact motivates us to explore the astrophysical origin of such massive stellar-mass BBH population 
in low-metallicity environments 
(below 50 \% of solar metallicity or possibly less) \cite{LIGO_ApJ}. 
So far, a number of authors have proposed formation channels; evolution of low-metallicity isolated binaries 
in the field \cite{Belczynski_2004,Dominik_2012,Belczynski_2016,Mapeli_2016}, 
dynamical processes in dense cluster systems (globular clusters, nuclear stellar cluster, or compact gaseous 
disks in active galactic nuclei) \cite{Portegies Zwart_2000, O'Leary_2009,BaeKimLee, Rodriguez_2016a, Antonini_2016, Stone_2017, Park_2017}, 
massive stars formed in extremely low-metallicity gas at high-redshift Universe (Population III stars, hereafter Pop III stars) 
\cite{Kinugawa_2014, Hartwig_2016, Inayoshi_2016,Inayoshi_2017}, and etc.
As an alternative non-astrophysical possibility, a primordial BH population in the extremely early universe (e.g., originated by phase transitions, a temporary softening in the EOS, quantum fluctuations) have been attracting attention \cite{Carr_2016,Sasaki_2016} (\cite{Sasaki:2018dmp} for a review). \\

\noindent {\it - Observations and measurements}


In order to reveal the astrophysical origin of stellar-mass BBH detected by LIGO and Virgo, we need to explore the properties 
of coalescing BBH, depending on their formation pathway and environment where they form. 
In particular, the effective spin parameters of BBH $\chi_{\rm eff}$ (the dimensionless spin components aligned or anti-aligned 
with the orbital angular momentum) are expected to be useful to discriminate the evolution models \cite{Kushnir_2016,Rodriguez_2016b}. 
In the field binary scenario, concerning binaries formed in a galactic plane, tidal torque exerting two stars (or one of them is a BH) in a close binary transport the orbital angular momentum into
stellar spins, resulting in $\chi_{\rm eff} \gtrsim 0$.
Since tidal synchronization occurs as quickly as $\dot{\Omega}\propto (R_\star/a)^6$, a binary with a short GW coalescence timescale 
(i.e., a small orbital separation) would be significantly spun to $\chi_{\rm eff} \sim1$.
Since BBH populations formed at high redshift take a long GW coalescence timescale, their BBH are hardly affected by 
tidal torque \cite{Hotokezaka_Piran_2017}. 
The underlying assumption is that 
the orbital angular momentum is not significantly changed via natal kicks that new-born BHs could receive during the core collapse of their progenitors \cite{Janka_2013}.
Although a strong natal kick $\gtrsim 500~{\rm km~s}^{-1}$ is required to affect the effective spin parameter, observations of low-mass X-ray binaries show no evidence of such strong natal kicks of BHs (see, e.g., \cite{Mandel_2016}).

On the other hand, in the dynamical formation scenario, concerning binaries formed in dense stellar environment, the distribution of effective spin would be isotropic at $-1\leq \chi_{\rm eff} \leq 1$, 
and negative values are allowed unlike the isolated binary scenario because the directions of stellar spins could be chosen randomly. 
As a robust result of the formation channel, the effective spin can have negative values at $-1\lesssim \chi_{\rm eff} \lesssim 0$.
Moreover, in the dynamical capture model, a small fraction of the binaries/BBH could gain significantly high eccentricities ($e\gtrsim 0.1$),
which will be imprinted in the GW waveform \cite{O'Leary_2009}. 

For PBHs formed in the early Universe (radiation dominated era), the spin is suppressed to $\chi_{\rm eff} \lesssim 0.4$~\cite{Chiba:2017rvs} or even smaller~\cite{DeLuca:2019buf} because they are formed from the collapses of density inhomogeneities right after the cosmological horizon entry. However, if the early matter phase exists after inflation, the spin can be large \cite{Harada:2017fjm}. Because of the uncertainty in the early Universe scenario, it would be difficult to distinguish the PBH origin from the astrophysical ones.

\begin{figure}[t]
\centering
\includegraphics[width=0.70\linewidth]{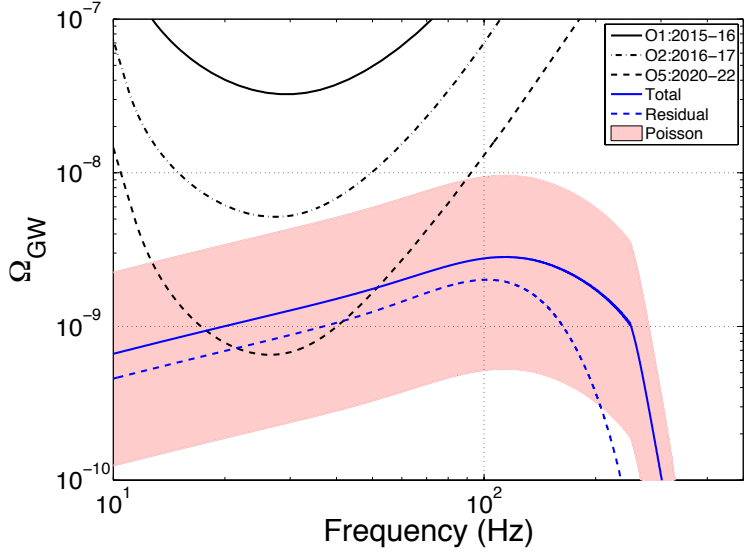}
\caption{
    Expected sensitivity of the network of aLIGO and AdV to a GWB from BBH in the fiducial model in \cite{LIGO_background}. Energy density spectra are shown in blue (solid for the total background; dashed for the residual background, excluding resolved sources, assuming final aLIGO and AdV sensitivity).
    The pink shaded region ``Poisson'' shows the 90\% CL statistical uncertainty, propagated from the local rate measurement, on the total background.
    The black power-law integrated curves show the $1\sigma$ sensitivity of the network expected for the two first observing runs O1 and O2, and for 2 years at the design sensitivity in O5. Adapted from \cite{LIGO_background}.
}
\label{fig:LIGO_background}
\end{figure}

In addition to the BBH population within a detection horizon of $z\lesssim 0.2$, 
BBH that coalesce at higher redshifts due to small initial separations are unresolved individually, 
but contribute to a GW background (GWB).
The spectral shape of a GWB caused by compact binary mergers is characterized by a single-power law of 
$\Omega_{\rm GW}\propto f^{2/3}$ at lower frequencies ($f< f_0$), is flattened from it at $f\simeq f_0$, and 
has a cutoff at higher frequencies ($f> f_0$) \cite{Phinney_2001}. 
Importantly, this result hardly depends on the merger history of the BBH population qualitatively. 
Therefore, a GWB produced from low-redshift, relatively less massive BBH population is robustly given by 
$\Omega_{\rm GW}\propto f^{2/3}$ in the frequency range. This GWB signal could be detected by future observing run O5 with aLIGO and AdV \cite{LIGO_background}, as shown in Fig.~\ref{fig:LIGO_background}. 
In fact, the spectral flattening occurs outside the GWB sensitive frequency range ($f_0 \gtrsim 100~{\rm Hz}$). 
On the other hand, a GWB caused by merger events of higher redshift, massive BBH population suggested by Pop III models are expected to be flattened at a lower frequency of $f_0 \simeq 30~{\rm Hz}$ \cite{Inayoshi_2016}, 
where LIGO/Virgo/KAGRA are the most sensitive. 
A detection of the unique flattening at such low frequencies will indicate the existence of a high-chirp mass, 
high-redshift BBH population, which is consistent with the Pop III origin. 
The detailed study of the GWB would enable us to explore the properties of massive binary stars at higher 
redshift and the epoch of cosmic reionization.\\

\noindent {\it - Future prospects}


LIGO-Virgo's sensitivities during O1 and O2 have been proved to be efficient to detect stellar-mass BBH with a mass range between ${\cal O}$(1-100) \citep{GWTC1}. The typical detection frequency of these stellar-mass BBH are between 30-500 Hz. The bKAGRA's sensitivity would be similar to aLIGO and AdV and we expect the observable sample for bKAGRA would be similar to those listed in GWTC-1 \citep{GWTC1}, i.e., stellar-mass BBH. The LIGO-Virgo-KAGRA (LVK) configuration would provide a better sky localization but without finding a host galaxy (via EM waves), it would be challenging to identify the location of a BBH (in a galactic disk or in a cluster, for example).

Toward the upgrades of bKAGRA, we expect the improved sensitivity toward the lower frequencies would be useful to increase the signal-to-noise ratio (SNR) for BBH populations. More observation means better number statistics in event rate estimation and distributions for underlying properties such as BH mass distribution. Furthermore, by observing earlier inspiral phase signals available at lower frequencies, it will be possible to achieve better accuracy in parameter estimation for individual masses and spin parameters. For a BBH population with small masses, say total mass below 10 $M_{\rm \odot}$, by improving the so-called `bucket' sensitivity around a few hundred Hz would be most effective to increase the detectability as well as the precision of parameter estimation.  

From the stellar-mass to supermassive populations, BHs are considered to have a broad mass spectrum. While a standard binary evolution scenario prefers a compact binary consisting of similar masses \cite{PostnovYungenson}, some binaries may have mass ratio $1/q\equiv m_{\rm 1}/m_{\rm 2}$ much larger than 1, where $m_{\rm 1}$ is the primary mass and $m_{\rm 2}$ is the secondary mass of the binary. Dense stellar system is considered to be a factory to generate compact binaries through dynamical interactions, e.g.~\cite{Benacquista}. However, a toy model for the dynamic formation scenario for compact binaries still prefers BBH with $1/q < 3$, e.g.~\cite{BaeKimLee}. 
Unequal-mass BBH with $1/q >$ a few are in particular interesting as the modulations in amplitude that is expected by post-Newtonian formalism in the inspiral phase become significant. Higher multiple modes (with the existence of spin(s)) are also more important for binaries with large mass ratios than equal-mass binaries~\cite{Pekowsky_2013}. Therefore, detections of unequal-mass binaries require the overall improvement in the detector sensitivity corresponding to the detection frequency band in the `bucket.' Indeed, the recent event, GW190412, has provided us the evidence to support the detection of the higher multiple models and proved that a broader frequency band play a crucial role for the detection~\cite{LIGOScientific:2020stg}.

In Table~\ref{tab:f_isco}, we present the innermost stable circular orbit (ISCO) frequencies \cite{Bardeen:1972fi} corresponding to stellar-mass BBH or BH-NS binaries with various masses. GW signals from binaries with the total mass of $\cal{O}$(10) M$_{\rm \odot}$ or larger would be spanned between $f_{\rm low}$ of an interferometer and around the estimated $f_{\rm ISCO}$ assuming the duration of ringdown signals would be much shorter than inspiral signals. The actual detectability will depend on the sensitivity of the interferometer within the frequency range given the binary.

\begin{table*}[t]
\begin{center}
\begin{tabular}{ccccc}
\hline \hline
$m_1$ ($M_{\odot}$)& $m_2$ ($M_{\odot}$) & $m_{\rm tot}$ ($M_{\odot}$) & $1/q=m_1/m_2$ ($M_{\odot}$) & $f_{\rm {ISCO,gw}}$ (Hz) \\
\hline 
 1.4 &  1.4 &  2.8 & 1   & 1574 \\
 5   &  1.4 &  6.4 & 3.6 & 1001 \\
10   &  1.4 & 11.4 & 7.1 &  387 \\
\hline
10   & 10   & 20   & 1   & 220 \\
30   & 10   & 40   & 3   & 110 \\
40   & 30   & 70   & 1.3 &  63 \\
\hline \hline
\end{tabular}
\end{center}
\caption{The ISCO frequencies for compact binary mergers with different mass components consisting of NSs and/or stellar-mass BHs. We assume no spin for BHs, i.e. $f_{\rm ISCO,gw}= (6^{3/2} \pi  m_{\rm tot})^{-1}$. The entries are sorted by the total mass of a binary.}
\label{tab:f_isco}
\end{table*}

In Table~\ref{tab:BBH30-PE-errors} and \ref{tab:BBH10-PE-errors}, the measurement errors of the binary parameters for equal-mass BBH with $30\,M_{\odot}$ and $10\,M_{\odot}$ at $z=0.1$ are estimated with the Fisher information matrix. We assume the detector networks composed of A+, AdV+, and bKAGRA or KAGRA+ (LF, HF, $40\,{\rm kg}$, FDSQZ, or combined). The waveforms we use are the spin-aligned inspiral-merger-ringdown waveform (PhenomD) for BBH. There are significant improvements from 2G detectors to 2.5G detectors (bKAGRA, LF, HF, 40kg, FDSQZ, Combined) in the SNR and the errors. But this is not solely due to KAGRA's contribution. The upgrade of bKAGRA to KAGRA+ among A+ and AdV+ modestly enhances the SNR and the sensitivities to the binary parameters in GW phase: the symmetric mass ratio $\eta$, the effective spin $\chi_{\rm eff}$, and the luminosity distance to a source $d_{\rm L}$. For the parameters in GW amplitude, the orbital inclination angle $\iota$ and the sky localization area $\Omega_{\rm S}$, the improvement of the errors is significant because the fourth detector is important to pin down the source direction and determine the other correlated parameters. The improvement factor depends on the configuration of the upgrade of KAGRA+. From the point of view to discriminate the formation scenarios, it would be better for KAGRA to improve the detector sensitivity at both low and middle frequencies. In addition, in order to increase the detectability of BBH coalescences at higher redshifts such as Pop III star binaries and primordial BH binaries, it is important to increase the horizon distance for BBH and analyze together with the data of a GWB measuring the spectral index and the spectral cutoff.\\


\begin{table*}[t]
\begin{center}
\begin{tabular}{lccccccc}
\hline \hline
quantities \;& 2G & bKAGRA & LF & HF & 40kg & FDSQZ & Combined \\
\hline \hline  
SNR & 44.4 & 76.4 & 75.5 & 76.0 & 81.2 & 81.5 & 89.5 \\
$\Delta \log \eta$ & 9.13 & 5.62 & 5.83 & 5.60 & 5.15 & 5.23 & 4.61 \\
$\Delta \chi_{\rm eff}$ & 15.6 & 9.75 & 9.41 & 9.58 & 9.35 & 9.13 & 8.59 \\
$\Delta \log d_{\rm L}$ & 11.7 & 6.82 & 8.67 & 7.19 & 6.19 & 7.32 & 5.79 \\
$\Delta \cos \iota$ & 8.89 & 4.92 & 6.35 & 5.60 & 4.68 & 5.25 & 4.36 \\
$\Delta \Omega_{\rm S}$ & 0.805 & 0.300 & 0.648 & 0.263 & 0.243 & 0.219 & 0.169 \\
\hline 
\end{tabular}
\end{center}
\caption{SNR, median errors of the mass ratio in the unit of $10^{-4}$, the effective spin $\chi_{\rm eff}$ in the unit of $10^{-3}$, the luminosity distance in per cents, the inclination angle in per cents, sky localization area in ${\rm deg}^2$ for $30\,M_{\odot}$ BBH. BBH are at the distance of $z=0.1$. 2G denotes the detector network composed of the second-generation detectors, aLIGO at Hanford and Livingston, AdV, and bKAGRA. The networks, bKAGRA, LF, HF, 40kg, FDSQZ, and Combined are composed of A+, AdV+, and bKAGRA or KAGRA+ (low frequency, high frequency, $40\,{\rm kg}$, frequency-dependent squeezing, and combined), respectively.}
\label{tab:BBH30-PE-errors}
\end{table*}

\begin{table*}[t]
\begin{center}
\begin{tabular}{lccccccc}
\hline \hline
quantities \;& 2G & bKAGRA & LF & HF & 40kg & FDSQZ & Combined \\
\hline \hline 
SNR & 19.8 & 32.4 & 31.6 & 34.2 & 34.9 & 33.9 & 36.7 \\
$\Delta \log \eta$ & 24.2 & 14.9 & 15.2 & 13.2 & 13.9 & 13.7 & 12.6 \\
$\Delta \chi_{\rm eff}$ & 9.48 & 6.02 & 5.73 & 5.83 & 5.75 & 5.87 & 5.62 \\
$\Delta \log d_{\rm L}$ & 25.0 & 15.0 & 19.8 & 15.6 & 14.2 & 13.9 & 11.8 \\
$\Delta \cos \iota$ & 19.1 & 10.7 & 13.7 & 12.5 & 10.8 & 10.1 & 8.55 \\
$\Delta \Omega_{\rm S}$ & 1.96 & 0.821 & 1.44 & 0.404 & 0.654 & 0.487 & 0.451 \\
\hline 
\end{tabular}
\end{center}
\caption{SNR and the median errors of parameter estimation for $10\,M_{\odot}$ BBH as in Table~\ref{tab:BBH30-PE-errors}.}
\label{tab:BBH10-PE-errors}
\end{table*}

\section{Intermediate-mass binary black holes}
\subsection{Formation scenarios}

\noindent {\it - Scientific objective}

The direct detections of GW emission have revealed the existence of massive BHs with masses of $\gtrsim 20-50~M_\odot$, which are significantly heavier than those ever observed in X-ray binaries
\cite{GW150914PRL, LIGO_ApJ,GWTC1}.
On the other side, supermassive BHs with the order of $10^6-10^9~M_\odot$ almost ubiquitously exist 
at the centers of galaxies and are believed to be one of the most essential components of galaxies.
However, the existence of a (binary) BH population between stellar-mass and supermassive regime,
i.e., $100\lesssim M_{\rm BH}/M_\odot \lesssim 10^5$, has not been confirmed yet (though there are some candidates).
The lack of such intermediate-mass binary BH (IMBBH) population is one of the most intriguing unsolved puzzles in astrophysics.
Detection of GWs from IMBBH would be one of the best way to probe their existence and physical natures.\\

\noindent {\it - Observations and measurements}

A plausible formation pathway of IMBHs is runaway collisions of massive stars in dense stellar systems 
(e.g., globular clusters and/or nuclear stellar clusters) \cite{Portegies_Zwart_2002,Freitag_2006}.
In a dense young star cluster, massive stars sink down to the center due to mass segregation and begin to physically collide.
During the collision processes, the most massive one gains more masses and grow to a very massive star (VMS) in a runaway fashion,
and thus an IMBH with a mass of $M_{\rm IMBH}\simeq 100-10^4~M_\odot$ is left after gravitational collapse of the VMS.
Even after IMBH formation, stellar-mass BHs (SBHs) are migrating to the central region and could form a binary system with the IMBH,
which is so-called intermediate-mass ratio inspirals (IMRIs) \cite{Amaro-Seoane_2010,Fragione_2018}.
GW emission from such IMRI systems can be detected not only by space-borne GW observatories such as LISA
but also by ground-based detectors whose configurations are specifically designed to reach a better sensitivity at lower frequencies (e.g., A+ or KAGRA LF).
For $M_{\rm IMBH}\sim 100-10^3~M_\odot$ and $M_{\rm SBH}\sim 10~M_\odot$, GWs produced from the IMRIs can be detected 
up to a distance of a few Gpc \cite{Shinkai_2017}.
Since IMRIs systems are likely to have high eccentricities due to the formation process, the GW energy distribution is shifted to higher frequencies,
increasing the SNR \cite{Amaro-Seoane_2018}.
Although the IMRI coalescence rate is still highly uncertain, $R\simeq 1-30$ events yr$^{-1}$ would be expected
from several theoretical studies \cite{Shinkai_2017, Fragione_2018}.
In addition, \cite{Gurkan_2006} discussed the possibility that two IMBHs are formed in a massive dense cluster via 
runaway stellar collisions, and would form an IMBBH at a lower rate \cite{Amaro-Seoane_2010}.

Alternatively, formation of VMSs in extremely low-metallicity environments (Pop III stars) 
can initiate IMBHs with masses of $\gtrsim 100~M_\odot$ \cite{Hirano_2014}.
Since a molecular cloud forming massive Pop III stars would be very unstable against its self-gravity,
the cloud would be likely to fragment into massive clumps with $\gtrsim 10~M_\odot$ and leave
a very massive binary system that will collapse into IMBBH with $\sim 100+10~M_\odot$ \cite{Stacy_Bromm_2013, Susa_2014,Inayoshi_Haiman_2014}.
Assuming a Salpeter-like initial stellar mass function and the merger delay-time distribution to be $dN/dt \propto t^{-1}$,
the merger event rate for such IMBBH is inferred as $\sim 0.1-1~{\rm Gpc}^{-3}~{\rm yr}^{-1}$, and thus $R\sim$ a few events yr$^{-1}$ \cite{Kinugawa_2014}.\\

\noindent {\it - Future prospects}

The possible target source that is detectable with ground-based detectors is IMRIs. See the next section~\ref{sec:science-IMRI}.

\subsection{Intermediate mass-ratio binaries}
\label{sec:science-IMRI}

\noindent {\it - Scientific objective}

The existence of IMBHs in globular clusters is suggested both by
observational evidence and theoretical prediction
\cite{Miller:2003sc}.
An IMBH in a globular cluster may capture a stellar-mass compact
object surrounding it through some process (e.g. two-body relaxation),
and may form an intermediate
mass-ratio binaries (IMRB) (where ``intermediate'' mass-ratio
means the range between the comparable mass ratio ($\sim 1$) and
extreme-mass ratio ($\lesssim 10^{-4}$)).

The captured object in an IMBR orbits the IMBH many times during the
inspiral phase before it plunges into the BH. Therefore,
 GW signals from IMRBs contain information on the
geometry around the IMBHs, which can be used to test the general
relativity in the strong field, for example, the
no-hair theorem of BHs and the tidal coupling between the central BH
and the orbit of the captured object \cite{Brown:2006pj}.
Also, if the observation of IMRBs are accumulated, it may give
a constraint on the event rate and population of IMBHs.\\

\noindent {\it - Observations and measurements}

\begin{figure}[t]
 \centering
 \includegraphics[width=0.6\textwidth]{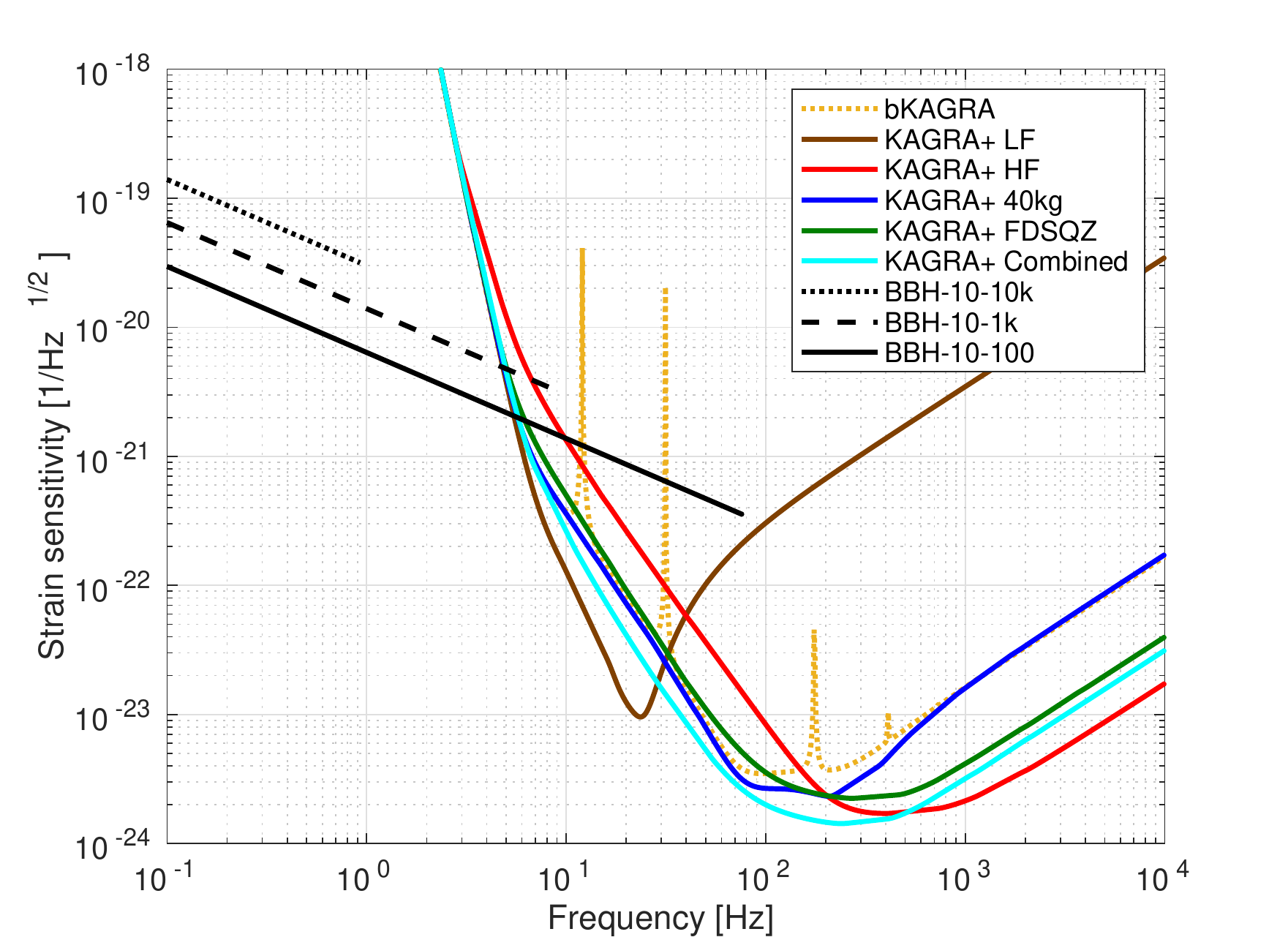}
 \caption{The strain sensitivity of KAGRA and strain amplitude, $2\sqrt{f}|\tilde{h}|$, of GWs from IMRBs.}
 \label{fig:strain-IMRI}
\end{figure}

The GW waveform from the inspiral phase of an IMRB can be
estimated in the stationary phase approximation by
\begin{equation}
\tilde{h}(f) = {\cal A} f^{-7/6} e^{i \Psi(f)},
\quad
{\cal A} =
 \frac{1}{\sqrt{30}\pi^{2/3}} \frac{{\cal M}^{5/6}}{D_L},
\end{equation}
where ${\cal M}$, $D_L$ and $\Psi(f)$ are the redshifted chirp mass of the binary,
the luminosity distance to the source, and the GW phase, respectively
\cite{Berti:2004bd, Isoyama:2018rjb}.
The black (solid, dash, dot) lines in Fig.\ref{fig:strain-IMRI} show
the strain amplitude of GWs, $2\sqrt{f}|\tilde{h}|$, from IMBRs for three
different cases, in which a BH with $10 M_\odot$ spirals into heavier
BHs with masses of $10^2\,M_\odot$, $10^3\, M_\odot$, and $10^4\,M_\odot$ (these masses include the redshift factors)
and the same spin parameter of $0.7$ at the luminosity distance of 100~Mpc.
The inspiral phase ceases around certain cutoff frequency, which
corresponds to the frequency just before the plunge phase.
For circular orbit cases, the cutoff is estimated by the frequency
for an ISCO \cite{Bardeen:1972fi}.
The ISCO frequency, $f_{{\rm ISCO}}$ mainly depends on the mass and
spin of the central BH. If the mass decreases or the spin increases,
the ISCO frequency becomes higher and the overlap with the sensitive
band of ground-based detectors becomes larger.

After the inspiral phase, the orbit of the captured object changes
to the plunge phase through the transition phase. The shift of the
frequency during the transition is roughly estimated by
$\Delta f/f_{{\rm ISCO}} \sim \eta^{2/5}$, where $\eta$ is the mass
ratio \cite{Ori:2000zn}. The plunge of the object to the
central IMBH will induce ringdown GWs, whose frequency of the
dominant mode is given by
$f_{{\rm RD}}\sim 32.3\,\mbox{Hz} \times (m_2/10^3M_\odot)^{-1} g(\chi_2)$,
where $m_2$ and $\chi_2$ are the redshifted mass and spin of the IMBH
and $g(\chi_2)$ shows the spin dependence (see Ref.\cite{Berti:2005ys}
for details).
Since the contribution from the transition phase and the ringdown
phase is smaller than that from the inspiral phase, here we do not
consider them (An analysis including the transition and ringdown
is shown in \cite{Huerta:2010un}).\\

\noindent {\it - Future prospects}

Since IMRBs with the redshifted mass of $10^2$-$10^4M_\odot$ are expected to be
GW sources in $0.4-40$ Hz band \cite{Miller:2008fi},
some of them may be possible targets for the low frequency band of
ground-based detectors.
To see the dependence of the detectability on the sensitivity in
lower band, here we consider the detectable luminosity distance
$D_L^{{\rm detect}}$ for the design sensitivity of the bKAGRA and KAGRA+ in the same way in \cite{Isoyama:2018rjb}.
\begin{figure}[t]
 \includegraphics[width=0.5\textwidth]{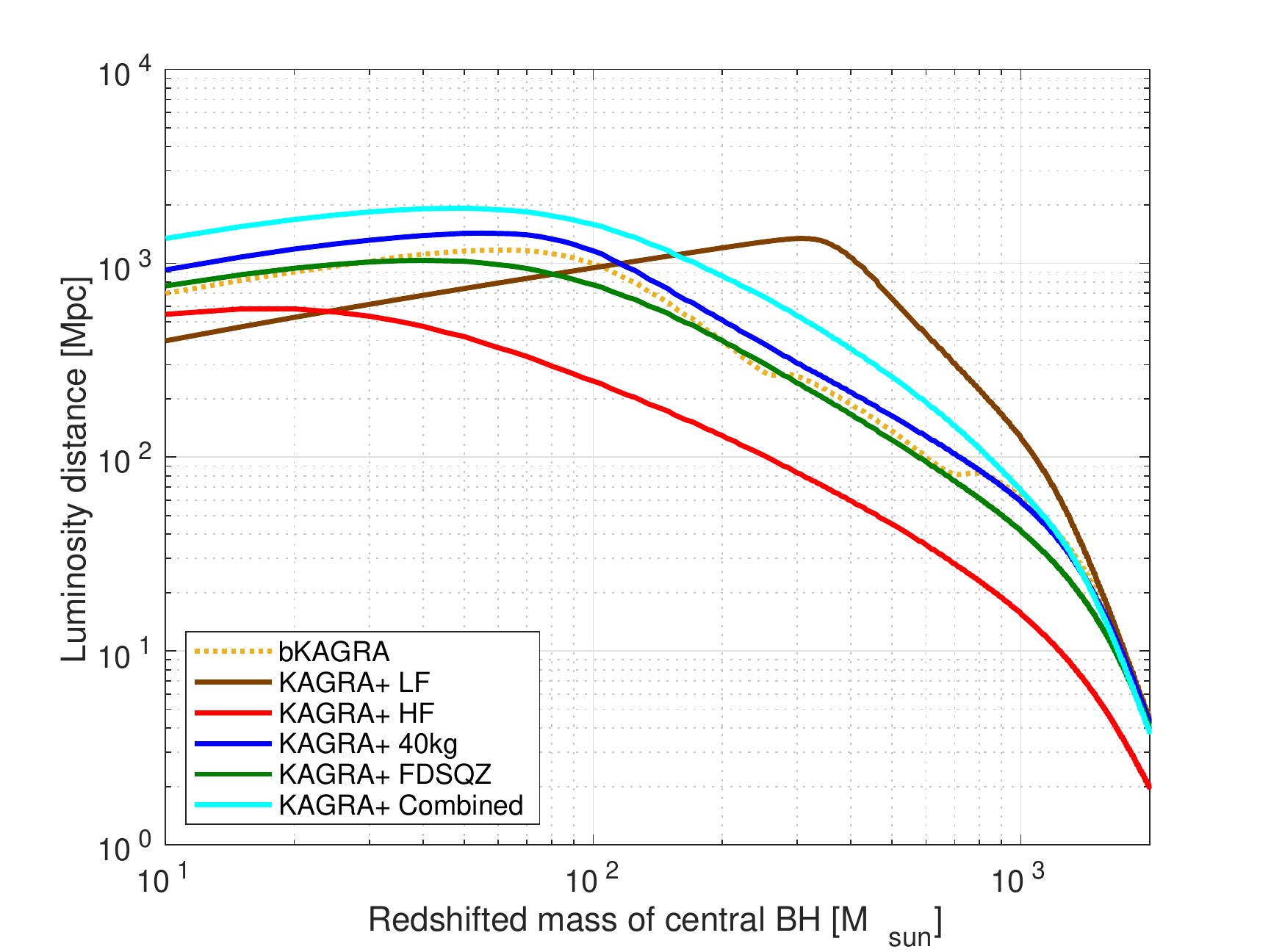}
 \includegraphics[width=0.5\textwidth]{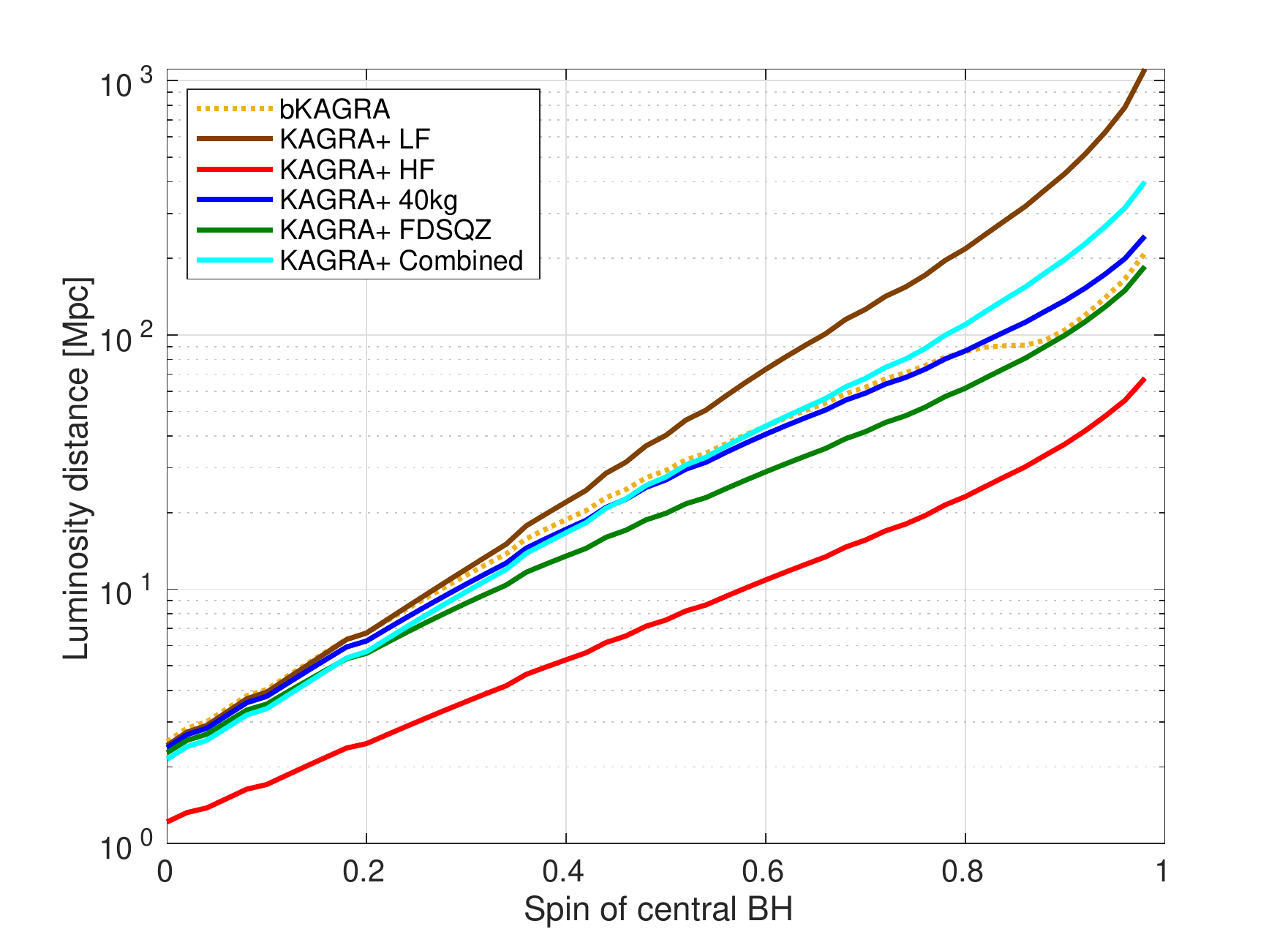}
 \caption{The detectable luminosity distance of IMRIs.}
 \label{fig:DL-IMRI}
\end{figure}
The left panel in Fig.\ref{fig:DL-IMRI} shows the dependence of
$D_L^{{\rm detect}}$ on the redshifted mass of the central IMBH
($m_2$) with fixed spin parameter ($\chi_2=0.7$).
The right panel shows the dependence of $D_L^{{\rm detect}}$ on
the spin parameter with fixed mass ($m_2=10^3M_\odot$).
In both figures, we fix the mass ($m_1=10M_\odot$) and spin parameter
($\chi_1=0$) of the captured object, and the threshold of ${\rm{SNR}}=8$.
For example, the values of $D_L^{{\rm detect}}$ with
$(m_1, m_2)=(10M_\odot, 1000M_\odot)$, $(\chi_1, \chi_2)=(0, 0.7)$
and SNR$=8$ are given as
62, 126, 16, 59, 42, 67\,Mpc for bKAGRA, LF, HF, 40kg, FDSQZ, and Combined, respectively.
From the figures, we find that the detectability of IMBRs with
$\sim 300M_\odot$ can be improved significantly by the LF,
while it is difficult to detect IMBRs by the HF.
To detect an event, the detection volume is a crucial quantity because it is roughly proportional to the event rate.
Using the detector network composed of two A+, AdV+, and bKAGRA or
KAGRA+ (LF, HF, 40kg, FDSQZ, and Combined),
the improvement factors of the detection volume are LF 8.29, 40kg 0.85, FDSQZ 0.30, HF 0.02, Combined 1.27
for $10^3\,M_{\odot}$ - $10\,M_{\odot}$ BBH with spin 0.7.\\

\section{Neutron star Binaries}
\subsection{Binary evolution}
\label{sec:BNS-evolution}

\noindent {\it - Scientific objective}


Formation scenarios of BNS are similar to those of stellar-mass BBH. One is based on the standard binary evolution in a galactic disk (e.g., \cite{Tauris:2017, Chruslinska:2018}) and the other involves dynamical interactions in dense stellar environments such as globular clusters (e.g., \cite{Pooley:2003}). By precision measurements of the NS masses and orbital parameters such as eccentricity by GW observations, it would be possible to shed lights on the origin of the BNS formation as pointed out many previous works on the BNS population including \cite{Tauris:2017}. Although the expected kick involved in the BNS formation is small \cite{Tauris:2017}, the role of the natal kick from a supernova explosion can be relatively more significant for the BNS formation than that of BBH formation. The evidence of the natal kick is observed in the known BNS in our Galaxy, such as the Hulse-Taylor binary pulsar where the eccentricity is 0.67 \cite{WeisbergNiceTaylor}. Even though a binary's orbit could be significantly eccentric at the time of formation, the BNS orbit becomes circularized quite efficiently \cite{Peters} and typical BNS within the detection frequency band of KAGRA is expected to have an eccentricity below $10^{-4}$. Similar to BBH, if there are eccentric NS binaries within the KAGRA frequency band, they are more likely to be formed through stellar interactions in dense stellar environment \cite{BaeKimLee, Rodriguez_2016a}. If eccentric binaries exist, measuring eccentricity would be one useful probe to distinguish the field population (formed by a standard binary evolution scenario) and the cluster population (formed by stellar interactions). We note that the measurement of eccentricity requires a much better sensitivity toward the lower frequency band below 20 Hz, therefore it is more plausible that a third generation detector may be better suited for searching eccentric BNSs.

Another difference between NS and BH populations is the mass range. While BH mass spectrum spans over nine orders of magnitudes, the NS mass range is narrow. For example, the NS mass range should depend on the formation mechanism but is estimated, assuming the Gaussian distribution, to be $1.33\pm 0.09\,M_{\odot}$ for double NSs, $1.54\pm 0.23\,M_{\odot}$ for recycled NSs, and $1.49\pm 0.19\,M_{\odot}$ for slow pulsars, which are likely to be NSs right after their birth \cite{ozelFreire:2016}. These are based on the known NS-pulsar binaries in our Galaxy before 2016 and do not include a possibly heavy NS recently discovered by GWs \cite{Abbott:2020khf}, but even if included, the mass range of NSs is much narrower than that of BHs, which is advantageous of establishing a template bank. In addition, spin distribution would be something to be compared between NS and BH populations. NS spin distribution for those in BNS is not well constrained and we expect GW observations would shed light on the NS spin distribution by parameter estimation.\\

\noindent {\it - Observations and measurements}


As of early 2019, there are fifteen BNS known in the Galactic disk (e.g., see Table 1 in \citep{PML2019}), where all systems consisting of at least one active radio pulsar \cite{ATNFPulsarCatalogue}. Eight of them are expected to merge within a Hubble time. Measurements of binary's orbital decay and other post-Keplerian parameters clearly showed the effects of gravitational radiation in these NS-pulsar binaries.

GW170817 is the first BNS discovered by GW observation. It is also the first extragalactic BNS. The binary is identified as BNS based on the mass estimation, where both $m_1$ and $m_2$ estimates are consistent with expected NS masses \cite{Abbott:2018wiz}. The detection of GW170817 also makes it possible to constrain the BNS merger rate. Considering the first and second observing runs, the merger rate solely based on GW observation is estimated to be 110 - 3840$\,{\rm Gpc}^{-3}\, {\rm yr}^{-1}$ at 90\% confidence interval \cite{GWTC1}. GW observation can observe extragalactic BNS, which are not accessible with current radio telescopes. The GW and radio observation of BNS would be complimentary to reveal underlying properties of the BNS.

In addition to the BNS population, GW observation would be able to provide rich information about neutron star interior (see Secs.~\ref{sec:NS-EOS} and \ref{sec:NS-remnant}) or the formation of young NSs (see also Secs.~\ref{sec:science-stellar-oscillation} and \ref{sec:science-supernova}) if late-inspiral, merger, or even ringdown phases are observed by future detectors.\\


\noindent {\it - Future prospects}


In terms of an observation, improving sensitivities toward lower and higher frequencies have significant implications for understanding astrophysics of BNS. A typical chirp mass of a BNS is $\approx 1.13\,M_{\odot}$ assuming $m_1=m_2=1.3\, M_{\odot}$. This implies the frequencies of merger and ringdown phases would be around 2 kHz where the current generation of GW detectors are not sensitive. However, the merger and ringdown phases of BNS coalescence would provide crucial hints on the remnant of the merger. On the other hand, the lower cutoff frequency of a detector is crucial to measure the early inspiral phase.

Effects of improved sensitivity of a detector can be considered in two folds: (a) At higher frequencies, better sensitivity would allows us to put stronger constrains for the NS EOS than what aLIGO and AdV could do for GW170817~\cite{Abbott:2018exr}, not to mention improving accuracy for mass and spin parameters at higher post-Newetonian (PN) orders. Therefore, better sensitivities in high frequencies will be useful to have the observed sample of BNS via GWs  as complete as possible in the mass-spin parameter space. (b) Improved sensitivity toward the lower frequencies below 20 Hz allows us observing early-inspiral signals. This would be crucial to constrain the orbital eccentricity and determine the origin of the binary formation.

BH-NS binaries are expected to exist and their evolution would be similar to those of stellar-mass BBH and BNS.
As mentioned above, as there are several parameters necessary to be measured in order to discriminate the formation scenarios, we estimate with the Fisher information matrix the measurement errors of the binary parameters for a $10\,M_{\odot}$ BH-NS binary is at $z=0.06$ and a BNS at $z=0.03$. The waveform we use is the spin-aligned inspiral waveform up to 3.5PN in phase. The maximum frequency for the inspiral part is set to the ISCO frequency. We assume the detector networks composed of A+, AdV+, and bKAGRA or KAGRA+ (LF, HF, $40\,{\rm kg}$, FDSQZ, or combined). The measurement errors of the binary parameters are shown in Table~\ref{tab:BHNS-PE-errors} and \ref{tab:BNS-PE-errors}. There are significant improvements from 2G detectors to 2.5G detectors (bKAGRA, LF, HF, 40kg, FDSQZ, Combined). But this is not solely due to KAGRA's contribution. The upgrade of bKAGRA to KAGRA+ among A+ and AdV+ modestly enhances the SNR and the sensitivities to the binary parameters in GW phase: the symmetric mass ratio $\eta$, the effective spin $\chi_{\rm eff}$, and the luminosity distance to a source $d_{\rm L}$. For the parameters in GW amplitude, the orbital inclination angle $\iota$ and the sky localization area $\Omega_{\rm S}$, the improvement of the errors is significant because the fourth detector is important to pin down the source direction and determine the other correlated parameters. The improvement factor depends on the configuration of the upgrade of KAGRA+. From the point of view to discriminate the formation scenarios, it would be better for KAGRA to improve the detector sensitivity at both low and middle frequencies.

\begin{table*}[t]
\begin{center}
\begin{tabular}{lccccccc}
\hline \hline
quantities \;& 2G & bKAGRA & LF & HF & 40kg & FDSQZ & Combined \\
\hline \hline  
SNR & 13.8 & 22.3 & 21.9 & 22.3 & 23.6 & 22.6 & 25.4 \\
$\Delta \log \eta$ & 9.47 & 8.93 & 8.86 & 8.86 & 8.86 & 8.87 & 8.71 \\
$\Delta \chi_{\rm eff}$ & 98.2 & 80.9 & 80.5 & 78.2 & 78.7 & 78.6 & 73.3 \\
$\Delta \log d_{\rm L}$ & 42.6 & 24.6 & 27.0 & 26.7 & 22.0 & 22.8 & 20.8 \\
$\Delta \cos \iota$ & 32.2 & 19.6 & 20.8 & 21.5 & 16.8 & 16.4 & 15.0 \\
$\Delta \Omega_{\rm S}$ & 9.04 & 4.26 & 9.02 & 3.48 & 3.57 & 3.42 & 2.71 \\
\hline 
\end{tabular}
\end{center}
\caption{SNR, median errors of the symmetric mass ratio in the unit of $10^{-2}$, the effective spin $\chi_{\rm eff}$ in the unit of $10^{-2}$, the luminosity distance in per cents, the inclination angle in per cents, sky localization area in ${\rm deg}^2$ for BH-NS binaries. The mass of a NS is $1.4\,M_{\odot}$. The distance of $10\,M_{\odot}$ BH-NS is at $z=0.06$. 2G denotes the detector network composed of the second-generation detectors, aLIGO at Hanford and Livingston, AdV, and bKAGRA. The networks, bKAGRA, LF, HF, 40kg, FDSQZ, and Combined are composed of A+, AdV+, and bKAGRA or KAGRA+ (low frequency, high frequency, $40\,{\rm kg}$, frequency-dependent squeezing, and combined), respectively.}
\label{tab:BHNS-PE-errors}
\end{table*}

\begin{table*}[t]
\begin{center}
\begin{tabular}{lccccccc}
\hline \hline
quantities \;& 2G & bKAGRA & LF & HF & 40kg & FDSQZ & Combined \\
\hline \hline 
SNR & 14.8 & 22.9 & 23.6 & 24.1 & 24.5 & 24.3 & 27.7 \\
$\Delta \log \eta$ & 8.50 & 5.91 & 5.48 & 5.64 & 5.76 & 5.74 & 5.18 \\
$\Delta \chi_{\rm eff}$ & 5.65 & 3.96 & 3.67 & 3.79 & 3.87 & 3.87 & 3.49 \\
$\Delta \log d_{\rm L}$ & 37.1 & 21.5 & 26.4 & 21.4 & 18.6 & 18.7 & 17.4 \\
$\Delta \cos \iota$ & 30.1 & 15.2 & 18.2 & 17.1 & 14.3 & 13.7 & 13.8 \\
$\Delta \Omega_{\rm S}$ & 3.14 & 1.60 & 2.39 & 0.80 & 1.28 & 1.05 & 0.790 \\
\hline 
\end{tabular}
\end{center}
\caption{SNR and the median errors of parameter estimation for BNS at $z=0.03$ as in Table~\ref{tab:BHNS-PE-errors}.}
\label{tab:BNS-PE-errors}
\end{table*}

\subsection{Neutron star equation of state and tidal deformation}
\label{sec:NS-EOS}

\noindent {\it - Scientific objective}

BNS mergers provide us rich information to study nuclear astrophysics.
Since NSs consist of ultra-dense matter \cite{Lattimer:2015nhk}.
In the BNS system, a NS is deformed by the tidal field generated by the companion star in the late-inspiral stage
\cite{Flanagan:2007ix, Hinderer:2007mb, Damour:2012yf}.
The tidal deformability originating from the matter effect encodes the information of NS EOS.
Therefore, the measurement of the tidal deformabilities from GWs of the BNS mergers provides the information about nuclear physics
\cite{Wade:2014vqa, Hinderer:2009ca, Hotokezaka:2011dh}. In this section, we investigate the ability to measure the tidal deformability of NSs with current and future detector networks including bKAGRA and KAGRA+ in addition to the A+  and AdV+. \\

\noindent {\it - Observations and measurements}

During the late stages of the BNS inspiral, 
at the leading order, the induced quadrupole moment tensor $Q_{ij}$ is proportional to 
the external tidal field tensor ${\cal E}_{ij}$ as $Q_{ij}=-\lambda {\cal E}_{ij}$.
The information about the NS EOS can be quantified by the tidal deformability parameter $\lambda=(2/3)k_2 R^5/G$,
where $k_2$ is the second Love number and $R$ is the stellar radius~\cite{Flanagan:2007ix, Hinderer:2007mb}.
The leading-order tidal contribution to the GW phase evolution appears 
through the binary tidal deformability~\cite{Wade:2014vqa}
\begin{eqnarray}
 \tilde{\Lambda} = \frac{16}{13} \frac{(m_1+12m_2)m_1^4\Lambda_1+(m_2+12m_1)m_2^4\Lambda_2}{(m_1+m_2)^5},
\end{eqnarray}
which is a mass-weighted linear combination of the both component tidal parameters,
where $\Lambda_{1,2}$ is the dimensionless tidal deformability parameter $\Lambda=G\lambda [c^2/(Gm)]^5$.
It is first imprinted at relative 5PN order.

The first detection of a GW signal from a BNS system, GW170817 \cite{GW170817PRL},
provides an opportunity to extract information about NSs via measuring the tidal deformability.
The LIGO-Virgo collaboration has placed an upper bound on the tidal deformability as $\tilde{\Lambda} \leq 800$ 
when restricting the magnitude of the component spins \cite{GW170817PRL} 
using the restricted \texttt{TF2} model \cite{Buonanno:2009zt,Blanchet:2013haa}
(this limit is later corrected to be $\tilde{\Lambda} \leq 900$ in Ref. \cite{Abbott:2018wiz}).
In Ref. \cite{Abbott:2018wiz, GWTC1}, the updated analysis by the LIGO-Virgo collaboration by using a numerical-relativity (NR) calibrated waveform model, 
\texttt{NRTidal} \cite{Dietrich:2017aum} have been reported.
By using EOS-insensitive relations among several properties of NSs \cite{Yagi:2016bkt}, 
the constraints on $\tilde{\Lambda}$ can be improved \cite{Abbott:2018exr}
(but see also \cite{Kastaun:2019bxo}).
As found in Refs.~\cite{Abbott:2018wiz, LIGOScientific:2019eut}, the stiffer EOSs (large radii) 
yielding larger tidal deformabilities such as MS1 and MS1b \cite{Mueller:1996pm}
are disfavored by the data of GW170817.
Independent analyses have also been done
by assuming the common EOS for both NSs \cite{De:2018uhw, Capano:2019eae}.
In Ref. \cite{Narikawa:2019xng}, the authors indicate that there is a difference in estimates of $\tilde{\Lambda}$ 
for GW170817 between NR calibrated waveform models (the \texttt{KyotoTidal} and \texttt{NRTidalv2} models).
Here, the \texttt{KyotoTidal} model is one of NR calibrated waveform models of inspiraling BNS 
\cite{Kiuchi:2017pte, Kawaguchi:2018gvj} and \texttt{NRTidalv2} model is an upgrade of the \texttt{NRTidal} model \cite{Dietrich:2019kaq}.
Several other studies have also derived constraints on the NS EOS via measuring tidal deformability from GW170817 \cite{Margalit:2017dij, Bauswein:2017vtn, Ruiz:2017due, Annala:2017llu, Zhou:2017pha, Fattoyev:2017jql, Paschalidis:2017qmb, Nandi:2017rhy, Most:2018hfd, Raithel:2018ncd, Landry:2018prl}. A review of these and other results are available in~\cite{Baiotti:2019sew}.\\

\noindent {\it - Future prospects}

In order to investigate the expected error in the measurement $\tilde{\Lambda}$,
we use the Fisher matrix analysis with respect to the source parameters $\{{\rm ln}d_L, t_c, \phi_c, {\cal M}, \eta, \tilde{\Lambda}\}$, where $d_L$ is the luminosity distance, $t_c$ and $\phi_c$ are the time and phase at coalescence, ${\cal M}$ is the chirp mass, and $\eta$ is the symmetric mass ratio.
We use the restricted \texttt{TF2\_PNTidal} model, which employ
the 3.5PN-order formula for the phase and only the Newtonian-order evolution for the amplitude
as the point-particle part \cite{Buonanno:2009zt,Blanchet:2013haa}
and the 1PN-order (relative 5+1PN-order) tidal-part phase formula (see e.g., \cite{Damour:2012yf}).
We take the frequency range from 23 Hz to the ISCO frequency.
Here, we consider non-spinning binary for simplicity.

In Table \ref{tab:Lambda_multi}, we show the fractional errors in the tidal deformability $\Delta\tilde{\Lambda}/\tilde{\Lambda}$ in the cases of some NS EOS models, using the inspiral PN waveform, the \texttt{TF2\_PNTidal} model.
We assume future detector networks including bKAGRA and KAGRA+ in addition to two A+ and AdV+.
We assume 1.35$M_\odot$-1.35$M_{\odot}$ BNS located at a distance of 100 Mpc and 
study three different NS EOS models~\cite{Takami:2014tva}: APR4~\cite{Akmal:1998cf} ($\tilde{\Lambda}$=321.7), SLy~\cite{Douchin:2001sv} ($\tilde{\Lambda}$=390.2),
and high-$\tilde{\Lambda}$ ($\tilde{\Lambda}$=1000).
In one example, for the SLy, the fractional error on the tidal deformability are 
0.069, 0.070, 0.061, 0.068, 0.067, and 0.062 
for bKAGRA, LF, HF, 40kg, FDSQZ, and Combined, respectively.
We find that the measurement precision of the tidal deformability can be slightly improved for the HF, 
while cannot be done for the LF.
We also present the fractional errors in the tidal deformability for the single detector case in Table \ref{tab:Lambda_single} to see how much KAGRA can contribute to the global networks.

In Ref.~\cite{Narikawa:2018yzt}, the authors have found a discrepancy in the tidal deformability of GW170817 between Hanford and Livingston detectors of aLIGO.
While the two distributions look consistent with each other and also consistent with what we
would expect from noise realization, 
the Livingston data are not very useful to determine the tidal deformability for GW170817.
Their results suggest that the measurement of the tidal deformability
by the third detector such as AdV or bKAGRA might be helpful
in order to improve the constraint on the tidal deformability.
In Ref. \cite{Narikawa:2019xng}, 
their results indicate that the systematic error for the NR calibrated waveform models 
will be significant to measure $\tilde{\Lambda}$ from upcoming GW detections.
It is needed to improve current waveform models for the BNS mergers.

\begin{table*}[!h]
\centering
\caption{\label{tab:Lambda_multi}
The fractional errors in the tidal deformability $\Delta\tilde{\Lambda}/\tilde{\Lambda}$ for a 1.35$M_\odot$-1.35$M_{\odot}$ BNS located at a distance of 100 Mpc, 
for three different NS EOS models: APR4 ($\tilde{\Lambda}$=321.7), SLy ($\tilde{\Lambda}$=390.2),
and high-$\tilde{\Lambda}$ ($\tilde{\Lambda}$=1000).
The networks, bKAGRA, LF, HF, 40kg, FDSQZ, and Combined, are composed of A+, AdV+, and bKAGRA or KAGRA+ (low frequency, high frequency, 40 kg, frequency-dependent squeezing, and combined), respectively.
}

\begin{tabular}{ccccccccc}
\hline
\hline
 &bKAGRA &LF  &HF  &40kg  &FDSQZ &Combined \\
\hline
\hline
APR4 ($\tilde{\Lambda}=321$)  &0.834 &0.0840 &0.0735 &0.0826 &0.0795 &0.0750 \\
SLy ($\tilde{\Lambda}=390$)  &0.0689 &0.0695 &0.0608 &0.0683 &0.0658 &0.0620 \\
High-$\tilde{\Lambda}$ ($\tilde{\Lambda}=1000$) &0.0276 &0.0278 &0.0244 &0.0273 &0.0263 &0.0248 \\
Network SNR &105 &102 &105 &108 &106 &115 \\
\hline
\end{tabular}
\end{table*}

\begin{table*}[!h]
\centering
\caption{\label{tab:Lambda_single}
The same results as Table \ref{tab:Lambda_multi} but for a single detector. 
}
\begin{tabular}{cccccccc}
\hline
\hline
Source &aLIGO &bKAGRA &LF  &HF  &40kg  &FDSQZ &Combined \\ 
\hline
\hline
APR4 ($\tilde{\Lambda}=321$) &0.289 &0.659 &59.7 &0.231 &0.589 &0.270 &0.180 \\ 
SLy ($\tilde{\Lambda}=390$) &0.239 &0.545 &49.3 &0.192 &0.487 &0.223 &0.149 \\ 
High-$\tilde{\Lambda}$ ($\tilde{\Lambda}=1000$) &0.0939 &0.217 &19.3 &0.796 &0.194 &0.0903 &0.0598 \\ 
SNR &33.9 &26.7 &12.4 &27.4 &35.4 &31.4 &53.9 \\ 
\hline
\end{tabular}
\end{table*}

\subsection{Neutron star remnants}
\label{sec:NS-remnant}

\noindent {\it - Scientific objective}

Investigating the evolution of the object that forms after the merger of a BNS system is more difficult, but also, possibly, rewarding (see \cite{bai17, Duez2019, Baiotti:2019sew} for recent reviews). In addition to the connection between such a merged object and ejecta powering the kilonova (which will be treated in detail in Sec.~\ref{sec:kilonovae}), strong interest in observations of the post-merger phase comes also from the fact that these would at some point yield information about the EOS of matter at densities higher (several times the nuclear density) than typical densities in inspiralling stars and also at much higher temperatures (up to $\sim 50$ MeV).

Interestingly, the post-merger phase may be luminous in GW emission,
perhaps more so than the preceding merger phase, but, since post-merger
GW frequencies are higher (from 1 to several kHz [116, 118]), the SNR
for current and projected detectors is smaller than the pre-merger
phase. However, when analyzed in different ways, such GW emission gives
us a novel opportunity to prove for a hypermassive NS (HMNS) in the
immediate post-merger phase (see below in this subsection).
Also gravitational collapse to a rotating BH, should it occur, opens an additional window to emission from a BH-torus system at relatively lower frequencies~\citep{van19b}.

Theoretical estimates from numerical simulations of emission from the post-merger HMNS are less reliable than those of the inspiral, because such simulations are intrinsically more difficult to carry out accurately. This is due to the presence of strong shocks, turbulence, large magnetic fields, various physical instabilities, neutrino cooling, viscosity and other microphysical effects. Currently there exist no reliable determinations of the phase of post-merger gravitational radiation, but only of its spectrum (see, e.g., \cite{Takami:2014zpa, Maione2017, Paschalidis_and_Stergioulas_2017, Breschi2019}).

Measuring even just the main frequencies of the post-merger waveform from the putative HMNS may, however, give precious hints on the supranuclear EOS. It has been found, in fact, that the frequencies of the main peaks of the post-merger power spectrum strongly correlate with properties (radius at a fiducial mass, compactness, tidal deformability, etc.) of a zero-temperature spherical equilibrium star (see, among others, \cite{Bauswein2011, Takami:2014zpa, Maione2017, Paschalidis_and_Stergioulas_2017} and, for applications, \cite{Bose2017, Chatziioannou2017, Torres-Rivas2019, Breschi2019}). For example, finding the tidal deformability of the post-merger object to be very different from that estimated from the inspiral may hint at phase transitions in the high-density matter (see \cite{Baiotti:2019sew} for a review). A similar hint would be given by an abrupt change in the post-merger main frequency \cite{Weih2020, Bauswein2020}. The complicated morphology of these post-merger signals makes constructing accurate templates challenging and thus matched-filtering less efficient \cite{Breschi2019}.

In addition to predictions on the GW spectrum, simulations also show, in most cases, delayed gravitational collapses to a rotating BH with dimensionless spin parameter $a/M\simeq 0.76-0.84$ \citep{bai08}, surrounded by a disc. Strong magnetic fields \citep{rez11} and baryon-rich disk winds are also predicted, though with inferior accuracy, and they are prescient to kilonova light curves \citep{kiu15}. 

However, simulations cannot accurately predict the time of delay in gravitational collapse of the putative HMNS to a BH, representative for the lifetime of the HMNS initially supported against collapse by differential rotation. The lifetime of the HMNS may be measured in EM-GW observations. Furthermore, because of constraints in computational resources, simulations of post-merger BH-disk or torus systems are limited to tens of ms \citep[e.g.][]{bai17,rad18}. Yet, rotating BHs provide a window to powerful emission over a secular Kelvin-Helmholtz time-scale $\tau_{\rm KH}=E_J/L_H$, where $E_J=2Mc^2\sin^2(\lambda/4)$ ~ $\left(\sin\lambda=a/M\right)$ is the spin-energy in angular momentum $J$ of a Kerr BH of mass $M$ \citep{ker63}, $c$ is the velocity of light and $L_H$ is the total BH luminosity most of which may be irradiating surrounding matter \citep{van99}. For gamma-ray bursts, canonical estimates show $\tau_{\rm KH}$ to be tens of seconds \citep[][]{van19b}.

EM-GW observations promise to fill the missing links in our picture of BNS mergers, namely the lifetime of the HMNS in delayed gravitational collapse to a BH \cite[e.g.][]{dep19a,kli19}. If finite, continuing emission may extend over the lifetime of BH spin by $\tau_{\rm KH}$ above. For GW170817, if the HMNS experienced delayed collapse, the merger sequence 
\begin{eqnarray}
\mbox{NS}+\mbox{NS}\xrightarrow[a]{}
\mbox{HMNS}\xrightarrow[b]{}\mbox{BH}+\mbox{GW} + \mbox{GRB170817A}+\mbox{AT2017gfo}
\label{EQN_MS}
\end{eqnarray}
offers a unique window to post-merger EM-GW emission powered by the energy reservoir $E_J$ of the remnant compact object. Crucially, gravitational collapse $b$ to a BH in the process (\ref{EQN_MS}) can increase $E_J$ significantly above canonical bounds on the same at formation $a$ of the progenitor HMNS \citep{hae09} in the immediate aftermath of the merger.

In the following subsection, we will focus only on calorimetric studies of the
post-merger object, leaving the other topics to the above-mentioned
reviews \cite{bai17, Duez2019, Baiotti:2019sew}.\\

\noindent {\it - Observations and measurements}

Post-merger EM-calorimetry on GRB170817A and the kilonova AT 2017gfo \cite{con17,sav17,moo18a,moo18b} 
show a combined post-merger energy in EM radiation limited to ${\cal E}_{\rm EM}\simeq 0.5\%M_\odot c^2$. This output is well-below the scale of total energy output in GW170817, and is insufficient to break the degeneracy between a HMNS or BH remnant. 

Gravitational radiation provides a radically new opportunity to probe a transient source which, ``if detected, promises to reveal the physical nature of the trigger"~\cite{cut02}. GW-calorimetry builds on earlier applications of indirect calorimetry to, notably, PSR1913+105 \cite{hul75} and pulsar wind nebulae \cite{wei78}. Power-excess methods \cite{GW170817:GRB,abb19a} 
have thus far proven ineffective, however, by a threshold of ${\cal E}_{\rm gw}= 6.5\,M_\odot c^2$~\cite{sun19}, given a total mass $M<3M_\odot$ of the BNS progenitor and hence the mass of its remnant compact object.

Independent EM and GW observations corroborate
the process (\ref{EQN_MS}) with a lifetime $t_b-t_a$ of the HMNS of less than about one second \citep{van19b,gil19}. In EM, the lifetime of the HMNS is inferred from an initially blue component in the kilonova AT 2017gfo ~\cite{sma17,pia17,rad18,poo18,gil19,luc19}, whereas the same is inferred from the time-of-onset $t_s$ of post-merger GW-emission in time-frequency spectrograms produced by butterfly filtering, developed originally to identify broadband Kolmogorov spectra of light curves of long GRBs in the BeppoSAX catalogue \cite{van14,van17}. Application to the LIGO snippet of 2048\,s of H1-L1 O2 data covering GW170817 serendipitously shows observational evidence of $\sim 5\,$s post-merger gravitational radiation \cite{van19a}. Subsequent signal injection experiments indicate an output ${\cal E}_{\rm gw}=(3.5\pm1)\%M_\odot c^2$ in this extended emission \citep{van19b}. Its time-of-onset $t_s<1$\,s post-merger falls in the 1.7\,s gap between GW170817 and GRB170817A, satisfying causality at birth of a central engine of GRB170817A. 

${\cal E}_{\rm gw}$ exceeds the maximal spin-energy of a (HM)NS \cite{hae09}, yet it readily derives from $E_J$ of a BH following gravitational collapse thereof at $b$ in the process (\ref{EQN_MS}). Moreover, aforementioned ${\cal E}_{\rm EM}$ is quantitatively consistent with model predictions of ultra-relativistic baryon-poor jets and baryon-rich disk winds from a BH-torus system, the size of which derives from the frequency of its gravitational-wave emission \cite{van19b}.

At current detector sensitivities, witnessing the formation and evolution of the progenitor HMNS in the immediate aftermath of a merger is extremely challenging. Null-results on GW170817 \citep{GW170817:PMS}
are entirely consistent with the energetically moderate and relatively high-frequency ($>1$kHz) spectra expected from numerical simulations \citep{bai17}.\\

\noindent {\it - Future prospects}

Once the BH forms, it has no memory of its progenitor except for total mass and angular momentum. This suggests pursuing GW-calorimetry in (\ref{EQN_MS}) to catastrophic events more generally, including mergers of a NS with a BH companion and nearby core-collapse supernovae (CCSNe), notably the progenitors of type Ib/c of long gamma-ray bursts (GRBs) \cite{van19c}. 
While the latter is rare, the fraction of failed GRB-supernovae that are nevertheless luminous in gravitational radiation may exceed the local GRB rate and hence the rate of mergers involving a NS \cite[e.g.][]{van14}. Blind all-sky butterfly searches may thus produce candidate signals with and without merger precursors from events within distances on par with GW170817, that may be followed up by time-slide correlation analysis and optical-radio signals from existing transient surveys of the local Universe. It also appears opportune to consider directed searches for events in neighboring galaxies \citep[e.g.][]{heo16}, notably M51 ($D\simeq7$Mpc) and M82 ($D\simeq 3.5$Mpc) each offering about one radio-loud CCSN per decade. By their close proximity, such appears of interest also independent of any association with progenitors of GRBs \citep{and13,aas14}.

For a planned KAGRA upgrade, the above suggests optimizing the broadband window of 50-1000\,Hz to pursue GW-calorimetry on extreme transient sources using modern heterogeneous computing, in multi-messenger approaches involving existing and planned high-energy missions such as THESUES \cite[][]{ama18a,stra18a}.
Jointly with LIGO and Virgo, KAGRA is expected to give us a window to modeled and unmodeled transient events out to tens of Mpc. \\

\section{Accreting binaries}
\subsection{Continuous GWs from X-ray binaries}

\noindent {\it - Scientific objective}

There are several classes of continuous GWs (CWs). One major population of CW sources is systems involving a spinning NS that has an asymmetry with respect to its rotation axis \cite{Riles2017}. Potential candidates include pulsars, NSs in supernova remnants, and NSs in binary systems. In particular, an accreting NS in a low-mass X-ray binary is a promising target. In this system, the central NS accretes materials from an orbiting late-type companion star. Finite quadrupole moment of an accreting NS can be possibly induced by various processes such as the laterally asymmetric distribution of accreted material, elastic strain in the crust as well as magnetic deformation \cite{Ushomirsky2000}.

Sco X--1 is the prime target for directed searches of CWs from X-ray binaries because it is the brightest persistent X-ray binary. The source is relatively nearby ($\sim 2.8$ kpc) and its X-ray luminosity suggests that the source is accreting near the Eddington limit for a NS. If the torque-balance (i.e. balance between accretion-induced spin-up torque and the total spin-down torque due to gravitational and EM radiation) can be maintained in the binary system, Sco X--1 is a promising target for CW signals \cite{Bildsten}.

Theoretical understanding of torque-balance and spin wandering effects of X-ray binaries are poorly understood \cite{Bildsten,Mukherjee2018}. By combining CWs from X-ray binaries and multi-wavelength observations, we may track the evolution of spin torques and orbit, shedding light on disk dynamics and maximum rotation frequency of a NS in a binary system \cite{Riles2017}.

No CW signal from X-ray binaries has been observed so far. During LIGO S6, upper limits have been estimated for Sco X--1 and XTE J1751--305 while the O1 observation of Sco X--1 yields a null detection as well \cite{Meadors2017,Abbott2017a,Abbott2017b}. In O2, by using a hidden Markov model (HMM) to track spin wandering, a more sensitive search for Sco X--1 has been performed \cite{Abbott2019scox1}. No evidence of CW can be found in the frequency range of 60 - 650~Hz. An upper limit (95\% confidence) of $h\sim3.5\times10^{-25}$ is placed at 194.6~Hz, which is the tightest constraint has been placed on this system so far \cite{Abbott2019scox1}.\\

\noindent {\it - Observations and measurements}

CW signals from accreting X-ray binaries are extremely small comparing to all known compact binary mergers. Assuming a torque-balance, the GW amplitude can be linked with the X-ray flux presumably associated with the accretion rate \cite{Bildsten}:
\begin{equation}
h \approx 4\times10^{-27} \left(\frac{F_X}{10^{-8} \mathrm{\,erg\, cm}^{-2} \mathrm{s}^{-1}}\right)^{1/2} \left(\frac{\nu_s}{300 \,\mathrm{Hz}}\right)^{-1/2} \left(\frac{R}{10 \,\mathrm{km}}\right)^{3/4} \left(\frac{M}{1.4 M_{\odot}}\right)^{-1/4}
\end{equation}

Sco X--1 is the brightest persistent X-ray binary ($F_X\sim4\times10^{-7}$ erg cm$^{-2}$ s$^{-1}$) and the expected GW signals are of the order of $h\sim10^{-25}$ or smaller. To search for such a weak signal, we need to integrate the data over a long period of time. One major challenge for Sco X--1 is that the spin period is unknown.  We therefore require to search for a broad range of frequencies.  Moreover, because of the spin wandering effect due to changing accretion rates onto the NS \cite{Mukherjee2018}, it is difficult to integrate the signals with a long coherence time (hence better sensitivity). If we could discover the pulsation of Sco X--1 via X-ray observations in the future, it will narrow the parameter space enhancing our chance to make the first detection of CWs from an X-ray binary.

Apart from Sco X--1, there are other luminous X-ray binaries such as GX 5--1 and GX 349+2. However, their X-ray flux is at least an order of magnitude lower than that of Sco X--1, making them even more difficult to be detected with current GW detectors. Although Sco X--1 is still the only promising persistent X-ray binary for CW signals, we may detect CWs from a very luminous outburst of a NS X-ray binary in the future. For example, the X-ray outbursts from Cen X--4 are even brighter than Sco X--1 although they only lasted for about a month \cite{Kaluzienski1980}.\\

\noindent {\it - Future prospects}

In order to optimise the search for CWs from X-ray binaries and Sco X--1 in particular, the best frequencies should be from a few tens to a few hundreds Hz. A stable (with high duty cycle) long-term (months to years) observing run is required to integrate the data for the weak signals of the order of $h\sim10^{-25}$ or smaller. Unless we know the spin frequency of the NS from EM observations, we have to search for a large frequency range implying a high demand of computation time. In any case, better computing algorithms have to be developed along with implementation of GPU computation in order to increase the detection efficiency \cite{Messenger2015,Meadors2018}. If we define the SNR ratio averaged over relevant frequencies, we can quantitatively compare performances of various configurations of possible KAGRA upgrades. We define the ratio $r_{\alpha/\beta} \equiv r_{\alpha}/r_{\beta}$ for the configurations 
$\alpha$ and $\beta$ where 
\begin{align}
r_{\alpha} &\equiv  
\sqrt{\int_{f_{\rm low}}^{f_{\rm high}} \frac{1}{S^{\alpha}_h(f)}df}
\label{eq:SNR-ratio}
\end{align}
and a similar equation holds for the $\beta$ configuration. In the case of an accreting X-ray binary, for a typical frequency range between $f_{\rm low} = 30$~Hz and $f_{\rm high} = 200$~Hz, the SNR ratios of KAGRA+ to bKAGRA are LF 0.04, 40kg 1.62, FDSQZ 1.38, HF 0.89, Combined 2.25.

Recently, a novel methodology of searching CWs by deep learning has been proposed \cite{Dreissigacker2019}. Since the CW is expected to be weak, it is necessary to integrate the data with very long time span to result in SNR above the detection threshold. This poses a computational challenge to the traditional coherent matched-filtering search. First, it is difficult to apply such method on a data span longer than weeks. Moreover, as we do not know the spin period and period derivatives precisely, we have to perform blind search in large parameter space at the same time which makes the problem more computationally demanding. On the other hand, once a deep learning network has been trained, prediction on any inputs can be executed very fast which is very favorable for searching CWs. Although the sensitivity of the first proof-of-principle deep learning CW search is far from being optimal \cite{Dreissigacker2019}, it does demonstrate an excellent ability in generalization. Further investigation on this methodology (e.g. exploring which network architecture is optimal for CW search) can be fruitful.

\section{Isolated neutron stars}
\subsection{Pulsar ellipticity} 
\label{sec:pulsar-ellipticity}

\noindent {\it - Scientific objective}

GWs that last more than $\sim$ 30 minutes, when we search for them using ground-based GW detectors, are susceptible to Doppler modulation due to the 
Earth spin and orbital motion. Those long-lasting GWs are called ``continuous" GWs, or CWs. 

Spinning stars with non-axisymmetric deformations may emit CWs.  
Pulsars may have such non-axisymmetric deformations and are good candidates for GW observatories. 
It is estimated that there are roughly $\sim$ 160,000 normal pulsars and $\sim$ 40,000 millisecond pulsars 
in our galaxy. As of writing this paper, the ATNF pulsar catalogue \cite{ATNFPulsarCatalogue} includes $\sim$ 2800 pulsars among which $\sim$ 480 pulsars 
($\sim$ 17 \%) spin at faster than 10 Hz. While $\sim$ 10 \% of pulsars are in binaries, the fraction increases to  
more than 57 \% when limited to pulsars with  $f_{\rm spin} \ge 10 $Hz. The Square Kilometer Array (SKA) is expected to 
find much more pulsars in near future \cite{Smits_2011}. 

A rapidly spinning NS may emit GWs (roughly) at its spin frequency $f_{\rm spin}$, 
 4/3, and/or twice of it, depending on emission mechanisms. 
If NS emits at $\sim f_{\rm spin}$, it may mean the NS is ``wobbling" (freely precessing). 
Detection of ``wobbling mode"  CW gives information on interaction between crust and fluid core of the NS. 
If a NS has a non-axisymmetric mass quadrupole deformation, it may emit CWs at $2f_{\rm spin}$.
CW frequency can be different from twice the spin frequency estimated from an EM observation,    
if a component producing EM radiations does not completely couple with that producing GWs. 
We sometimes call the $2f_{\rm spin}$ mode ``mountain mode".  ``Mountains" on a star may be due to, {\it e.g.},  
fossil deformations developed during NS formation supported by crustal strain and/or strong magnetic field within the NS, or 
thermal gradient (in case of an accreting NS in a binary).  If both GWs of the wobbling mode and the mountain mode are detected from a single NS, one may be able to determine its mass \cite{OnoEdaItoh}. 
R-mode may be unstable within a young NS and it may emit CWs at $\sim 4f_{\rm spin}/3$. 
Detection of ``r-mode"  CW gives us information on evolution history of the star, as damping time-scale depends on interior temperature. See, e.g., \cite{Paschalidis_and_Stergioulas_2017,Glampedakis_and_Gualtieri_2018,Sieniawska_and_Bejger_2019}, for recent reviews on physics of possible mountain mode, r-mode, and wobbling mode CWs.

Due to the lack of the space, we  
focus on isolated NSs that emit mountain mode CWs in this section. 
LVC has been searching for mountain mode CWs \cite{LVC_CW_O1CW_APJ_2017,LVC_CW_O2CW_APJ_2019}. Readers may be referred 
to \cite{LVC_CW_O2CW_APJ_2019} for a wobbling mode CW search (or more generally, $m=1$ mode CWs) and \cite{LVC_CW_SNR_APJ_2019} for a r-mode search.
Mountain mode CWs typically last for more than observation time $T_0$ ($\sim$ year). 
As such, the detectable CW amplitude $h$ at frequency $f_{\rm gw}$ scales as  
\begin{align}
h &= C \sqrt{ \frac{S_h(f_{\rm gw})}{T_0} } 
\end{align}
where $C$ depends on  a search method and a pre-defined threshold for detection. 
The LSC Bayesian time-domain search for known pulsars \cite{LVC_CW_O1CW_APJ_2017} adopts $C\simeq 10.8$, where 
``known" means their timing solutions as well as their positions on the sky are known to sufficient accuracies  
and there is little, if any, need to search over parameter space in the search. \\

\noindent {\it - Observations and measurements}

A mountain mode CW depends on  strain amplitude $h$, GW frequency $f_{\rm gw}$, its higher order time derivatives, 
direction to the source, inclination angle between the line of sight and the spin angular momentum, GW initial phase, and 
polarization angle. As a persistent source, a single detector can detect and locate a CW source. Since the detector beam pattern changes 
during the observation time, one can test general relativity (GR) by searching for CWs having nontensorial polarizations \cite{LVC_CW_NonTensorialCW_PRL_2018}.   
GW frequency and its higher time derivatives together with $h$ may tell us how the source loses its energy and spin angular momentum. 

The amplitude $h$ of the dominant mountain mode CW depends on  $\ell = m = 2$ mass quadrupole deformation $Q_{22}$, 
distance to the source $r$, and the GW frequency as  
\begin{align}
h &\simeq 1.4\times10^{-27}\left(\frac{Q_{22}}{10^{38} {\rm g\cdot cm^2} }   \right)
\left(\frac{r}{1 {\rm kpc} }\right)^{-1}
\left(\frac{f_{\rm gw}}{100 {\rm Hz} } \right)^2.
\label{eq:Itoh_GW_amplitude}
\end{align}
$Q_{22}$ is related with the stellar ellipticity $\epsilon$ in the literature as  
$\epsilon {\cal I}_3 = \sqrt{8\pi/15}Q_{22}$
where ${\cal I}_3$ is the moment of the inertia of the star with respect to its spin axis.

Roughly speaking, $Q_{22,{\rm max}} \propto \mu \sigma R^6/M$  where $\mu$ is the shear modulus, 
$\sigma$ is the breaking strain of the crust, $R$ is the stellar radius, and $M$ is its mass. As it depends on the stellar radius and mass, 
maximum possible $Q_{22,{\rm max}}$ depends on EOS \cite{Johnson-McDaniel_2013a,Johnson-McDaniel_2013b}. For a normal NS 
\cite{Pitkin_2011},  
\begin{align}
Q_{22,{\rm max}} \simeq 2.4\times 10^{39} {\rm g\cdot cm^2} 
\left(\frac{\sigma}{0.1}\right)
\left(\frac{R}{10{\rm km}}\right)^{6.26}
\left(\frac{M}{1.4 M_{\odot} }\right)^{-1.2}. 
\end{align}
If we find  some ``NS" has $Q_{22}$ much larger than $Q_{22,{\rm max}}$, 
it means that  that particular star may not be ``a normal NS" (or we do not understand 
``the normal NS" good enough to predict $Q_{22,{\rm max}}$).

Real NSs may have $Q_{22}$ much smaller than $Q_{22,{\rm max}}$.
The LIGO-Virgo collaboration
reported upper limits on $Q_{22}$  for more than 200 pulsars, many of which already 
surpassed the theoretical upper limit significantly.  For those pulsars, even if their GWs are detected, 
we may not be able to obtain information on the NS EOS from measurement of $Q_{22}$.\\

\noindent {\it - Future prospects}

Figures of merit for isolated pulsar search may be the $h$ value given by Eq. (\ref{eq:Itoh_GW_amplitude}) for possible $Q_{22}$ values, and 
the so-called spin down upper limit. Many pulsars show spin-downs which indicate the rate of the loss of the stellar rotational energy.
Assuming a part of the rotational energy is radiated as GWs, we can estimate the maximum possible GW amplitude, called the spin down 
upper limit on GW amplitude. 

\begin{align}
h_0^{sd} &=
\left(
\frac{5 \eta G {\cal I}_3 \dot f_{\rm spin}}{2c^3 f_{\rm spin} r^2 }
\right)^{1/2}
\notag\\
&\simeq 8.06\times 10^{-19} \eta^{1/2}
\left(  \frac{{\cal I}_3}{10^{45} {\rm g\cdot cm^2}} \right)^{1/2}
\left( \frac{r}{1 {\rm kpc}} \right)^{-1}
\sqrt{
\frac{ | (\dot f_{\rm spin}/{\rm Hz \cdot s^{-1}}) | }  { (f_{\rm spin}/{\rm Hz} ) } 
}
\end{align}
where this equation assumes that $100 \eta \%$ of energy is radiated as GWs. 

\begin{figure}[htp]
\begin{center} 
 \includegraphics[clip,width=11.0cm]{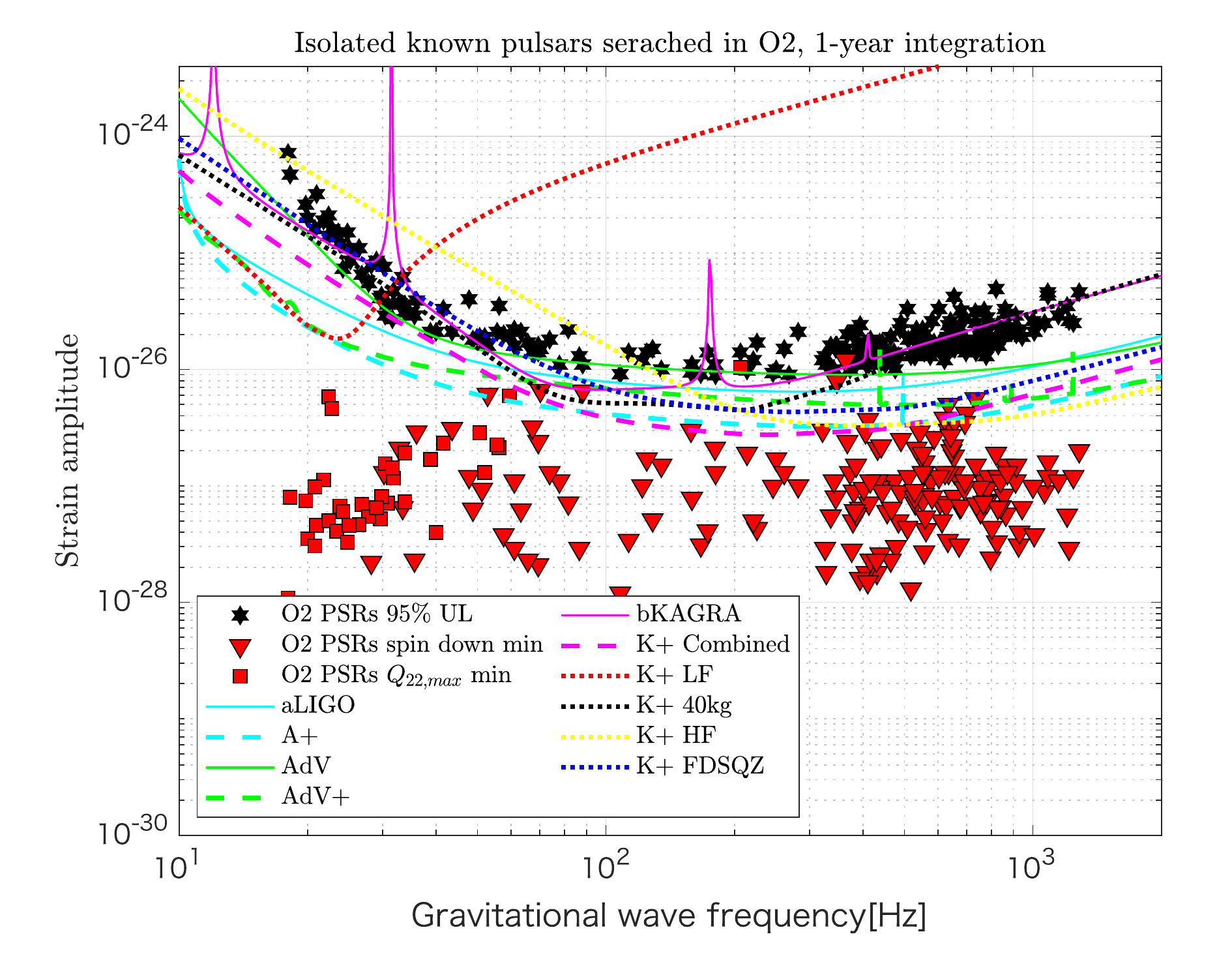}
\end{center}
\caption{The detector sensitivity curves as well as the figures of merit for the pulsars that 
the LIGO-Virgo
collaboration has searched for using O2 data \cite{LVC_CW_O2CW_APJ_2019}.  
The black stars are the 95\% upper limits by the 
LIGO-Virgo
O2 search. 
For each pulsar that 
the LIGO-Virgo collaboration
searched for using O2 data, smaller of the spin-down upper limit or the upper limit assuming 
$Q_{22,{\rm max}}$  is plotted: if the spin-down upper limit is smaller for the pulsar, then an inverted triangle is plotted, if not,  
then a red box is plotted. Coherent 1 year integration is assumed.
}
\label{Fig:Itoh_CW_IsolatedO2Pulsars}
\end{figure}

Figure \ref{Fig:Itoh_CW_IsolatedO2Pulsars} shows the detector sensitivity curves for CW search (assuming 1 year integration, coherent search) as well as the figures of merit for the pulsars that 
the LIGO-Virgo
collaboration has searched for using O2 data \cite{LVC_CW_O2CW_APJ_2019}.  It is clear that KAGRA+ HF is more useful for CW search. 

\begin{figure}[htbp]
\begin{center} 
 \includegraphics[clip,width=10.0cm]{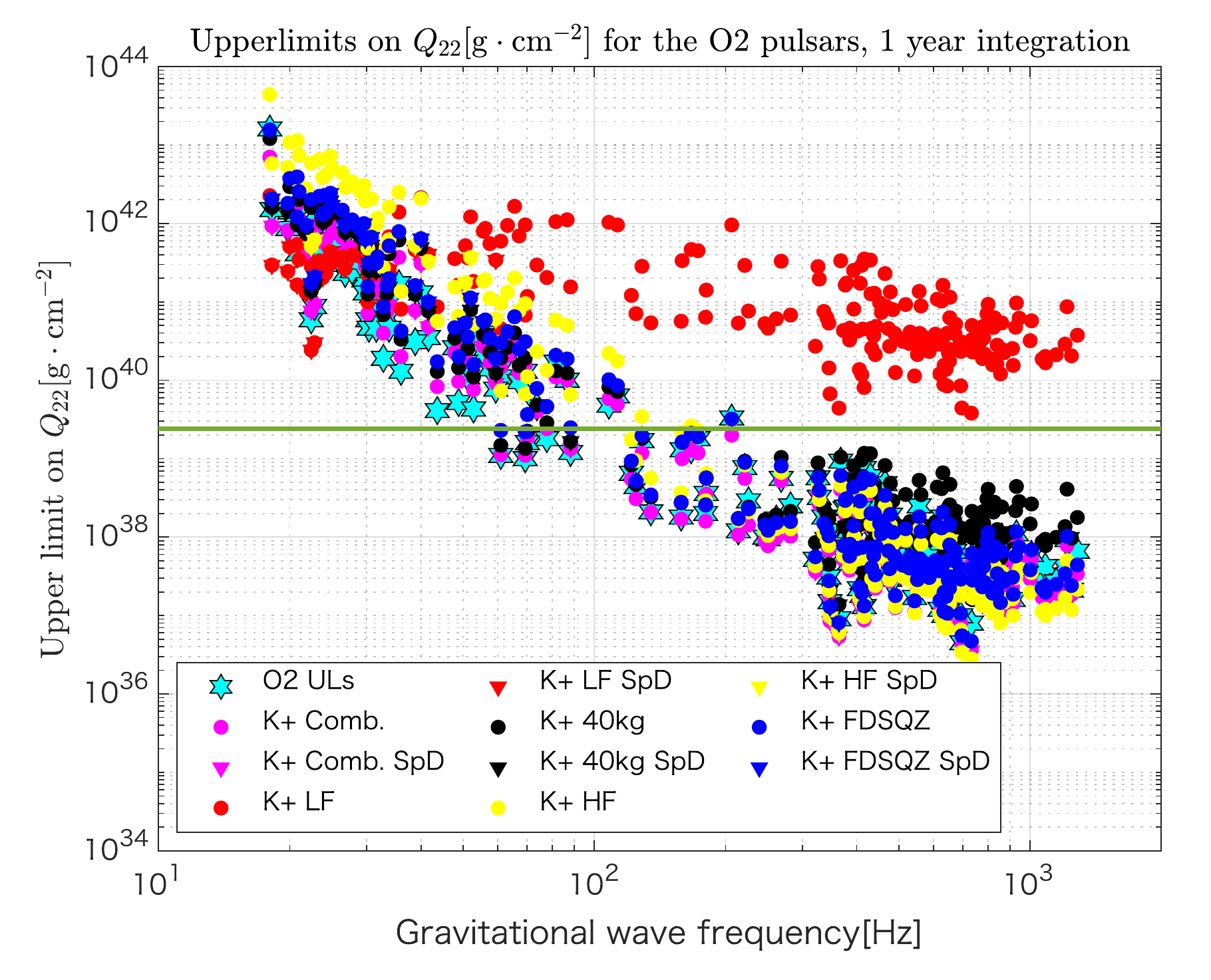}
\end{center}
\caption{The upper limits on $Q_{22}$ that 
LIGO and Virgo
obtained in their O2 search (stars) \cite{LVC_CW_O2CW_APJ_2019}, as well as 
possible upper limits that are expected by using various configurations of KAGRA+ for the same pulsars. 
Coherent 1 year integration is assumed. The horizontal thick line indicates the theoretical maximum for $Q_{22}$ for a normal NS.
LIGO and Virgo have already beaten $Q_{22,{\rm max}}$ at higher frequencies than $\sim 100$~Hz.
KAGRA+ LF does not have any power to explore normal NS CWs.
}
\label{Fig:Itoh_CW_Q22IsolatedO2Pulsars}
\end{figure}

Figure \ref{Fig:Itoh_CW_Q22IsolatedO2Pulsars} shows 
the upper limits on $Q_{22}$ that 
LIGO and Virgo
obtained in their O2 search (stars), as well as 
possible upper limits that are expected by using various configurations of KAGRA+ for the same pulsars.
The horizontal thick line indicates the theoretical maximum for $Q_{22}$ for a normal NS.
The LIGO-Virgo collaboration
has already beaten $Q_{22,{\rm max}}$ at higher frequencies than $\sim 100$Hz.
KAGRA+ LF does not have any power to explore normal NS CWs.

\begin{figure}[htbp]
\begin{center} 
 \includegraphics[clip,width=10.0cm]{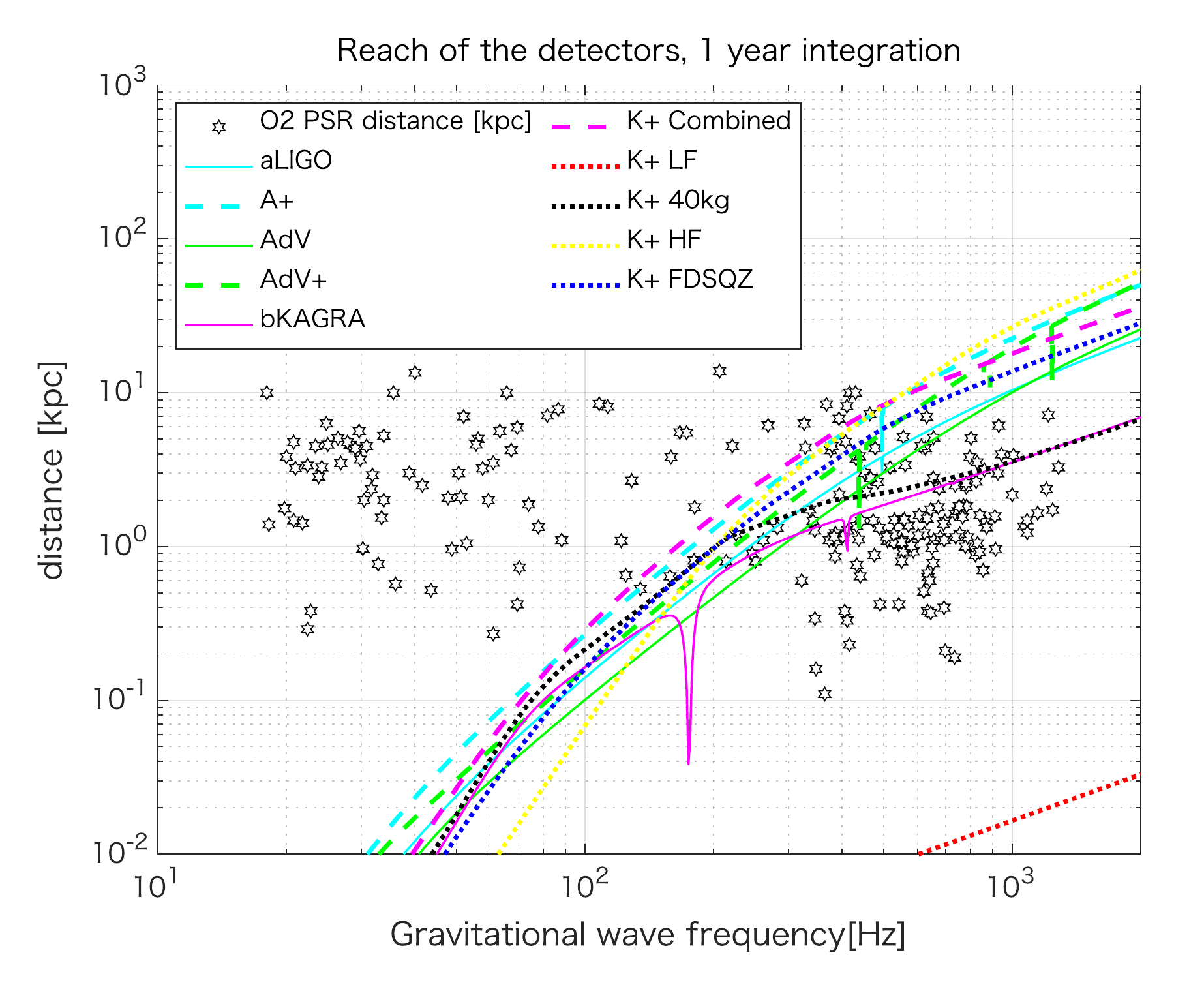}
\end{center}
\caption{
This figure shows possible reaches of detectors of various configurations 
where $\epsilon = 10^{-7}$ (or $Q_{22} = 7.7\times 10^{37} {\rm g\cdot cm^{-2}}$) is assumed. 
Stars indicate the distances estimated for the pulsars that LIGO and Virgo has searched for using O2 data.}
\label{Fig:Itoh_CW_Reach}
\end{figure}

It is possible to conduct unknown pulsar search (e.g., all-sky search, wide-band frequency search). 
In this case, a possible figure of merit is reach of a detector, which is shown for various detector configurations 
in  Fig. \ref{Fig:Itoh_CW_Reach}, assuming coherent 1 year integration (although it is impossible in practice to conduct a fully coherent unknown pulsar search due to 
computational resource limitation). 
This plot again shows that we may be able to have more chance of detection 
if we put more emphasis on higher frequencies.


We can quantitatively compare performances of various configurations of possible KAGRA upgrades 
by introducing the gain in the SNR averaged over relevant frequencies. Namely, we define the ratio $r_{\alpha/\beta} \equiv r_{\alpha}/r_{\beta}$ for the configurations 
$\alpha$ and $\beta$ where $r_{\alpha}$ is defined in Eq.~(\ref{eq:SNR-ratio}) and a similar equation holds for the $\beta$ configuration.
In the case of a continuous wave search, we may be interested in the sensitivity at higher frequency, say, $f_{\rm low} = 500$~Hz and $f_{\rm high} = 1$~kHz where the possible constraints on $Q_{22}$ would be the tightest. For $\alpha=$``KAGRA+" and $\beta=$``bKAGRA",  
$r_{\alpha/\beta}$'s then are 1.04 (``40kg"), 3.27 (``FDSQZ"), 5.89 (``HF"), and 4.53 (``Combined").

\subsection{Magnetar flares and pulsar glitches}
\label{sec:magnetar-flares}

\noindent {\it - Scientific objective}

Giant flares in soft gamma-ray repeaters (SGR) which are a class of strongly magnetized NSs, so-called magnetars, were observed: March 5, 1979 (SGR 0526-66); August 27, 1998 (SGR 1900+14); December 27, 2004 (SGR 1806-20).
Event rate is quite rare, but huge amount of energy is known to be radiated in EM X and gamma bands.
For example, in the last SGR 1806-20 hyperflare, 
the peak luminosity was $10^{47}$ erg/s and total energy 
released in a spike within 1 second was estimated as $10^{46}$ ergs~\cite{2005ApJ...628L..53I}.
Quasi-periodic oscillations were also observed in the burst tail. Their frequencies are 20-2000~Hz, and are identified by seismic modes excited in magnetar flare.
The (quasi)-periodic oscillations are separately discussed in Sec.~\ref{sec:science-stellar-oscillation}. 
The property and mechanism of the flare are not certain at present, owing to limited number of the events. Especially, the energy carried by GWs is unclear.
Theoretical estimate of GW energy associated with flares ranges from 
$\sim 10^{40}$ erg~\cite{2001MNRAS.327..639I,2011PhRvD..83j4014C}  to $\sim 10^{49}$ erg~\cite{2011MNRAS.418..659L,2012PhRvD..85b4030Z}.
Dipole field-strength derived by stellar spin and its time-derivative is normally used for the energy deposited before a flare, but much stronger fields might be hidden inside. Thus, the problem will be solved only by actual observations due to many uncertain factors.
Actually, GW energy at the SGR 1806-20 hyperflare was limited by LIGO. At that time, one interferometer at Hanford was operated.
Upper limit on the GW amplitude is $ 5\times 10^{-22}$~Hz$^{-1/2}$ and GW energy is less than $8\times 10^{46}$ erg~\cite{Abbott:2007zzb}. 
It is very important to further proceed the similar argument to each flare observation.

Another interesting astrophysical event is a pulsar glitch. 
Sudden change in spin angular frequency is 
$\Delta \Omega /\Omega \sim 10^{-6}$ in a typical giant glitch and 
rotation energy changes by 
$I \Omega^2/2 \times \Delta \Omega /\Omega \sim 10^{41}$ ergs 
as for a maximum value.
The change is likely to be much smaller, since the rearrangement of stellar structure
is partial in the transition.\\

\noindent {\it - Observations and measurements}

These burst events will be probably first detected in EM signs, and GW signals should be carefully searched at the burst epoch.
There is no reliable waveform, so that GW signals are identified by combinations of multiple detectors. 
The burst is a dynamical event in a timescale of millisecond, and 
therefore a sharp peak at kHz range is expected.
Coincidence among different GW/EM detectors is crucial.
The observational strategy of GW analysis may depend on the EM information. 
When QPO frequencies are identified in EM data, we should search for GW counterparts in the same frequency range.
GW signals in LIGO O2 data\cite{2019ApJ...874..16} were searched for magnetar bursts , which are less energetic events. Two software packages, X-Pipeline and STAMP were used: Clusters of bright pixels above a threshold in each time-frequency map are searched in the X-Pipeline, and directional excess power due to arrival time-delay between detectors in the STAMP.\\

\noindent {\it - Future prospects}

As for the future extension of bKAGRA, improvements in a higher frequency band are favored for these events. If we fix the frequency of the signal to 1 kHz, SNR ratios of KAGRA+ to bKAGRA at 1 kHz are 40kg 1.03, FDSQZ 3.98, HF 7.47, Combined 5.01. However, it is not clear at present to predict the detection level of meaningful signals by improvements. We will catch them or upper limits by chance, when bursts will happen. The stable operation is desirable.

\subsection{Stellar oscillations} 
\label{sec:science-stellar-oscillation}

\noindent {\it - Scientific objective}

In order to observationally extract the properties of astronomical objects, asteroseismology is a very powerful technique, where one could get the information about the objects via their specific frequencies. This is similar to seismology in the Earth and helioseismology in the Sun. For this purpose, GWs must be the most suitable astronomical information, owing to their high permeability. In fact, several modes in GWs are expected to be radiated from the compact objects~\cite{KS99}, and each mode depends on different aspect of internal physics. The GWs have complex frequencies, because the GWs carry out the oscillation energy, where the real and imaginary parts correspond to the oscillation frequency and damping rate, respectively. So, by identifying the observed frequencies (and damping times) of GWs to the corresponding specific modes, one would get the physics behind the phenomena. In general, the GWs are classified into two families with their parity. The axial parity (toroidal) oscillations are incompressible oscillations, which do not involve the density variation, while the polar parity (spheroidal) oscillations involve the density variation and stellar deformation. Thus, from observational point of view, the polar type oscillations are more important, although those are coupled with the axial type oscillations on the non-spherical background.\\

\noindent {\it - Observations and measurements}

Some of the GWs from NSs are classified in the same way as the oscillations of usual (Newtonian) stars. That is, the fundamental ($f$) and the pressure ($p_i$) modes are excited as acoustic oscillations of NSs, while the gravity ($g_i$) modes are excited by the buoyancy force due to the existence of the density discontinuity or composition gradients inside the star. The Rossby ($r$) modes are also excited by the Coriolis force in the rotating stars, although they are axial parity oscillations. In addition to these modes associated with the fluid motions, the oscillations related to the spacetime curvature are also excited, i.e., the so-called GW ($w_i$) modes, which can be discussed only in the relativistic framework. The typical frequencies of $f$, $p_1$, $g_1$, and $w_1$ modes for standard NSs are around a few kHz, higher than $4-7$ kHz, smaller than a hundred Hz, and around 10 kHz, respectively~\cite{KS99}. With these specific modes from the NSs, it has been discussed to extract the NS information via the direct detection of GWs.

One of the good examples of the GW asteroseismology may be the results shown by Andersson and Kokkotas~\cite{AK96,AK98}, where they found the universal relations for the frequencies of the $f$ and $w_1$ modes as a function of the average density of star and the compactness, respectively, almost independently of the EOS for NS matter. This is because the frequency of $f$ mode, which is an acoustic wave, should depend on the crossing time inside the star with the sound velocity, while the frequency of $w$ mode is strongly associated with the strength of the gravitational field. So, if one would simultaneously detect two modes in GWs from the NS, one could get the average density and compactness of the source object, which tells us the NS mass and radius. On the other hand, $g$ mode GWs may tell us the existence of the density discontinuity due to the phase transition~\cite{F87,STM01,MPBGF03,SYMT11}, or they may become more important in the newborn NSs or protoneutron stars (PNSs)~\cite{RG92,FGP07,PAH16}. The nonaxisymmetric $r$ mode oscillations may also be interesting due to their instability against the gravitational radiations~\cite{A98,FM98}. That is, for highly rotating case, the GW amplitude can exponentially grow up due to the instability, which may become a detectable signal. Eventually, GW amplitude would be suppressed via nonlinear coupling~\cite{AFMSTW03}. 
We note that even the $f$ mode oscillations may become unstable in rapidly rotating relativistic NSs~\cite{GGKZ11}. In any case, the detection of GWs from the isolated (cold) NS may be quite difficult with the current GW detectors, because most of their frequencies are more than around kHz.

On the other hand, the detection of GWs from the PNSs would be more likely if they are formed within our galaxy. This is because, since the PNSs are initially less massive and larger radius, i.e., their average density is smaller than the cold NSs, one can expect the frequency of $f$ mode GW would be smaller than the typical one for cold NSs. In practice, the $f$ mode frequency of PNS is around a few hundred Hz in the early phase after core-bounce~\cite{ST16}. However, the studies of the asteroseismology in PNSs are very few, unlike cold NSs. This is because of the difficulty for providing the background models. That is, the cold NSs can be constructed with the zero-temperature EOS, i.e., the relation between the energy density and pressure, while one has to consider the distribution of the electron fraction and entropy per baryon together with the pressure distribution as a function of density to construct the PNS models. But, such distributions can be determined only after the numerical simulation of core-collapse supernovae. Even so, the study of asteroseismology in PNSs is becoming possible with the numerical data of simulation. 

Up to now, two different approaches are proposed for providing the PNS models, i.e, one is that the PNS surface is defined with a specific density around $10^{10-12}$ g/cm$^3$ \cite{ST16,SKTK17,MRBV18,SKTK19,Sotani19}, and the other is that the oscillations inside the shock radius are considered \cite{TCPF18,TCPOF19}, where the different boundary conditions should be adopted according to the approaches. In contrast to the cold NSs, matter widely exists outside the PNSs, which makes difficult to treat the boundary of the background models. Anyway, even with either approach, it seems that the mode frequencies of GWs are expected to range in wide frequencies and that such frequencies would change with time. If one could identify the frequencies of GWs with a specific mode, it must be very helpful for understanding the physics of PNS cooling via understanding of the PNS properties. In practice, the time evolution of the identified frequencies is considered as a result of the changing of the PNS mass and radius, i.e., the mass increases due to the accretion and the radius decreases due to the both of a relativistic effect and neutrino cooling, with time. So, in the same way as for the cold NSs, if one could simultaneously identify the frequencies with the $f$ and $w_1$ modes GWs from PNSs, one would get the information of the average density and compactness at each time, which enables us to know the evolution of mass and radius of PNSs \cite{SKTK17}. That is, unlike the cold NSs, in principle one can expect to make a severe constraint on the EOS even with one event of GW detection. It should be emphasized that the frequencies of $w_1$ mode GWs from the PNSs becomes typically around a few kHz, because the compactness is smaller than that of cold NSs.\\

\noindent {\it - Future prospects}

\begin{figure}[t]
\begin{center}
\includegraphics[scale=0.7]{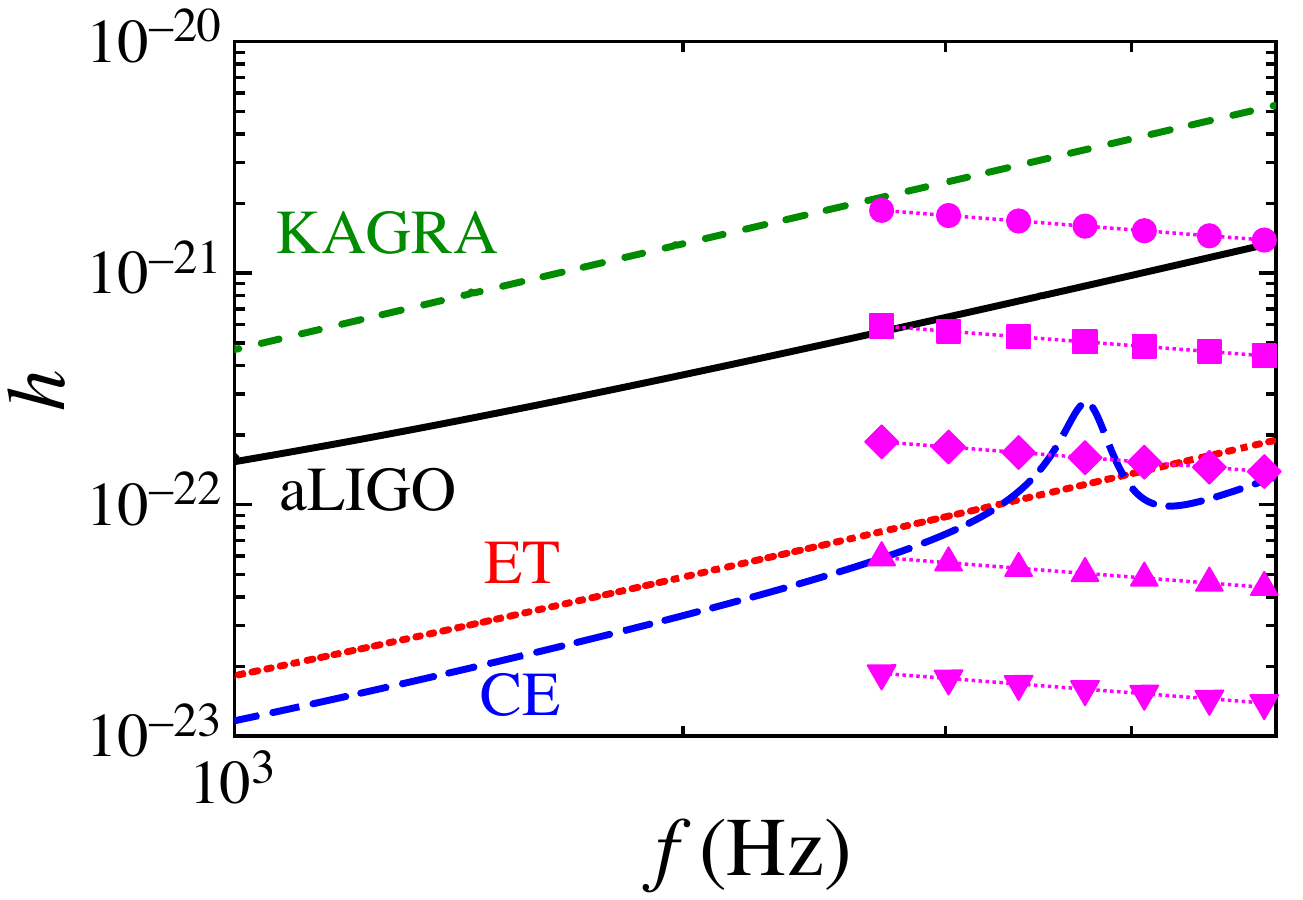}
\end{center}
\caption{
Effective amplitude of $w_1$-mode in GWs from the PNS model constructed with SFHx EOS is shown in the sensitivity curves of KAGRA, AdVanced LIGO (aLIGO), Einstein Telescope (ET), and Cosmic Explorer (CE). Taken from Ref.~\cite{SKTK17}.
}
\label{fig:w-sensitivity}
\end{figure}

We make a comment about the detectability of these modes in GWs. The GW amplitudes have been discussed in Ref. \cite{AK96,AK98} for isolated cold NSs, assuming the radiation energy of GWs with each mode. Since the radiation energy with each mode can not be estimated as far as one examines the GW asteroseismology with linear perturbation analysis, one should eventually estimate such radiation energy with numerical simulation to see the detectability. The situation in PNSs born after the core-collapse supernova is the same. Even so, adopting the formula shown in Ref. \cite{AK96,AK98}, the effective gravitational amplitude of $w_1$ mode from the PNSs may be estimated with its energy, $E_{w_1}$, at each time step as
\begin{equation}
  h_{\rm eff}^{(w_1)} \approx 7.7\times 10^{-23} \left(\frac{E_{w_1}}{10^{-10} M_\odot}\right)^{1/2}\left(\frac{4\ {\rm kHz}}{f_{w_1}}\right)^{1/2}\left(\frac{10\ {\rm kpc}}{D}\right),
\end{equation}
where $f_{w_1}$ and $D$ denote the frequency of considering $w_1$ mode and the distance to the source object, respectively \cite{SKTK17}. The total radiation energy of the $w_1$ mode in the GWs from PNS, $E_{w_1}^T$, may be estimated as $E_{w_1}^T\approx E_{w_1} T_{w_1}/\tau_{w_1}$, where  $\tau_{w_1}$ denotes the damping time of $w_1$ mode with the frequency of $f_{w_1}$ at each time step. In this estimation, we assume that $w_1$ mode GW emission would last for the duration of $T_{w_1}$, as the frequency of $f_{w_1}$ evolves, where we simply adopt $T_{w_1}=250$ ms. Then, the effective amplitude is plotted as in Fig.~\ref{fig:w-sensitivity}, where the circles, squares, diamond, triangles, and upside-down triangles correspond to the expectations with $E_{w_1}^T/M_\odot=10^{-4}$, $10^{-5}$, $10^{-6}$, $10^{-7}$, and $10^{-8}$, respectively.

To detect the GW signal due to the NS oscillations, improvements in a higher frequency must be more suitable, although the amount energy of GWs are unknown at present. If we fix the frequency of the signal to 1 kHz as in the previous section~\ref{sec:magnetar-flares}, the SNR ratios of KAGRA+ to bKAGRA at 1 kHz are 40kg 1.03, FDSQZ 3.98, HF 7.47, Combined 5.01. The improvements even in a middle frequency band may be suitable for detecting the $f$-mode GWs from the PNS, but it must be acceptable only in the early phase after the core-bounce, because the frequency increases with time and eventually becomes $\sim$ kHz.\\

\section{Supernovae}
\subsection{The explosion mechanisms and the GW signatures}
\label{sec:science-supernova}

\noindent {\it - Scientific objective}

Accumulating observational evidence from multi-wavelength EM observations such as of ejecta morphologies and 
spatial distributions of heavy elements have all pointed 
toward CCSNe being 
generally aspherical (e.g., \cite{casA,tanaka17} and references therein).
However, these signals are secondary for probing 
the multidimensionality of the supernova engine
  because they are only able to provide images of optically 
  thin regions far away from the central core. The GWs from CCSNe are the
  primary observables, which imprint a live information of the central engine.


From a theoretical point of view, understanding the origin of the explosion
 multidimensionality is indispensable for
 clarifying the yet uncertain CCSN mechanisms. 
After a half-century of continuing efforts, 
theory and neutrino radiation-hydrodynamics simulations 
are now converging to a point that multi-dimensional (multi-D)
 hydrodynamics instabilities play a pivotal role in the 
neutrino mechanism (\cite{bethe}), the most favoured scenario 
to trigger explosions.
In fact, self-consistent 
simulations in three spatial dimensions (3D)
 now report shock revival of the stalled bounce shock into explosion
 by the multi-D neutrino mechanism
(see \cite{bernhard16,janka16,burrows13,Kotake12} for recent reviews).\\ 

\noindent {\it - Observations and measurements}


The most distinct GW emission process commonly seen in
 recent self-consistent three-dimensional (3D) models is associated with
 the excitation of core/PNS oscillatory modes
(e.g., \cite{radice18,andresen18,andresen17,kuroda16}). 
The dominant
 modes are the quadrupolar ($\ell =2$) deformations of the PNS excited by
  the non-spherical mass accretion onto it (see also Sec.~\ref{sec:science-stellar-oscillation}). The characteristic GW frequency increases
almost monotonically with time due to an accumulating
accretion onto the PNS, which ranges from $\sim 500$ to 1000Hz (see \cite{astone18}
 for the dedicated scheme to detect the ramp-up GW signature from the noises).

 In addition to the prime GW signature,
the Standing Accretion Shock Instability (SASI) (e.g., \cite{thierry15})
has been also considered as a key to generate GW emission
in the postbounce supernova core. It has become apparent \cite{radice18,andresen18,kuroda16} that 
the predominant GW emission from CCSNe is produced in the same
 region behind the postshock region where both the SASI and neutrino-driven
 convection vigorously develop 
(e.g., \cite{Kotake13} for a review). The GW emission from this region--from
just above the neutrinospheres at densities of roughly 
10$^{11}$ g/cm$^3$ and inward--is determined by the effects of the super-nuclear EOS,
as well as the impact of 
hydrodynamic instabilities present in the post-shock flow (e.g., \cite{andresen17}). 
The typical frequency of the SASI-induced GWs is in the range of $\sim$100 -- 250 Hz,
which is in the best sensitivity range of the currently running interferometers.

Rapid rotation in the iron core
leads to significant rotational flattening of the collapsing and bouncing core,
 which produces a time-dependent quadrupole (or higher) GW emission (e.g.,
 \cite{ott_rev} for a review).
For the bounce signals having a strong and characteristic 
signature, the iron core must rotate rapidly enough\footnote{Although rapid rotation
  and the strong magnetic fields in the core
 is attracting great attention as the key to solve
  the dynamics of collapsars and magnetars, one should keep in mind that recent
 stellar evolution calculations predict that such an extreme condition can be
  realized only in a special case \cite{woos06} ($\lesssim$ 1\% of massive star population).}. 
 The GW frequency associated with the rapidly rotating collapse and bounce is in
 the range of $\sim 600 - 1000$~Hz.




Another distinguished feature of GWs from a CCSN is a circular polarization. The importance of detecting circular GW polarization from CCSNe was first discussed
 by \cite{Hayama16} in the context of rapidly rotating core-collapse.
 More recently \cite{Hayama18} presented analysis of the GW
 polarization using results from 3D general-relativistic simulations of a non-rotating
 $15 M_{\odot}$ star \cite{kuroda16}.
 The amplitude of the GW polarization was shown to become larger for 
 the 3D general-relativistic model that exhibits strong SASI activity. 
 The non-axsymmetric flows associated with spiral SASI give rise to the circular polarization.
If detected, it will provide us a new probe into the SASI activity in the preexplosion supernova core.
 Note that the spiral SASI develops mainly outside of the PNS, so that the circular polarization from the
spiral SASI has different features in the spectrogram comparing to the stochastic, weak circular polarization
 from the PNS oscillations that have no preferential direction to develop \cite{Hayama18}.
 By estimating the SNR of the GW polarization, 
they pointed out the possibility that the 
detection horizon of the circular polarization could extend by more than a 
factor of several times farther comparing to that of the GW amplitude.  
In order to detect the GW circular polarization with a high SNR, the joint GW observation
 with KAGRA in addition to two LIGOs and Virgo is indispensable.\\

\noindent {\it - Future prospects} 

Each phase of a CCSN has a range of characteristic GW signatures and
 the GW polarization that can provide diagnostic constraints on the evolution
 and physical parameters of a CCSN and on the explosion hydrodynamics.
 The GW signatures described in this section commonly appear in the frequency range of $100$ - $1000$ Hz in recent simulations of the CCSN.
 Among them, the PNS oscillatory modes are currently recognized as the model-independent
 GW signature.
The characteristic frequencies of the fundamental ($f/g$) modes are predominantly dependent on the PNS mass ($M$) and the radius ($R$), however, in a non-trivial manner \cite{sotani17}.
In order to break the degeneracy, the detection of other eigen-modes ($p,w$ modes)
of the PNS oscillations is mandatory. In fact,  the GW asteroseismology of the PNS
has just started \cite{forne19,morozova18,sotani17} (see also Sec.~\ref{sec:science-stellar-oscillation}). The outcomes of which should reveal
the requirement of KAGRA (and the next-generation detectors) to detect the whole of the
eigenmodes (presumably extending up to several kHz) for the future CCSN event. This could be done in the next five years.

For the detectability of GWs from PNSs, see Sec.~\ref{sec:science-stellar-oscillation}. The GWs due to SASI for a Galactic event may be detected with the SNR of the GW signals predicted
in the most recent 3D CCSN models in the range of $\sim$ 4 -- 10 for the aLIGO \cite{andresen17}. With the third-generation detectors online (e.g., Cosmic Explorer), these signals would be never missed for the CCSN events throughout the Milky way. A current estimate of SNR for the GWs from rapidly rotating collapse and bounce based on predictions from a set of 3D models using a coherent network analysis
 shows that these GW signals could be detectable up to
 about $\sim 20$~kpc for the most rapidly rotating model (\cite{hayama15}, see also \cite{powell19,gossan16}).
If the spectra of the supernovae are assumed to be flat in the frequency range of 100 Hz - 1 kHz, we can quantitatively compare performances of various configurations of possible KAGRA upgrades by defining the SNR ratio $r_{\alpha/\beta} \equiv r_{\alpha}/r_{\beta}$ for the configurations 
$\alpha$ and $\beta$ where $r_{\alpha}$ is defined in Eq.~(\ref{eq:SNR-ratio}) and a similar equation holds for the $\beta$ configuration. For the frequency range between $f_{\rm low} = 100$~Hz and $f_{\rm high} = 1$~kHz, the SNR ratios of KAGRA+ to bKAGRA averaged over relevant frequencies are 40kg 1.51, FDSQZ 2.35, HF 3.43, Combined 3.55. However, since the spectral evolution of the CCSN GWs has not been completely clarified yet and it does depend upon the details of CCSN modeling (such as on the initial rotation rates and the progenitor masses), it is too early to make a quantitative discussion about the best-suited sensitivity curves for KAGRA+. Because recent CCSN simulations indicate that the important GW frequency to unveil the physics of CCSNe will be $\geq 100$Hz, all the candidate upgrades except for LF should help enhance the chance of detection.\\

\section{The early Universe}
\subsection{GWs from inflation}

\noindent {\it - Scientific objective}

Inflation, an accelerated expansion phase of the very early Universe,
was proposed in the 1980s to solve problems in the standard model of
the Big Bang theory, such as the horizon, flatness, and monopole
problems \cite{inflation1,inflation2,inflation3,inflation4}. A key
feature of inflation is that the rapid expansion stretches quantum
fluctuations in a scalar field to classical scales and generates
primordial density fluctuations
\cite{scalar1,scalar2,scalar3,scalar4,scalar5}, which become the seeds
of galaxies. The nearly scale invariant spectrum of scalar
perturbations, predicted by the inflationary theory, is strongly
supported by cosmological observations of the cosmic microwave
background and galaxy surveys. The same mechanism also applies to
quantum fluctuations in space-time, which produce a stochastic
background of GWs over a wide range of frequencies
with a scale-invariant power spectrum \cite{tensor1,tensor2,tensor3}. The detection of inflationary GWs is a key to test
inflation and distinguish a number of models since the amplitude and
spectral index strongly depend on the model. \\

\noindent {\it - Observations and measurements}

The amplitude of the GWs is typically far below the sensitivity of the ground-based detectors. The dimensionless parameter, $\Omega_{\rm GW}$, which describes the energy density of GWs per logarithmic frequency
interval, is roughly given by $\Omega_{\rm GW}\simeq 10^{-15}(H_{\rm
  inf}/10^{14}{\rm GeV})^2$ for a scale invariant primordial spectrum
$\Omega_{\rm GW}\propto f^0$, where $H_{\rm inf}$ is the Hubble
expansion rate during inflation. Inflation predicts a slightly
red-tilted spectrum, which further reduces the amplitude at high
frequency. For example, the model predicts $\Omega_{\rm GW}\sim
10^{-16}$ at $100$Hz for chaotic inflation and $\Omega_{\rm GW}\sim
10^{-17}$ for $R^2$ inflation, while the cross correlation between
LIGO, Virgo, and KAGRA 
with their final design sensitivities is expected to reach a stochastic background of $\Omega_{\rm GW}\sim 10^{-9}$. 
For details on the shape of the spectrum, see \cite{Kuroyanagi:2008ye,Kuroyanagi:2014qaa}. 

Note that a search for a stochastic GW background is performed by cross-correlating data from different detectors over a long time period. The sensitivity is determined by noise curves of the multiple detectors as well as the observation time. Since the two detectors of LIGO have advantages in both sensitivity and observation time, the sensitivity of the four-detector network $\Omega_{\rm GW}\sim 10^{-9}$ will be dominated by LIGO detectors, while KAGRA with the current design sensitivity would not contribute to increase the overall sensitivity.
Currently LIGO-Virgo O2 data provide an upper limit of $\Omega_{\rm GW} < 6.0 \times 10^{-8}$ for a scale invariant spectrum \cite{LIGOScientific:2019vic}.

Although the ground-based detectors are not sensitive enough to detect standard inflationary GWs, some models going beyond the standard picture
predict a blue-tilted spectrum, e.g, string gas cosmology
\cite{Brandenberger:2006xi}, super-inflation models
\cite{Piao:2004tq,Baldi:2005gk}, G-inflation \cite{Kobayashi:2010cm},
non-commutative inflation \cite{Calcagni:2004as,Calcagni:2013lya},
particle production during inflation
\cite{Cook:2011hg,Mukohyama:2014gba}, Hawking radiation during
inflation \cite{Mohanty:2014kwa}, a spectator scalar field with small
sound speed \cite{Biagetti:2013kwa}, massive gravity
\cite{Fujita:2018ehq}, and so on. The amplitude of the inflationary
GW spectrum is constrained at low frequencies by the
cosmic microwave background $\Omega_{\rm
  GW} \lesssim 10^{-15}$, but the amplitude can be large at the
frequency of $\sim 100$~Hz if the spectrum is blue-tilted. There would be a chance for the ground-based detectors to detect inflationary GWs if the
spectral index of the spectrum $n_T$ is larger than $\sim 0.5$.

Another possibility for the ground-based detectors is to probe a non-standard Hubble
expansion history of the early Universe
\cite{Peebles:1998qn,Giovannini:1998bp,Giovannini:1999bh,Giovannini:1999qj,Tashiro:2003qp,Giovannini:2008tm}.
The information on the expansion rate, which is determined by the
dominant component of the Universe, is imprinted in the spectrum of
inflationary GWs. For a scale invariant primordial
spectrum, the spectral dependence of the stochastic background becomes
$\Omega_{\rm GW}\propto f^{2(3w-1)/(3w+1)}$, where $w$ is the EOS of the Universe. In the standard cosmological model, the
matter-dominated ($w=0$) and radiation-dominated ($w=1/3$) phases
yield $\Omega_{\rm GW}\propto f^{-2}$ and $f^0$, respectively. When
one considers a kination dominated phase ($w=1$) soon after inflation,
the spectral amplitude is enhanced at high frequency by the dependence
$\Omega_{\rm GW}\propto f$ and peaks at the frequency corresponding to
the size of the Universe at the end of inflation, which is typically
higher than $\sim 100$~Hz. Although it is difficult to find a
case where GWs are detectable by the ground-based
experiments satisfying the Big-Bang nucleosynthesis (BBN) bound \cite{Figueroa:2018twl,Ahmad:2019jbm}, $\Omega_{\rm GW} \lesssim
10^{-5}$ for $f \gtrsim 10^{-10}$~Hz, the upper bound on the stochastic
background is still helpful to explore early-Universe physics beyond
the standard cosmology.\\

\noindent {\it - Future prospects}

The SNR for a stochastic GW background is computed by taking cross-correlation between detector signals \cite{Allen:1997ad}, and given by 
\begin{equation}
{\rm SNR}_{IJ}=\frac{3H_0^2}{10\pi^2} \sqrt{2T_{\rm obs}}
\left[\int^{\infty}_{0}df\frac{|\gamma_{IJ} (f)|^2
\Omega_{\rm GW} (f)^2}{f^6S_{n,I} (f)S_{n,J} (f)}\right]^{1/2},
\end{equation}
where $H_0$ is the Hubble constant, $T_{\rm obs}$ is the observation time, and $S_{n,I} (f)$ is the noise spectral
density with $I$ and $J$ labeling two different detectors. 
The overlap reduction function $\gamma_{IJ}(f)$ is given by the detector responses of the two polarization modes $F^+$ and $F^\times$ as 
\begin{equation}
\gamma_{IJ} (f)\equiv\frac{5}{8\pi}\int d\hat{\bf \Omega}
 (F^+_IF^+_J+F^{\times}_IF^{\times}_J)
e^{-2\pi if\hat{\bf \Omega}\cdot ({\bf x}_I-{\bf x}_J)}\,,
\end{equation}
where ${\bf x}$ describes the position of the detector and $\hat{\bf \Omega}$ specifies the propagation direction of GWs. 
For a network of $N$ detectors, the total SNR is given by
\begin{equation}
{\rm SNR}=
\left[\sum^N_{I=1}\sum^N_{J<I}{\rm SNR}_{IJ}^2\right]^{1/2}.
\end{equation}

As seen in the overlap reduction function,
the SNR depends on the distance and relative orientation of the detectors. For KAGRA, correlating with AdV+ is better than with A+ because of the relative orientation. Assuming the flat spectrum of $\Omega_{\rm GW}(f)$, the sensitivities of the pair of KAGRA and AdV+ to $\Omega_{\rm GW}$ with ${\rm SNR}=5$ are $1.1\times 10^{-7}$ for bKAGRA, $3.4\times 10^{-8}$ for LF, $8.3\times 10^{-8}$ for 40kg, $1.1\times 10^{-7}$ for FDSQZ, $3.3\times 10^{-7}$ for HF, and $5.0\times 10^{-8}$ for Combined.
The lower frequency upgrades give better sensitivities. However, the sensitivity between two A+ is $4.0\times 10^{-9}$, which is an order of magnitude better than those with KAGRA because of much closer separation of the detector pair. Therefore, KAGRA is not good for detection but will be able to play an important role for determining the spectral shape and measurig anisotropy, non-Gaussianity and polarizations, which are discussed in Sec.~\ref{sec:distinguishing-GWB}.

\subsection{GWs from phase transition}

\noindent {\it - Scientific objective}

The thermal history of the Universe after BBN,
which takes place at a temperature scale of order of 1~MeV,
has been well established thanks to amazing progress made by cosmological
observations.
On the other hand, our knowledge about the Universe earlier than BBN is
quite limited.
If we extrapolate our understanding of particle physics,
several phase transitions should occur
in such early stages of the Universe.
These include the QCD phase transition,
the electroweak phase transition, the GUT phase transition and
some phase transitions of new models beyond the Standard Model of particle
physics.
The nature of a phase transition is classified according to
the evolution of the order parameter: first-order or second-order.
In addition, the case where the order parameter changes smoothly
and no phase transition occurs is referred to as a crossover.
For a first-order phase transition, a latent heat is released,
supercooling emerges and phases are mixed like boiling water.
Non-uniform motion of vacuum bubbles during the first-order phase transition,
GWs are produced.
Therefore, one can survey new physics models
involving a first-order phase
transition by observing GWs~\cite{Grojean:2006bp}.\\

\noindent {\it - Observations and measurements}

GWs produced by the first-order phase transition
in the early Universe are stochastic: 
isotropic, stationary and unpolarized.
They are characterized only by the frequency.
There are three known sources for GWs from
the first-order phase transition:
collisions of vacuum bubble walls, sound waves in the plasma
and plasma turbulence.
In general, one has to sum these contributions.
For typical non-runaway bubbles, the contribution from sound waves is dominant
and the peak energy density~\footnote{The energy density of gravitational
  waves $\Omega_{\rm GW}^{}$ and the power spectral density $S_h^{}$ is
  related through $\Omega_{\rm GW}^{} H_0^2=(2 \pi^2/3)f^3 S_h^{}$.} is roughly~\cite{Caprini:2015zlo}
\begin{eqnarray}
  \Omega_{\rm GW,peak}^{} h^2 \sim 10^{-6}~\kappa^2 \alpha^2
  v^{} \left( \frac{\beta}{H_*^{}} \right)^{-1}  ,
  \label{eq:omegah2}
\end{eqnarray}
and the peak frequency is
\begin{eqnarray}
  f_{\rm peak}^{} \sim 10^{-5}\times ~v^{-1}
  \left( \frac{\beta}{H_*^{}} \right)
  \left( \frac{T_*^{}}{100~\rm{GeV}} \right) ~{\rm Hz}.
\end{eqnarray}
Here, $T_*^{}$ is the transition temperature,
$\alpha$ is the ratio of the released energy (latent heat)
to the radiation energy density, $\beta$ is the inverse of the time duration
of the phase transition, $H_*^{}$ is the Hubble expansion parameter at $T_*^{}$,
$\kappa$ is the ratio of vacuum energy converted into fluid motion, and
$v$ is the bubble wall velocity.
If one selects a new physics model and chooses a model parameter set,
one can compute the finite temperature effective potential, from which
the above quantities are derived.
Namely, the above quantities are functions of the underlying model parameters.
Therefore, the number of free model parameters can be generally reduced by measuring the stochastic GW spectrum.\\

\noindent {\it - Future prospects}

Among new physics models that involve phase transitions,
models with an extended Higgs sector achieving the electroweak phase transition
are of most importance and should be addressed.
Although the Higgs boson has been discovered at the CERN Large Hadron Collider
and the spontaneous breaking of the electroweak symmetry has been established,
the dynamics of the electroweak symmetry is still unknown.
The baryon asymmetry of the Universe may be accounted for
by electroweak baryogenesis, which requires
a strongly first-order electroweak phase transition.
Since the transition temperature of the electroweak phase transition is
of the order of 100~GeV,
unfortunately the peak frequency of the predicted GWs spectrum
falls in the range of $10^{-3}$ Hz to 0.1 Hz, regardless of the details of
the extended Higgs models.
Therefore, the detection of GWs from the electroweak phase
transition and determination of the model parameters of extended Higgs models
are challenging for KAGRA upgrade and need future space-based interferometers,
such as LISA and DECIGO~\cite{Kakizaki:2015wua,Hashino:2018wee}.
Conversely, one can envisage and construct
hypothetical models with the phase transition
temperature of $10^5$~GeV to $10^7$~GeV so that the peak frequency matches
with the KAGRA frequency band~\cite{Dev:2016feu}.
Even in such models, an enormously large latent heat and/or an very long
transition duration are necessary for GW signals
to hit the sensitivity curves for KAGRA upgrade candidates.
Another issue concerns uncertainties in computing
the spectrum of generated stochastic GWs at a first-order
phase transition.
In particular, numerical simulations of the dynamics of nucleated bubbles
are rather cumbersome.
With the development of computational techniques,
the result in Eq.~(\ref{eq:omegah2}) might be revised by one order of magnitude or more. Therefore, although it is difficult to predict the detectability of GWs from a phase transition, KAGRA will play an important role in constraining models of
first-order phase transitions that take place
at temperatures of around $10^5$~GeV to $10^7$~GeV.


\subsection{Distinguishing the origins of the GW backgrounds}
\label{sec:distinguishing-GWB}

\noindent {\it - Scientific objective}

GWs generated in the early Universe are typically in
the form of a stochastic background, which is a continuous noise-like
signal spreading in all directions, while numerous astrophysical
GWs also become a stochastic background by overlapping
one another. Even if a stochastic background is detected, it may be
difficult to identify its origin among many cosmological and
astrophysical candidates.\\

\noindent {\it - Observations and measurements}

A standard data analysis provides
information only on the amplitude and the tilt of the spectrum. In
order to discriminate the origin of the GW background,
we need to go beyond conventional data analysis and extract more
information, such as the spectral shape \cite{Kuroyanagi:2018csn},
anisotropy \cite{Allen:1996gp,TheLIGOScientific:2016xzw}, polarization
\cite{Nishizawa:2009bf, Abbott:2018utx}, and non-Gaussianity (or
so-called popcorn background) \cite{Drasco:2002yd, Thrane:2013kb,
  Martellini:2014xia}. Anisotropy and non-Gaussianity could arise in a GW background formed by overlapping GW bursts. Typical examples are the ones originating from astrophysical sources and cosmic strings. Additional GW polarizations appear when the generation process is related to physics beyond general relativity (see Sec.~\ref{sec:gw-pol}). Therefore, they can be used as indicators of specific generation mechanisms.

The spectral shape of the stochastic background depends strongly on its origin and it can be generally used to distinguish a wide range of generation models.
Many models predict a spectral shape spanning a broad range of frequencies. In this case, the ground-based detectors observe a part of the spectrum because of the limited sensitivity range of frequency, and multi-band observations are needed to determine the whole spectral shape. However, some models predict a peaky spectral shape; one good example testable by the ground-based detector network itself is a GW background generated by superradiant instabilities \cite{Brito:2017wnc}.
One way to obtain the spectral shape in data analysis is to use broken
power-law templates instead of the currently used single power-law
templates \cite{Kuroyanagi:2018csn,Bose:2005fm}. The uncertainties in
the parameter determination decrease in proportional to the inverse of
the SNR, and at least SNR $\gtrsim 10$ is
necessary to obtain useful information on the entire spectral shape.
If detection were made with higher SNR, the analysis could be extended
to the whole spectral shape reconstruction with multiple frequency
bins, which evaluates the spectral index in each bin (see
\cite{Figueroa:2018xtu} for the discussion for LISA, which could be
applied to ground-based experiments).\\

\noindent {\it - Future prospects}

In general, a higher SNR is necessary not only to determine the spectral shape
but also to obtain the above mentioned additional properties of the stochastic background. 
Since inflationary GWs have a red-tilted spectrum in terms of strain amplitude, an improvement of sensitivity at lower frequency would be helpful to test more properties of the stochastic background. On the other hand, as previously mentioned, the sensitivity of the ground-based detectors is not enough to detect a standard inflationary GW background, and a blue-tilted spectrum is necessary to enhance the amplitude for detection. If $n_T > 3$, where we define the spectral tilt as $\Omega_{\rm GW} \propto f^{n_T}$, an improvement of sensitivity at higher frequencies becomes important (see Fig. 1 of \cite{Kuroyanagi:2018csn}). Astrophysical GWs also have a blue-tilted spectrum $n_T=2/3$, but when the tilt is not so steep, sensitivity at mid-frequencies ($20-50$Hz) is more important.

Apart from sensitivity for the amplitude, multiple detectors facing different directions provide independent information about anisotropy and polarization, and thus the inclusion of KAGRA is essential.\\

\section{Test of gravity}
\subsection{Test of consistency with general relativity}

\noindent {\it - Scientific objective}

There are currently no alternatives to Einstein's GR that are physically viable.
As such, one cannot test GR by comparing the data with waveforms in alternative theories.
Instead, one can test for the consistency of the signal with GR.
In doing so, one must ensure that any deviation from GR cannot be attributed to noise artifacts, but instead are intrinsic to the GW signal itself.
However, this must be done without explicitly assuming an alternative model for gravity that governs the morphology of the GW signal.\\

\noindent {\it - Observations and measurements}

Existing efforts largely focus on comparing different portions or aspects of the signal and require consistency amongst them.
For example, one can verify the self-consistency of the post-Newtonian description of the inspiral phase, and from the phenomenological GR model of the merger-ringdown~\cite{LiEtAl:2012a,LiEtAl:2012b,Agathos:2013upa,Cornish:2011ys,Sampson:2013jpa,Meidam:2017dgf}.
In addition, one can also test the final remnant mass and spin, as determined from the low-frequency (inspiral) and high-frequency (post-inspiral) phases of the signal for mutually consistent~\cite{Ghosh:2016qgn,Ghosh:2017gfp}.
Furthermore, one can also compare the residual signal after subtracting the best-fitting GR waveform to the expected statistical properties of the noise.
For the first set of BBH detections, all tests so far have not revealed any significant deviations from GR~\cite{TheLIGOScientific:2016src,TheLIGOScientific:2016pea,Abbott:2018lct,LIGOScientific:2019fpa}.
Instead, constraints on various possible beyond-GR effects have been placed.
For example, the fractional deviations to the lowest post-Newtonian parameters have been constrained to tens of percent, whereas the higher post-Newtonian orders and the merger-ringdown parameters are constrained to around a hundred percent.
Moreover, the possibility of dipolar radiation has been constrained to the percent level, and the mass of the graviton has been constrained to $m_g \leq 4.7 \times 10^{-23}\,\mathrm{eV/c}^2$.\\

\noindent {\it - Future prospects}

In the future, as the LVK network operates at their full strengths, we can anticipate several effects that will improve our ability to constrain general relativity.
Firstly, the improved detection rate will allow us see many events for which the different aspects of their dynamics are covered by the network's sensitive frequencies.
Secondly, more detections also means that our constraints will be improved statistically. While we do not expect improvements of orders of magnitude, it will help to shrink the parameter space of possible deviations from GR.
Thirdly, a full LVK network will greatly enhance tests that require multiple detectors, including tests for alternative polarisations (see Sec.~\ref{sec:gw-pol}). These tests are expected to greatly improve as KAGRA, as the fourth detector, improves its sensitivity.
Finally, the prospect of observing novel source classes, including highly eccentric systems, highly precessing systems and even head-on collisions will open up a entirely new suite of tests.

In the era of KAGRA+, we expect more detections of BBH and BNS, and anticipate detections of BH-NS systems.
The precise way in which KAGRA+'s sensitivity develops will govern the way it will contribute to tests of GR.
For example, with the focus on low-frequency sensitivity, one improves the insight into the merger and ringdown of heavy BHs, where the gravitational fields are the strongest.
In contrast, high frequency sensitivity allows one to track lower mass binary merger for a longer period of time, which improves the precision of the consistency checks.
Additionally, by improving the sky coverage and increasing the multiple-detector live time, KAGRA+ will increase the number of detected events and thereby reducing the statistical errors due to detector noise by combining results from multiple events.

\subsection{GW generation in modified gravity}

\noindent {\it - Scientific objective}

We can raise several motivations to study modified gravity: a) GR is not ultraviolet (UV) complete theory. It should be modified to describe 
physics near Planck scale. b) Cosmology based on GR requires dark components to explain the 
observed Universe, such as dark matter and dark energy. It is attractive if we can 
naturally explain the observed Universe by a well-motivated modification of gravity 
without introducing cosmological constant. c) To test GR, it is not enough to know the prediction based on GR 
when we use GWs as a probe. Since GW data is noisy, 
an appropriate projection of the data in the direction predicted by the possible modification 
of gravity is necessary to obtain a meaningful constraint.

As possible modifications of gravity, there are several categories. 
One is to add a new degree of freedom like a scalar field in scalar-tensor gravity. 
Such a kind of extra degrees of freedom is strongly constrained by existing observations, 
since they introduce a fifth force. There are various ways, however, to hide the fifth force. 1) Sufficiently small coupling to the ordinary matter. 1-a) Coupling through higher derivative terms, which becomes important only under strong field environment. 1-b) Some non-linear mechanism to reduce the coupling like Chameleon/Vainstein mechanism 
\cite{Khoury:2003aq,Vainshtein:1972sx}. 2) Giving a sufficiently large mass to the field that mediates the fifth force.

There is also a possibility to modify gravity without introducing a new degree of freedom. 
One example is so-called minimal extension of massive gravity. Phenomenologically, 
we can give small mass to graviton in this kind of model, but the graviton mass at the 
present epoch is already strongly constrained by the observation of GW170817. \\

\noindent {\it - Observations and measurements}

We focus on compact binary inspirals. 
In this context, 
the most probable modification of the generation mechanism
of GWs is the one caused by the change in orbital evolution. 
The orbital evolution is governed by two aspects. One is the conservative part of 
the gravitational interaction and the other is the dissipative effects of radiation.
It is not so obvious which effect is more important even in the weak field regime. 
Hence, we need a little more careful investigation of each model.

Whichever effect dominates, the main purpose of this study is the same. 
What we want to clarify is whether or not there exists a new effective propagating 
degree of freedom which appears only when we consider extremely compact objects 
like BHs and NSs.  

There might be various possibilities. From the observational point of view 
it is convenient to introduce the parametrized post-Einstein (PPE) framework~\cite{YunesPretorius2009}. 
The waveform of quasi circular inspiral in Fourier domain will be given in the form of
\begin{equation}
 h(f)\approx A(f) f^{-7/6} e^{i\Psi(f)}\,,
\end{equation}
in GR. The waveform is characterized by the functions $A(f)$ and $\Psi(f)$ and 
they are modified in the extended theories like
\begin{equation}
  A\left(f\right) \to \left(1+\sum_i \alpha_i u^{a_i}\right) A_{\rm GR}\left(f\right)\,, 
\qquad
  \Psi\left(f\right) \to \left(\Psi_{\rm GR}\left(f\right)+\sum_i \beta_i u^{b_i}\right) \,, 
\end{equation}
with $u:=\pi M f$. 
Since the test of gravity using GW data is more sensitive to 
the modification in the phase $\Psi(f)$, we will focus on it. 

The smallest non-vanishing value of $b_i$ is related to the leading 
post-Newtonian order. The $n$-th post Newtonian 
correction appears as $b=(2n-5)/3$. 
Here we list representative possible modifications of gravity with the post-Newtonian order of the leading correction in phase in Table \ref{tab:modifiedG}. 
BH evaporation is negligible in the ordinary scenario, but 
can be accelerated if there are many massless hidden degrees of freedom. 
For the dipole radiation to be emitted, charge not proportional to the mass of each constituent object is required. 
One typical example is the scalar tensor theory. 
Although BHs are thought not to bear a scalar hair in many cases, 
they can have a scalar charge in some models like in Einstein-Gauss-Bonnet-dilaton theory \cite{Yagi:2011xp}.
If we naively consider that gravitons have non-vanishing mass, the dispersion relation is also modified to give 1PN order correction, although this is not the effect in the generation of gravitational waves. 
However, if we consider covariant theories of massive gravity, we need to consider a bigravity model~\cite{Hassan:2011zd}.
As a modification that appears at the 1PN order, one may think of a model in which 
the mass of each constituent object depends on the Newton potential. 

\begin{table*}[t]
\begin{center}
\begin{tabular}{cccccc}
\hline \hline
modification & post-Newtonian order & References \\
\hline \hline  
variable Newton's constant & -4PN & \cite{Yunes:2009bv}
\\ \hline
 BH evaporation 
 & -4PN & \cite{Yagi:2011yu}
\\ \hline  
dipole radiation & -1PN & \cite{Will:1989sk} 
\\ \hline  
non-covariant massive graviton & 1PN & \cite{Will:1997bb}
\\ \hline 
dynamical Chern-Simons gravity & 2PN & \cite{Yagi:2012vf}
\\ \hline 
\end{tabular}
\end{center}
\caption{List of representative possible modifications in gravitational wave generation. The leading post-Newtonian order 
of the correction in phase relative to the quadrupole formula.}
\label{tab:modifiedG}
\end{table*}

As a typical modification of GR, the scalar-tensor theory is often discussed. 
If the binary members have some scalar charges coupled to the extra degree of freedom, 
we have dipolar radiation. Although the overall magnitude also depends on the coupling strength, 
if we focus only on the frequency dependence, the flux due to 
dipolar radiation is larger at a low frequency by the factor $(v/c)^{-2}\sim (m\omega)^{-2/3}$ 
compared with that due to GW radiation. On the other hand, the extra scalar 
degree of freedom changes the conservative part of the force, but there is no such enhancement 
in the sense of post Newtonian order.   
Hence, at low frequency, where  $v/c\ll 1$ 
the modification to the conservative gravitational interaction can be neglected. 

For the modification that predicts negative post-Newtonian order, the deviation 
form GR is relatively larger at lower frequencies, compared with GR contributions. Therefore, it is not so easy 
for GW observations to compete with the constraint from the 
observations at lower frequencies, such as the pulsar timing. 

There are some more detailed aspects that cannot be expressed by the  PPE framework. One possibility is the spontaneous scalarization. When we consider BNS, each NS in isolation may not have a scalar charge. However, there is a region 
in the model parameter space in which instability occurs as the binary separation 
decreases and both NSs get charged. Another possibility is that the mediating scalar field is massive. This is quite likely 
to avoid the instability similar to the spontaneous scalarization to occur 
in the whole Universe after the temperature of the Universe becomes $10\,{\rm MeV}$ or lower. 
If the instability occurs, the success of the big-bang nucleosynthesis will be significantly 
damaged. 
This gives the constraint on the mass of the mediating field to be greater than $10^{-16}$~eV
or so~\cite{Ramazanoglu2016}, which is comparable to the  constraint from pulsar timing (PSRJ0348+0432). 
On the other hand, ground-based detectors are sensitive to modification if the mass is smaller than $10^{-12}$~eV. If the mass is in the range between these two, the existence of 
the extra degree of freedom might be probed only by GWs. 

Basically we check the deviation of the observed GW signal from GR templates. 
For this purpose, it is better to have a better theoretical template based on each candidate of 
modified gravity. In particular this demand is higher when we consider the models 
whose leading order modification appears at the positive post-Newtonian order. 
In such models, we have more chances to observe the largest deviation at 
the final stage of inspirals. In many cases, this task remains to be a challenge for the theorists. \\

\noindent {\it - Future prospects}

As long as we consider BHs with the mass larger than a few times the solar mass and NSs, 
it seems difficult to use a high frequency band to give constraints on modified gravity. In the case of 
BHs, we do not have much signals at very high frequencies. In the case of binary NSs, high 
frequency signal can be expected but it might be difficult to disentangle the effects of modified gravity 
from the uncertainty of the complex physics of NSs. 

In general, when we wish to give a tighter constraint on the parameters contained in the 
waveform template, broader observational sensitivity band is advantageous. A broadband observation 
is not necessarily realized by a single detector. It would work well if a wider range of the frequency band 
is covered by plural detectors as a whole. In this sense, it would be necessary to 
negotiate among the various plans of future upgrade of detectors. 

In near future we will establish a network composed of LIGO, Virgo and KAGRA. 
On the point of view of detecting modification in the gravitational wave generation, the 
improvement of the constraints depends on the increase of statistics and finding of 
some golden events with appropriate binary parameters and a large signal-to-noise.

\subsection{GW propagation test}
\label{sec:gw-pro}

\noindent {\it - Scientific objective}

The properties of GW propagation are changed when a gravity theory is modified from GR. Particularly, such modifications are paid much attentions at cosmological distance because the extended theories of gravity may explain the accelerating expansion of the Universe. In other words, testing GW propagation can constrain the possible modification of gravity at cosmological distance and give us an implication about the physical mechanism of the cosmic accelerating expansion~\cite{Lombriser:2015sxa}. According to~\cite{Saltas:2014dha}, the modifications of GW propagation in the effective field theory is expressed in general as
\begin{equation}
h^{\prime\prime}_{ij}+(2+\nu){\cal H} h^{\prime}_{ij} + (c_{\rm T}^2 k^2 + a^2 \mu^2) h_{ij} = a^2 \Gamma \gamma_{ij} \;,
\label{eq:GW-propagation}
\end{equation}
and are characterized by four properties: propagation speed $c_{\rm T}$, amplitude damping rate $\nu$, graviton mass $\mu$, and a source term $\Gamma$. Each modification appears differently in a specific theory, but this framework allows us to test many classes of gravity theories at the same time. Among the modifications, the propagation speed is important to test the violation of the Lorentz symmetry and the quantum nature of spacetime (modified dispersion relation). The amplitude damping rate is equivalent to the variation of the gravitational constant for GWs and enables us to test the equivalence principle. The graviton mass is a characteristic quantity in massive gravity. If parity in gravity is violated, it modifies in general the propagation speeds and the amplitude damping rates in Eq.~(\ref{eq:GW-propagation}) differently for left-handed and right-handed polarizations  with opposite signs as shown in \cite{Nishizawa:2018srh,Zhao:2019xmm}.\\ 

\noindent {\it - Observations and measurements}

The coincidence detection of GW170817/GRB170817A \cite{GW170817PRL} brought us the first opportunity to measure the speed of a GW from the arrival time difference between a GW and an EM radiation\footnote{Before GW170817/GRB170817A, the measurements of the speed of a GW have done with GWs from BBH based on the arrival time difference between detectors~\cite{Blas:2016qmn,Cornish:2017jml}.}, as was predicted in \cite{Nishizawa2014PRD}, and constrained the deviation from the speed of light at the level of $10^{-15}$ \cite{GW170817:GRB}. Based in this constraint on GW speed, a large class of theories as alternatives to dark energy have already been almost ruled out~\cite{Baker:2017hug,Creminelli:2017sry,Sakstein:2017xjx,Ezquiaga:2017ekz}. The graviton mass $\mu$ has been tightly constrained from the observation of BBH mergers in the range of $\mu < 5.0\times 10^{-23}\,{\rm{eV}}$ \cite{LIGOScientific:2019fpa}, which is tighter than that obtained in the Solar system \cite{Bernus:2019rgl}. 
On the other hand, the amplitude damping rate $\nu$ has also been measured from GW170817/GRB170817A for the first time in \cite{Arai:2017hxj}, but the constraint is still too weak, $-75.3 \leq \nu \leq 78.4$. However, the anomalous amplitude damping rate is one of the prominent signatures of modified gravity, which is likely to be accompanied by the gravity modification explaining the cosmic acceleration, and plays a crucial role when we test gravity at cosmological distance \cite{Nishizawa:2019rra}. 

In generic parity-violating gravity including recently proposed ghost-free theories with parity violation~\cite{Crisostomi:2017ugk}, not only the amplitude damping rate but also the propagation speed are modified, in contrast to Chern-Simons modified gravity in which the GW speed is the speed of light. From GW170817/GRB170817A, since a GW is required to propagate almost at the speed of light, the parity violation in the GW sector of the theory has been constrained almost to that in Chern-Simons gravity, allowing only the amplitude damping rate to differ from GR \cite{Nishizawa:2018srh}.\\

\noindent {\it - Future prospects}

To forecast the abilities to measure the modified gravity parameters with a future detector network including bKAGRA and KAGRA+, we estimate the measurement errors of the model parameters in the modified GW waveform with the Fisher information matrix (For the detail of the analysis, see  \cite{Nishizawa:2017nef}). For the waveform, we consider the simplest one in which the amplitude damping rate $\nu$ and the graviton mass $\mu$ are assumed to be constant, taking $c_{\rm T}=1$ and $\Gamma=0$. Setting $c_{\rm T}=1$ is motivated by the existing constraint from GW170817/GRB170817A and $\Gamma=0$ is true in most gravity theories including the Horndeski theory \cite{Saltas:2014dha}. Under the assumptions above, the GW waveform is \cite{Nishizawa:2017nef}
\begin{equation}
h = (1+z)^{-\nu/2} e^{- i k \Delta T}  h_{\rm GR}\,, \qquad \Delta T = - \frac{\mu^2}{2k^2} \int_{0}^{z} \frac{dz^{\prime}}{(1+z^{\prime})^3 {\cal H}} \;,
\end{equation}
where $k$ is the wave number. The waveform $h_{\rm GR}$ is the one in GR. Then the luminosity distance observed by a GW is interpreted as an effective one, $d_{\rm L}^{\rm gw}(z) = (1+z)^{-\nu/2} d_{\rm L}(z)$, which differs from the standard one for EM waves, $d_{\rm L}$. To break a parameter degeneracy and determine $\nu$ separately, one needs to know a source redshift independently of a GW observation. We assume that the source redshift is obtained somehow from EM observations and impose a prior on $z$ with a standard deviation $\Delta z=10^{-3}$. We will generate 500 binary sources with random sky locations and other angular parameters, having the network SNR $\rho>8$, for each case with fixed masses and redshift. 

In Table~\ref{tab:errors}, the parameter estimation errors of $\nu$ and $\mu$ with detector networks including KAGRA are shown. The sensitivities are better for heavier binaries and best for $30\,M_{\odot}$ BBH. The upper limits on graviton mass are similar to the current constraint and are not expected to be improved significantly with the ground-based detectors in the future. Since the graviton mass affects GW phase more at lower frequencies, the constraints from the planned spaced-based detectors such as LISA \cite{Audley:2017drz} and DECIGO \cite{Sato:2017dkf} will be much stronger down to $\mu \sim 10^{-26}$ -- $10^{-25}\,{\rm{eV}}$ \cite{Yagi:2009zm, Yagi:2009zz}. On the other hand, there are modest improvements of the sensitivity to $\nu$ with the addition of bKAGRA or KAGRA+ to the global detector network. For a measurement of $\nu$ with $30\,M_{\odot}$ BBH, the improvement factors of the errors are LF 0.87, 40kg 1.16, FDSQZ 0.99, HF 0.98, Combined 1.35, reaching $\Delta \nu =1.09$ in the best case.

\begin{table*}[t]
\begin{center}
\begin{tabular}{llcccccc}
\hline \hline
source & quantile & bKAGRA & LF & HF & 40kg & FDSQZ & Combined \\
\hline \hline  
SNR & $30\,M_{\odot}$ BBH & 77.4 & 73.9 & 76.6 & 78.0 & 79.6 & 84.8 \\
& $10\,M_{\odot}$ BBH & 32.3 & 32.9 & 34.2 & 33.9 & 32.3 & 36.6 \\
& $10\,M_{\odot}$ BH-NS & 25.1 & 23.7 & 24.8 & 25.2 & 26.4 & 27.2 \\
& BNS & 24.3 & 24.2 & 24.7 & 24.7 & 24.9 & 27.5 \\
\hline
$\nu$ & $30\,M_{\odot}$ BBH & 1.47 & 1.69 & 1.50 & 1.27 & 1.48 & 1.09 \\
& $10\,M_{\odot}$ BBH & 3.13 & 3.85 & 3.56 & 2.74 & 3.08 & 2.56 \\
& $10\,M_{\odot}$ BH-NS & 7.86 & 9.53 & 7.96 & 6.91 & 7.32 & 6.24 \\
& BNS & 13.8 & 17.6 & 14.1 & 12.6 & 13.3 & 11.4 \\
\hline  
$\mu$ & $30\,M_{\odot}$ BBH & 4.76 & 4.78 & 4.53 & 4.86 & 4.77 & 4.87 \\
& $10\,M_{\odot}$ BBH & 11.9 & 11.7 & 11.6 & 11.9 & 11.9 & 11.6 \\
& $10\,M_{\odot}$ BH-NS & 15.7 & 15.4 & 15.9 & 15.9 & 15.5 & 15.4 \\
& BNS & 28.0 & 27.9 & 28.0 & 28.1 & 28.1 & 27.6 \\
\hline 
\end{tabular}
\end{center}
\caption{Median SNR, median errors of $\nu$, and median errors of $\mu$ in the unit of $10^{-23}\,{\rm eV}$. The BBH is at the distance of $z=0.1$, the $10\,M_{\odot}$ BH-NS is at $z=0.06$, and the BNS with masses $1.4\,M_{\odot}$ is at $z=0.03$. bKAGRA, LF, HF,
40kg, FDSQZ, and Combined denote the detector network composed of A+, AdV+, and bKAGRA or KAGRA+ (one of LF, HF, 40kg, FDSQZ, and Combined).}
\label{tab:errors}
\end{table*}

\subsection{GW polarization test}
\label{sec:gw-pol}

\noindent {\it - Scientific objective}

In GR, GWs have only two tensor polarization modes. However, general metric theory of gravity allows four nontensorial polarization modes in addition to two tensor modes \cite{Eardley1973}. The properties of GW polarization depend on the alternative theories of gravity. Thus, the observations of polarizations can be utilized to test the nature of gravity. The testable degrees of freedom of GWs are related to the number of GW detectors and it is expected that the participation of KAGRA in the network of detectors make it possible to search for more polarizations more accurately \cite{Takeda2018,Hagihara:2018azu}. Here, we shall consider KAGRA potential for polarization test of GWs from the most promising sources that is compact binary coalescences.\\

\noindent{\it - Observations and measurements}

In general, the I-th detector signal $h_I$ of GWs propagating in the direction $\hat{\Omega}$ can be expressed as
\begin{equation}
 \label{detector_signal}
 h_I(t,\hat{\Omega})=\sum_A F_I^{A}(t,\hat{\Omega})h_A(t).
 \end{equation}
 Here $F_I^{A}$ is the antenna pattern function of the I-th detector for the polarization $A$ and $h_A$ is the GW waveform for the polarization $A=+$ ,$\times$, $x$, $y$, $b$, and $\ell$ modes, called plus, cross, vector-x, vector-y, breathing, and longitudinal modes, respectively. Provided that there are enough detectors, we can test the theory of gravity by solving the inverse problem, separating and evaluating the waveforms of polarization modes directly. Currently there are observational constraints on the nontensorial polarization modes of GWs from binary systems by Bayesian inference with the simple substitution of the antenna pattern functions and the assumption of the waveform models \cite{GW170814PRL, Abbott:2018lct}. They found the Bayes factors of more than 200 and 1000 for GW170814 and a logarithm of the Bayes factors of 20.81 and 23.09 for GW170817 in favor of the pure tensor polarization against pure vector and pure scalar, respectively.

Another method of model-independent searches for nontensorial polarizations is constructing null streams, in which tensor signals are automatically canceled and only scalar and vector signals remain \cite{Chatziioannou:2012rf}. From three (four) detectors, one can obtain one (two) null stream to search for one (two) additional polarization. However, as proposed in \cite{Hagihara:2018azu,Hagihara:2019ihn,Hagihara:2019rny}, a fortune case that more polarizations can be searched exists if an EM counterpart pins down the sky direction of a GW source. In this case, one additional polarization signal happens to be canceled and one more polarization can be searched. Even with four detectors including KAGRA, five polarizations can be searched if an EM counterpart is coincidentally detected.\\

\noindent{\it - Future prospects}

We estimated the additional polarization amplitude parameters $\{A_{V_x}, A_{V_y}\}$, whose fiducial values are unity, in the following polarization model including additional two vector modes
\begin{equation}
h_I=\{\mathcal{G}_{T,I}+A_{V_x}\mathcal{G}_{V_x,I}+A_{V_y}\mathcal{G}_{V_y,I}\}h_{\rm{GR}},
\end{equation}
where $\mathcal{G}$ denotes the geometrical factors for each mode and includes the antenna pattern function, the Dopper phase, and the binary's orbital inclination angle \cite{Takeda2018}. We assume several detector network configurations; A+, AdV+, and bKAGRA or KAGRA near term upgrade candidates (LF, HF, 40kg, or FDSQZ), or KAGRA longer term upgrade candidate (Combined). 
In actual, the accurate waveforms of nontensorial modes depend on a theory one considers and its specific parameters. However, the accurate waveforms in the alternative theories, especially in the strong gravity regime, are not yet well known because of the complexity of the field equations. Therefore, we consider only the inspiral phase and assume for the nontensorial modes the same waveforms and cutoff frequency as the waveform in GR, $h_{\rm GR}$, and the ISCO frequency in GR, $f_{\rm ISCO}$. In other words, we consider the most pessimistic case in which separating polarization modes is most difficult in order to study fundamental separability. Table \ref{table_result_pol} shows the medians of parameter estimation errors for $10M_{\odot}-10M_{\odot}$ BBH at $z = 0.05$ and $1.4M_{\odot}-1.4M_{\odot}$ BNS at $z = 0.01$. A choice of networks does not affect the amplitude estimation drastically. However, this kind of search is possible only with four detectors including KAGRA.
Basically the errors of the additional polarization amplitude parameters are scaled by the network SNR depending on the detector network as long as a degeneracy among the amplitude parameters are broken. Thus, it would be best to separate polarizations with 40kg network among networks with KAGRA near term upgrade candidates. This is because we are assuming the promising sources with masses $\sim 1.4M_{\odot}$ or $\sim10M_{\odot}$ whose merging frequencies are around the bottom of the 40 kg noise curve. If we compute the rms errros of two vector modes, the improvement factors (better case of BBH and BNS is taken) are LF 0.80, 40kg 1.10, FDSQZ 1.04, HF 1.00, Combined 1.23. Therefore, the differences among KAGRA near term upgrade candidates are not large.

On the other hand, as for the polarization search with an EM counterpart, we do not need any information about GW waveforms of additional polarizations and do not have any preference on the KAGRA upgrades other than precise sky localization.

\begin{table*}
\begin{center}
\begin{tabular}{|c|c|c|c|c|c|c|c|} \hline
  					&& bKAGRA 	& LF		& HF		& 40kg 	& FDSQZ		& Combined	\\ \hline
BBH &	$\Delta A_{Vx} $ 	& 0.284	& 0.325	& 0.353 	& 0.258 	& 0.283		& 0.236			\\ 	
	&	$\Delta A_{Vy} $ 	& 0.344 & 0.456 	& 0.434 	& 0.312 	& 0.342		& 0.281			\\ \hline\hline
	
BNS &	$\Delta A_{Vx} $	& 0.157 	& 0.198 	& 0.161 	& 0.149 	& 0.151		& 0.127			\\ 
	&	$\Delta A_{Vy} $	& 0.195	& 0.244 	& 0.193 	& 0.180 	& 0.185		& 0.158			\\ \hline
	
\end{tabular}
\caption{Medians of parameter estimation errors of additional polarization amplitude parameters for $10M_{\odot}-10M_{\odot}$  BBH at $z = 0.05$ and $1.4M_{\odot}-1.4M_{\odot}$ BNS at $z = 0.01$ with six global detector networks. The lower cutoff frequency is $10 {\rm Hz}$ and the upper one is the twice the inner most stable circular orbit frequency for a point mass in Schwarzschild spacetime}
  \label{table_result_pol}
  \end{center}
\end{table*}

\subsection{Black-hole spectroscopy}

\noindent {\it - Scientific objective}

When we treat GWs in the inspiral phase of binaries,
there are many parameters to specify the binaries
and also a variety of modifications of the waveforms
(exotic compact objects, modification of gravity and so on).
Therefore, it will be difficult to test GR,
especially in a strong gravity regime
because the derivation of the theoretical waveforms
relies on the post-Newtonian approximation.
The data analysis of quasi-normal modes (QNMs) of BHs
observed in the ringdown phase
is the best way to identify a compact object
with the BH predicted by GR.

The current status of the ringdown data analysis
is that the least-damped ($n=0$) QNM
of the remnant BH has been observed
but the SNR is too weak
to determine the BH parameters precisely.
We find for GW150914 that the late time behavior
is consistent with the ($\ell=m=2$, $n=0$)
QNM~\cite{TheLIGOScientific:2016src}.
Furthermore, to verify the no-hair theorem in GR, 
we need to measure not only the loudest mode
but also multiple QNMs
(see, e.g., Refs.~\cite{Dreyer:2003bv,Berti:2005ys,Berti:2007zu}).
It should be noted that the determination of the initial time
of the ringdown phase is important in the data analysis
because this affects the estimation of BH parameters
(see, e.g., Refs.~\cite{Sakai:2017ckm,Carullo:2018sfu}).
We also note that an analysis of multiple QNMs for GW150914
has been done in Ref.~\cite{arXiv:1902.07527} recently.
Reference~\cite{Berti:2018vdi}
is the latest review on theoretical and experimental challenges
in BH spectroscopy.\\

\begin{table}[!t]
\centering
\begin{tabular}{c|cccccccc}
\hline
\hline
$M_{\rm rem}$ & bKAGRA & LF & HF & 40kg & FDSQZ & Combined & AdVLIGO & Aplus \\
$[M_{\odot}]$ & & & & & & & ZDHP & \\
\hline
\hline
 & & & & q=1 & & & & \\
\hline
60 &19.0 &0.0844 &45.4 &26.9 &36.6 &56.3 &22.0 &48.6 \\
\hline
700 &62.4 &335 &19.2 &71.0 &56.9 &128 &253 &403 \\
\hline
\hline
 & & & & q=2 & & & & \\
\hline
60 &17.7 &0.0795 &40.0 &25.3 &32.9 &50.9 &19.8 &43.5 \\
\hline
700 &51.0 &307 &15.6 &57.5 &45.9 &103 &208 &332 \\
\hline
\hline
\end{tabular}
\caption{\label{tab:QNM-SNR}
SNR of the QNM GWs in the cases of
$M_{\rm rem}=60~M_{\odot}$ and $700~M_{\odot}$. Here we assume the luminosity distance of $r=400\,{\rm Mpc}$.
}
\end{table}

\noindent {\it - Observations and measurements}

The (single-mode) ringdown waveform is given by 
\begin{equation}
h(t) = A e^{-(t-t_0)/\tau} \cos(2 \pi f_{\rm R} (t-t_0) - \phi_0) \,, 
\label{eq:ringdown}
\end{equation}
where $A$ is the amplitude,
$\tau$ and $f_{\rm R}$ are the damping time
and frequency, respectively.
$t_0$ and $\phi_0$ denote the initial time and phase,
respectively.
The damping time is related to the quality factor $Q$ as
$\tau=Q/(\pi f_{\rm R})$.
We note that it is not necessary to search the initial phase
in the matched filtering method~\cite{Mohanty:1997eu}.
In Table~\ref{tab:QNM-SNR},
we show the SNR of the QNM GWs in the cases of
$M_{\rm rem}=60~M_{\odot}$ and $700~M_{\odot}$
for various detector's configurations.
Here we assume the source distance of $r=400\,{\rm Mpc}$.

\begin{table}[t]
\centering
\begin{tabular}{c|ccc|ccc}
\hline
\hline
$(\ell,m)$ & $f_1$   &$f_2$   &$f_3$   &$q_1$   &$q_2$   &$q_3$ \\
\hline
$(2,2)$  &$1.5251$ &$-1.1568$ &$0.1292$ &$ 0.7000$ &$1.4187$ &$-0.4990$ \\
\hline
$(3,3)$  &$1.8956$ &$-1.3043$ &$0.1818$ &$ 0.9000$ &$2.3430$ &$-0.4810$ \\
\hline
$(4,4)$  &$2.3000$ &$-1.5056$ &$0.2244$ &$ 1.1929$ &$3.1191$ &$-0.4825$ \\
\hline
\end{tabular}
\caption{\label{tab:fitting}
Fitting coefficients for the QNM frequencies~\cite{Berti:2005ys}.
Here, we focus only on the lowest overtone ($n=0$).}
\end{table}

In the case where the ringdown waveform is described 
by the QNMs of a BH predicted by GR,
the frequency and quality factor are written in terms of the BH mass $M$
and (non-dimensional) spin $\chi$ as
\begin{eqnarray}
f_{\rm R} &=& \frac{1}{2\pi M} \left[f_1+f_2(1-\chi)^{f_3}\right] \,,\\
Q &=& q_1+q_2(1-\chi)^{q_3} \,.
\label{eq:fitting}
\end{eqnarray}
These are fitting formulae presented in Ref.~\cite{Berti:2005ys},
and $f_i, q_i$ are the fitting coefficients summarized in Table~\ref{tab:fitting}.

In general, it is not clear which QNMs will be excited.
Here, we restrict the excitation to that of merging BBH.
In this case, the ($\ell=m=2$) QNM becomes 
the least-damped, dominant mode,
and the other modes are the subdominant modes.
Furthermore, the mass ratio ($q=m_1/m_2$)
of binaries and the spins of each BH affect the excitation.
In the following analysis, we focus only on non-spinning BBH
for simplicity.
Then, the next-dominant mode becomes the ($\ell=m=3$) or ($\ell=m=4$) mode
(see Fig.~4 of Ref.~\cite{Baibhav:2017jhs}, and the ($\ell=3,m=2$) mode
arises from mode mixing).
As a summary, if we observe at least two different QNMs for an event,
we can test the no-hair theorem
since the remnant BH mass and spin are estimated from each QNM.

Although we focus only on the ($n=0$) QNMs in our analysis,
it should also be noted that there is a direction to
study overtones ($n>0$).
Here are some papers~\cite{Giesler:2019uxc,Isi:2019aib,Bhagwat:2019dtm,Ota:2019bzl}.\\

\begin{figure}[t]
 \centering
 \includegraphics[width=0.48\textwidth]{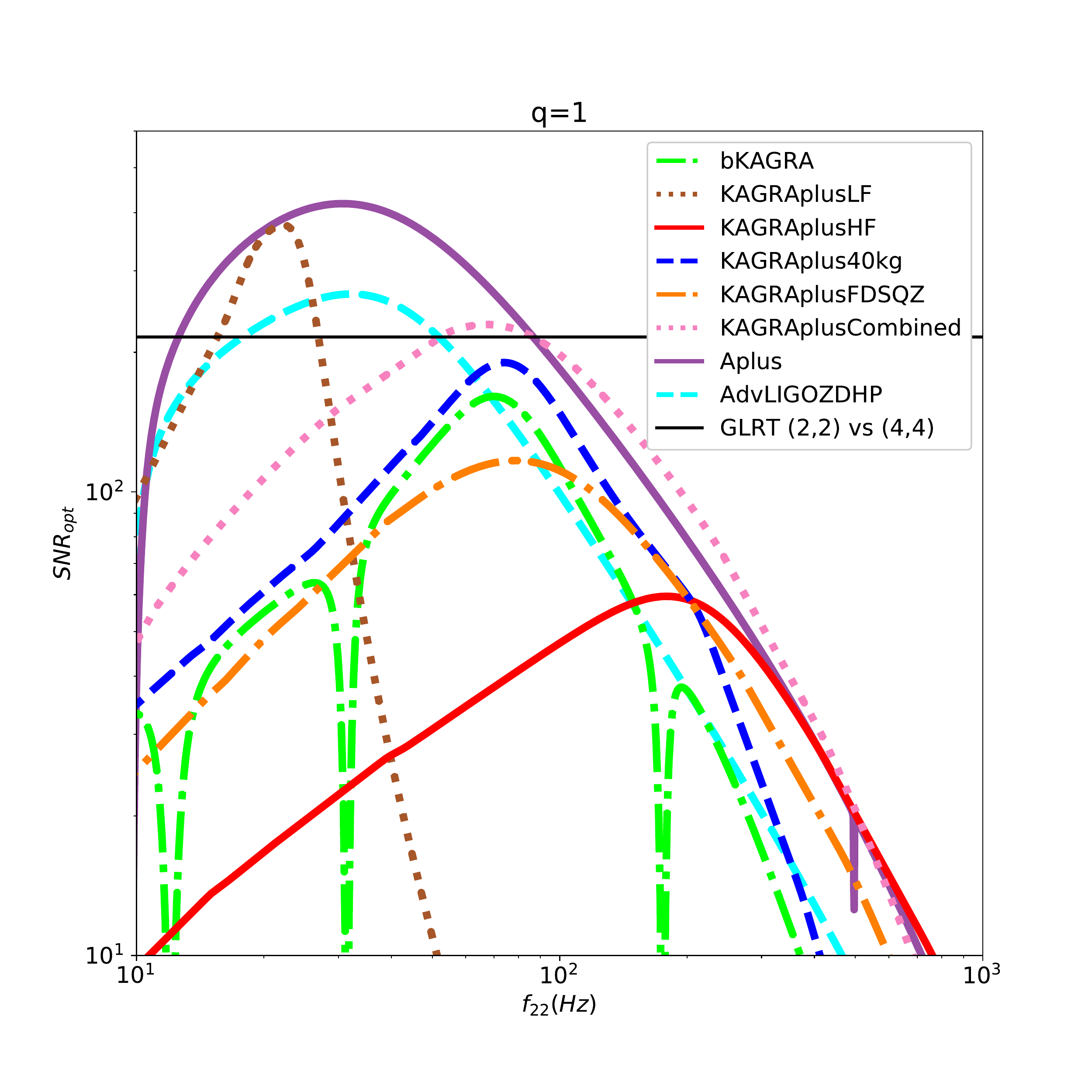}
 \includegraphics[width=0.48\textwidth]{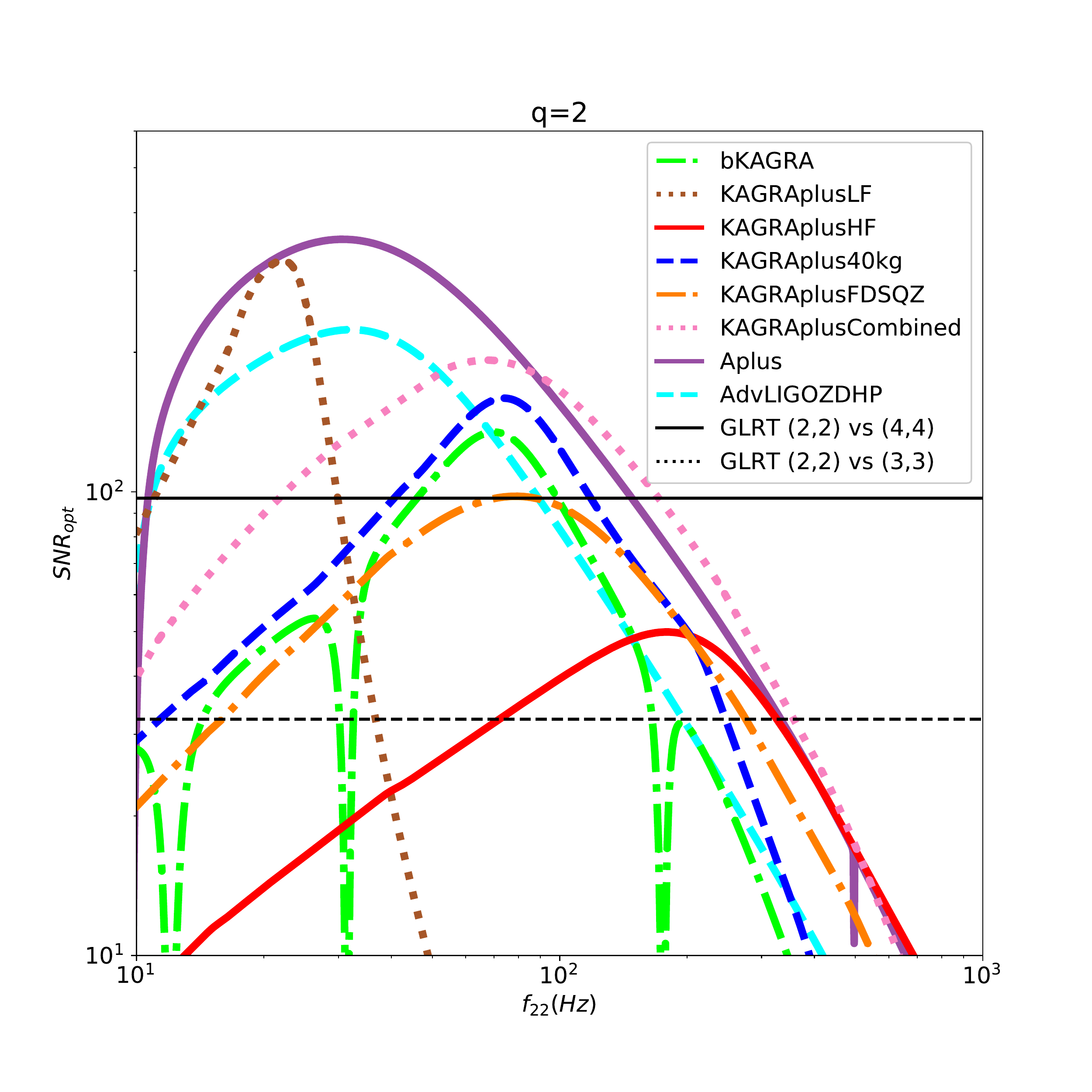}
\caption{SNRs vs. the real part of the ($\ell=m=2$) QNM frequency for the mass ratio $q=1$ (left) and $2$ (right),
assuming the optimal source direction.
Here, the luminosity distance is $r=400$~Mpc.
We have also plotted the critical SNR evaluated by the GLRT
with the following assumptions~\cite{Berti:2016lat}:
we have already evaluated the presence of one ringdown signal,
and known the parameters for the QNMs and the amplitude of the dominant mode.}
\label{fig:GLRT}
\end{figure}

\noindent {\it - Future prospects}

Based on Ref.~\cite{Baibhav:2018rfk}, 
we discuss which noise curve for KAGRA is best for
observing multiple QNMs
in the ringdown data analysis. First, to estimate the SNR of the QNM GWs, 
we use Eq.~(15) of Ref.~\cite{Baibhav:2018rfk},
\begin{equation}
\rho_{\ell m}^2 = \left(\frac{M {\cal A}_{\ell m} \Omega_{\ell m}}{r}\right)^2
\frac{\tau_{\ell m}}{2 S_h(f_{\ell m})} \,,
\label{eq:SNR_sq}
\end{equation}
where ${\cal A}_{\ell m}$ is the QNM amplitude related to
the radiation efficiency~\cite{Baibhav:2017jhs},
$\Omega_{\ell m}$ is the sky sensitivity, 
$r$ is the luminosity distance, and $\tau_{\ell m}$ and $f_{\ell m}$ denote $\tau$ and $f_{\rm R}$
of the ($\ell, m$) QNM.

To check the possibility of testing the no-hair theorem, 
we use the generalized likelihood ratio test (GLRT)~\cite{Berti:2007zu}
where the SNR greater than $\rho_{\rm GLRT}$ can be resolved from
the dominant ($\ell=m=2$) mode~\cite{Berti:2016lat}:
\begin{eqnarray}
\rho_{\rm GLRT}^{2,3} &=& 17.687 + \frac{15.4597}{q-1} - \frac{1.65242}{q} \,,\\
\rho_{\rm GLRT}^{2,4} &=& 37.9181 + \frac{83.5778}{q} + \frac{44.1125}{q^2} + \frac{50.1316}{q^3} \,,
\label{eq:SNR_GLRT}
\end{eqnarray}
for the ($\ell=m=3$) and ($\ell=m=4$) modes, respectively.
Here, $q>1$ and the ($\ell=m=3$) mode is not excited for the equal mass case.
In the following analysis, we treat the cases of $q=1$ and $2$.
The remnant BH spin becomes $\chi=0.6864$ for $q=1$ (RIT:BBH:0198)
and $\chi=0.6235$ for $q=2$ (RIT:BBH:0117)~\cite{Healy:2017psd,RITcatalog}.

In Fig.~\ref{fig:GLRT}, we show
the SNR of the ($\ell=m=2$) QNM in terms of $f_{22}$
in the case of $r=400$~Mpc and $\Omega_{22}=\Omega_{22}^{\rm max}$ by using Eq.~(\ref{eq:SNR_sq}).
We have also plotted the critical SNR evaluated by the GLRT
in Eq.~(\ref{eq:SNR_GLRT}) (note the caption).
It should be noted that 
the mass ratios of BBH estimated in the current observation
are less than $2$~\footnote{
Note that a recent observation, GW190412~\cite{LIGOScientific:2020stg}
has a larger mass ratio of $\sim 3.6$.} (the median values)~\cite{GWTC1},
although unequal mass BBH give a better opportunity to test the no-hair theorem.

\begin{figure}[t]
\centering
 \includegraphics[width=0.48\textwidth]{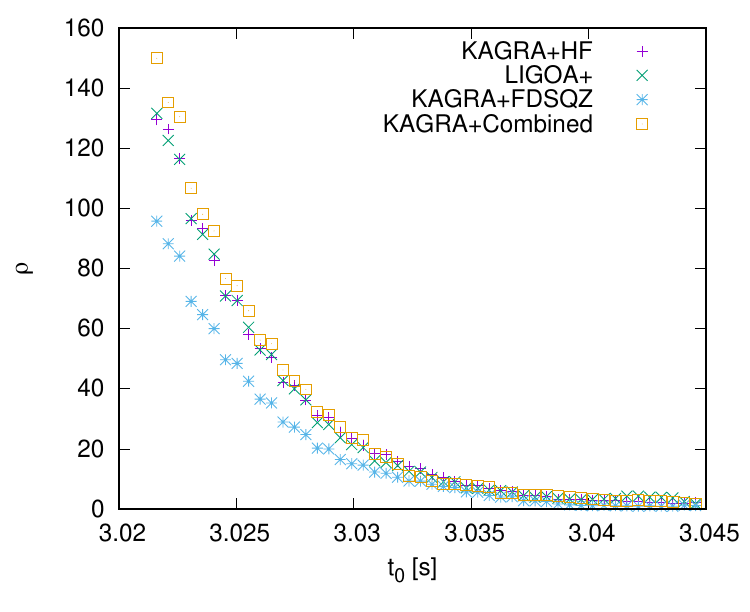}
\caption{Evaluated SNRs of the ringdown phase in colored Gaussian noises.
The input GW is an SEOBNRv2 (non-precessing) waveform with masses $45M_{\odot}$ and $15M_{\odot}$, and spin parameters $\chi_1=\chi_2=0.85$, starting at $t_0=0\,$s.
Note that the peak amplitude of the injected GW is around $t_0=3.02\,$s.}
\label{fig:curves1}
\end{figure}

\begin{figure}[t]
\centering
\includegraphics[width=0.48\textwidth]{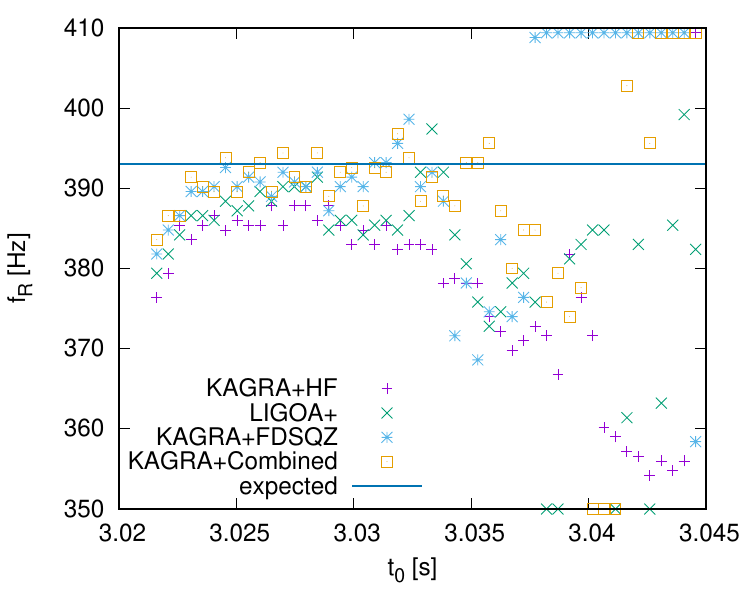}
\includegraphics[width=0.48\textwidth]{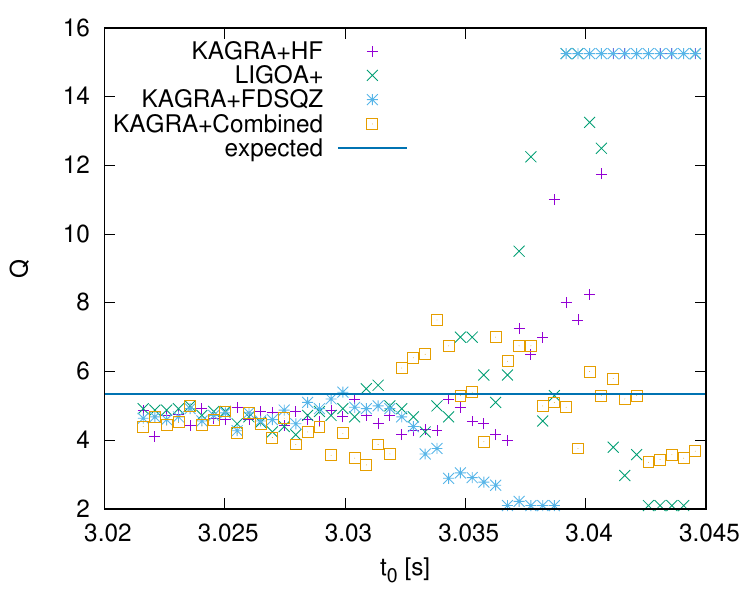}
\caption{Same as Fig.~\ref{fig:curves1}, but QNM parameter extraction. The dominant QNM has $f_R=393$Hz and $Q=5.35$.}
\label{fig:curves2}
\end{figure}

Next, we discuss effects of the shape of noise curves on the QNM parameter extraction.
Figures~\ref{fig:curves1} and \ref{fig:curves2} show an example only for the KAGRA+ noise curves, HF, FDSQZ, and Combined, because of the QNM frequency. 
In the ringdown data analysis, we can reduce the initial phase $\phi_0$ for Eq.~\eqref{eq:ringdown} analytically. Then, we assume that the initial time $t_0$ as a free parameter, and search $f_R$ and $Q$ to maximize the SNR against each $t_0$. Therefore, we obtain the SNR, $f_0$ and $Q$ for each $t_0$.
We see that there are some biases
which depend on the detector's configurations.
It is not clear what this means,
and we will give a more detailed analysis in the future~\cite{Uchikata_prep}.

\section{Late-time cosmology}
\subsection{Measurement of the Hubble constant}
\label{sec:Hubble-const}

\noindent {\it - Scientific objective}

A GW signal from a compact binary provides a unique way to measure the luminosity distance to  the source. Given a source redshift, a compact binary can be utilized for measuring the cosmic expansion \cite{Schutz:1986Nature} and is called {\it the standard siren} in this cosmological context \cite{Holz:2005ApJ}. Particularly, the standard siren is expected to play a crucial role in measuring the Hubble constant in the next five years, using relatively nearby ($z \lesssim 0.1$) GW sources, and to resolve the discrepancy problem of the Hubble constants between the cosmological measurements such as cosmic microwave background \cite{Planck2018cosmology} and baryon acoustic oscillation \cite{Schoneberg:2019wmt,Cuceu:2019for} and the local measurements with Cepheid variables applied to supernovae \cite{Riess:2018byc,Riess:2019cxk} at $7$ -- $10\,\%$ level. While the measurement by gravitational lensing time delay have given values consistent with that from the local measurements \cite{Collett:2019hrr,Wong:2019kwg,Chen:2019ejq}. Recently the local measurement calibrated by the red giant branch gave a value consistent with that from the cosmological measurements \cite{Freedman:2019jwv}. Therefore, pinning down the Hubble constant is an urgent issue to be solved for understanding the standard model of cosmology and could be a smoking gun of a new physics.\\

\noindent {\it - Observations and measurements}

The availability of the standard siren essentially depends on whether a source redshift is obtained or not. There are two ways to obtain a source redshift for an individual GW event observed by the current GW detectors: observing an EM counterpart \cite{mmaGW170817} or identifying a unique host galaxy without an EM counterpart \cite{Cutler:2009qv}. If compact binaries are BNS or BH-NS binaries, one can obtain the source redshift by observing an EM counterpart that occurs coincidentally with the GW event and identifying a unique host galaxy from the sky position of the EM counterpart. For BBH, since an EM counterpart cannot be expected, redshift information needs to be obtained by identifying its host galaxy and following-up with EM spectroscopic observations. To do that, the sources must be located at lower redshifts $z \lesssim 0.1$, having high SNRs, and must be measured with good determinations of distance and sky localization \cite{Nishizawa:2016ood}. Even if a unique host galaxy is not identified, information about the source redshift distribution based on galaxy catalogs is useful. By combining multiple BBH events, it is possible to select out a statistically consistent set of the cosmological parameters~\cite{MacLeod:2008PRD}.    

Current observational constraints from one BNS merger (GW170817) and several BBH mergers with statistical redshifts are at the precisions of $17\,\%$ \cite{Abbott:2017xzu} and $53\,\%$ \cite{Soares-Santos:2019irc}, respectively. As the luminosity distance error is a dominant error in the measurement of the Hubble constant and degenerates with the inclination angle of a compact binary, the error in the Hubble constant can be improved by determining the inclination angle independently of the GW observation. With the help of the radio imaging observation of the superluminal jet, the inclination angle of the BNS has been constrained and the Hubble constant has been measured at the precision of $\sim 7\,\%$ \cite{Hotokezaka:2018dfi}. However, these constraints are still weak to resolve the discrepancy problem and need to be improved further. To realize such a precision in the future, the modeling uncertainty of a jet, which is currently estimated to be a few percents~\cite{Hotokezaka:2018dfi}, also needs to be addressed.\\

\noindent {\it - Future prospects}

In all methods above, a good measurement of sky localization volume is essential for the identification of source redshifts and the use of the compact binaries as a cosmological probe. We estimate the measurement errors of the sky localization volume with the Fisher information matrix \cite{Nishizawa:2017nef}. The sky localization volume for each compact binary is defined by \cite{Nishizawa:2016ood}  
\begin{equation}
\Delta V \equiv \left[ V (d_{L,\rm max}) - V (d_{L,\rm min}) \right] \frac{\Delta \Omega_{\rm S}}{4\pi} \;.
\label{eq:Verr}
\end{equation}
Here $V(d_L)$ is the comoving volume of a sphere with the radius $d_L$. The maximum and minimum luminosity distances are determined by $d_{L,{\rm max}} = d_{L} (z_{\rm f}) + \Delta d_{L}$ and $d_{L,{\rm min}} = \max \left[ d_L(z_{\rm f}) - \Delta d_{L}, 0 \right]$, where $z_{\rm f}$ is a fiducial source redshift and $\Delta d_{L}$ is a 2$\sigma$-parameter estimation error of luminosity distance. In Table~\ref{tab:dV-errors}, the sky localization volumes for various GW sources with different masses and detector networks are shown.\\ 

\begin{table}[t]
\begin{center}
\begin{tabular}{llcccccc}
\hline \hline
source & quantile & bKAGRA & LF & HF & 40kg & FDSQZ & Combined \\
\hline \hline  
$30\,M_{\odot}$ BBH & top 1\% & 0.661 & 0.778 & 0.600 & 0.454 & 0.483 & 0.335 \\
 & median & 3.70 & 10.8 & 3.37 & 2.65 & 2.75 & 1.63 \\
$10\,M_{\odot}$ BBH & top 1\% & 3.47 & 3.61 & 1.89 & 2.62 & 2.15 & 1.59 \\
 & median & 21.8 & 52.6 & 11.6 & 15.4 & 12.5 & 8.53 \\
$10\,M_{\odot}$ BH-NS & top 1\% & 8.31 & 10.2 & 6.52 & 5.68 & 5.74 & 3.71 \\
 & median & 44.2 & 115 & 40.0 & 31.2 & 31.6 & 22.8 \\
BNS & top 1\% & 0.261 & 0.315 & 0.148 & 0.222 & 0.205 & 0.120 \\
& median & 1.79 & 3.76 & 0.967 & 1.33 & 0.995 & 0.743 \\
\hline 
\end{tabular}
\end{center}
\caption{Top 1\% and median errors of the sky localization volume in the unit of $10^3\,{\rm Mpc}^3$. The mass of a NS is $1.4\,M_{\odot}$. The BBH is at the distance of $z=0.1$, the $10\,M_{\odot}$ BH-NS is at $z=0.06$, and the BNS is at $z=0.03$. bKAGRA, LF, HF, 40kg, FDSQZ, and Combined denote the detector network composed of A+, AdV+, and bKAGRA or KAGRA+ (low frequency, high frequency, $40\,{\rm kg}$, frequency-dependent squeezing, and combined, respectively).}
\label{tab:dV-errors}
\end{table}

\begin{itemize}

\item{BNS and BH-NS}\\
Based on our results of the sky localization volume, BNS are much better than BH-NS binaries. Even considering the merger rate uncertainty of BH-NS and the effect of spin precession on parameter estimation, a typical error in the Hubble constant only with BH-NS binaries is still worse than that with BNS \cite{Vitale:2018wlg}. Using $N$ BNS observed by the detector network of two aLIGO and one AdV with redshift information from EM counterparts, the Hubble constant error is estimated to be $\Delta H_0/H_0 \sim 2 \sqrt{50/N}\, \%$ \cite{Seto:2017swx,Chen:2017rfc}. With the network of the KAGRA+ near-term upgrades, the network SNR is larger by about $1.5$ times and the Hubble constant error is $\Delta H_0/H_0 \sim 1.5 \sqrt{50/N}\, \%$. However, the number of sources accompanying with EM counterparts is still significantly uncertain and 50 sources roughly correspond to $\sim 10\%$ redshift identification, which is an intermediate value of theoretical predictions \cite{Gupte:2018pht,Howell:2018nhu,Mogushi:2018ufy}. Based on the comparison of the sky localization volume in Table~\ref{tab:dV-errors}, the improvement factors for BNS (median) are LF 0.476, 40kg 1.35, FDSQZ 1.80, HF 1.85, and Combined 2.41. The upgrades at higher frequencies lead to smaller localization volume and better sensitivity to the Hubble constant.\\

\item{BBH}\\
Seen from our results of the sky localization volume, $30\,M_{\odot}$ BBH are better than $10\,M_{\odot}$ BBH and are easier to identify a unique host galaxy (The number density of galaxies that covers roughly 90\% of the total luminosity in B-band \cite{Gehrels:2015ApJ} is $n_{\rm gal}=0.01\,{\rm Mpc}^{-3}$) and then obtain redshift information. Assuming the number density of galaxies, top 1\% events of $30\,M_{\odot}$ BBH at the distance closer than $z \lesssim 0.03$ can marginally identify a unique host galaxy. Extrapolating the result obtained for the bKAGRA detector network \cite{Nishizawa:2016ood} to the KAGRA+ near-term upgrade networks, the Hubble constant error is $\Delta H_0/H_0 \sim 0.8 \sqrt{10/N}\, \%$. Based on the comparison of the sky localization volume in Table~\ref{tab:dV-errors}, the improvement factors for $30\,M_{\odot}$ BBH (top1\%) are LF 0.85, 40kg 1.46, FDSQZ 1.37, HF 1.10, Combined 1.97. For BBH, the upgrades at middle frequencies lead to smaller localization volume and better sensitivity to the Hubble constant.

\end{itemize}

In conclusion, we cannot tell which source gives better determination of the Hubble constant because of uncertainties in the merger rates of BNS and BBH, the success fraction of redshift identification, and the mass function of BBH. In addition, there would be potential contributions from systematic errors, e.g.~the modeling of the EM counterparts and the calibration errors of GW detectors. Although all of these uncertainties must be reduced or improved, the standard siren has a potential to achieve the $1\%$-precision measurement of the Hubble constant in the future and resolve the discrepancy problem, for which KAGRA can contribute mainly from better sky localization.

\subsection{GW lensing}

\noindent {\it - Scientific objective}

Gravitational lensing is a well-established subject in the EM
observation of the Universe~\cite{Turner:1984ch, Treu:2010uj}.  For GWs, there are controversial claims that lensing was already
observed as well from LIGO-Virgo O1/O2 detections~\cite{Broadhurst:2018saj,
Broadhurst:2019ijv}, while the LIGO-Virgo collaboration claimed that the
observed events are fully consistent with unlensed GWs~\cite{GWTC1}. From an astrophysical viewpoint on the
lensing rate, it is unlikely that the observed events are multiple-imaged by
the stellar-mass objects~\cite{Christian:2018vsi}, intervening galaxy
population~\cite{Ng:2017yiu}, or massive galaxy clusters~\cite{Smith:2017mqu}.
However, for the second-generation ground-based GW detectors,
it is possible to observe several lensing events at design
sensitivity~\cite{Smith:2017mqu, Ng:2017yiu, Li:2018prc}, and almost for sure
for the third-generation detectors to detect lensed events~\cite{Li:2018prc,
Christian:2018vsi}.  The scientific reward from lensing of GWs
is huge~\cite{Finn:1995ah, Wang:1996as, Nakamura:1997sw}, just to name a few,
ranging from probing small structures such as compact dark matter or the
primordial BHs with mass $M \approx 10$--$10^5\,
M_\odot$~\cite{Jung:2017flg}, to discovering intermediate-mass BHs~\cite{Lai:2018rto}, to testing gravity theories and
cosmography~\cite{Schutz:1986Nature, Finn:1995ah}.\\

\noindent {\it - Observations and measurements}

The detection of possible lensing events has multiple implications for
measurement. First, the lensed event is magnified in its SNR
by $\sqrt{\mu}$ where $\mu$ is the magnification~\cite{Wang:1996as}, thus if
the possibility for lensing is not considered,  one will underestimate its
luminosity distance by $\sqrt{\mu}$, and overestimate the masses in the source
frame with a relevant redshift rescaling thereof~\cite{Wang:1996as, Nakamura:1997sw,
Takahashi:2003ix, Smith:2017mqu}.  Second, because of the magnification effect,
for events at large $z$ whose unlensed SNR is below the
detection threshold, lensing might lift it to be detectable~\cite{Wang:1996as}.
Therefore, lensing affects the inference on the event rate and underlying mass
distributions (and stellar evolution) of coalescing binaries. Third, if the
size of lens is comparable to or smaller than the GW
wavelength, diffraction introduces frequency-dependent time delays and
modulation of the waveform~\cite{Nakamura:1997sw, Takahashi:2003ix,
Christian:2018vsi}, thus reducing the matched-filter capability if unlensed
waveforms are used as template. Fourth, for GW events with
observable EM counterparts, it is important to associate them with
each other in a convincing manner which enables tests of the speed of gravity
and inference of cosmological parameters~\cite{Schutz:1986Nature, Finn:1995ah,
Collett:2016dey, Fan:2016swi}. A precise localization with multiple detectors
helps with such statistical association.\\

\noindent {\it - Future prospects}

For KAGRA, it will help in localizing
GW events significantly together with LIGO and Virgo
observations, thus increasing the possibility to detect lensed events. No
specific hardware requirement is additionally requested to achieve the
gravitational lensing science goal. However, advanced data analysis including the possibility of lensing events is required. For the operation, coincident observation
with LIGO and Virgo detectors is advantageous. The contribution of KAGRA data
to the cooperative parameter estimation with lensing effects encoded in the
waveform template will benefit the overall science output. Therefore, the same suggestion as the measurement of the Hubble constant with BBH in Sec.~\ref{sec:Hubble-const} is applied here.

\section{Multimessenger observations}

\subsection{Kilonovae/Macronovae}
\label{sec:kilonovae}

\noindent {\it - Scientific objective}

A kilonova (a.k.a. macronova) is the emission which has been expected to be associated with a NS-NS or BH-NS merger as the consequence of the mass ejection from the system, e.g.~\cite{Rosswog:1998hy,Hotokezaka:2012ze}. Since the ejected material is composed of neutron-rich matter, heavy radioactive nuclei would be synthesized in the ejecta by the so-called {\it r-process} nucleosynthesis~\cite{Lattimer:1974slx,Eichler:1989ve,Korobkin:2012uy,Wanajo:2014wha}, and EM emission could occur by radioactive decays of heavy elements~\cite{Li:1998bw,Kulkarni:2005jw,Metzger:2010sy,Kasen:2013xka,Tanaka:2013ana}. Kilonova emission is expected to be bright in the optical and near infrared wavelengths and lasts for $1\sim10$ days after the merger. We note that the event rate is currently highly uncertain, while it could be $\sim10^3\,{\rm Gpc}^{-3}{\rm yr}^{-1}$ if kilonovae associates with all NS-NS mergers. 

The lightcurve of kilonova reflects the merger process and the late time evolution of the merger remnant~\cite{Li:1998bw,Kasen:2013xka,Kasen:2014toa,Barnes:2016umi,Wollaeger:2017ahm,Tanaka:2017qxj,Tanaka:2017lxb}. The presence of kilonova emission suggests that substantial amount of material is ejected from the system, and this gives constraints on the binary parameters by combining with a theoretical prediction obtained by numerical simulations. Furthermore, detailed properties of kilonova lightcurves, such as those brightness and color evolution, provide information of the mass and the composition of the ejected material~e.g., \cite{Kasliwal:2017ngb,Cowperthwaite:2017dyu,Kasen:2017sxr,Villar:2017wcc,Perego:2017wtu,Tanvir:2017pws,Kawaguchi:2018ptg}. Such information is useful to understand the merger process of the binary and the late-time evolution of the merger remnant as well as the chemical evolution of the universe.\\

\noindent {\it - Observations and measurements}

The improvement in the source localization is important to identify the host galaxy of a GW event. The source localization only by the GW data analysis would not be tight enough to uniquely identify the host galaxy even after KAGRA has joined the observation~\cite{GBM:2017lvd} (see Tables~\ref{tab:BHNS-PE-errors} and~\ref{tab:BNS-PE-errors}). Thus, the detection of the EM counterparts will play an important role to identify the host galaxy. In particular, searching the kilonova could have an advantage for this purpose because its emission is expected to be approximately isotropic (c.f. a short GRB). However, the detection of the kilonova will be more challenging at large distances because the kilonova will be dimmer as the distance to the event increases and the number of galaxies in the localized area significantly increases. For example, assuming the same intrinsic brightness as in GW170817, the peak {\it r}-band brightness of the kilonova would be fainter than $\approx21\,{\rm mag}$ at $d_{\rm L}\geq$200 Mpc, which is dimmer than the typical 5$\sigma$ limiting magnitude of 1-m class telescopes with 60 s exposure~\citep{Nissanke:2012dj}. Thus, the improvement of the source localization via GW detection is still the key point to restrict the number of the host candidate and to achieve simultaneous detection of the kilonova of an event.

The low-latency GW alert is also a key point for the follow-up of the EM counterparts. It helps us to find the kilonova because the lightcurves, in particular in the wavelength shorter than optical bands, are most bright within a few days. Moreover, the lightcurves of the EM-counterparts at $< 1\,{\rm day}$ are in particular of interest. For example, in addition to the kilonova emission, the so-called cocoon emission, which is powered by internal energy deposited by relativistic jets, can contribute the early part of the lightcurve in the optical and UV wavelengths~\cite{Gottlieb:2017mqv}. It is pointed out that the observed data of lightcurves in $< 1\,{\rm day}$ can be indicative to distinguish these different components of emission~\cite{Arcavi:2018mzm,Matsumoto:2018gzy}.

Tighter constraint on the inclination angle has a great impact to maximize the scientific return from the observation of the kilonova. Lightcurves of a kilonova are expected to have dependence on the observation angle~\cite{Kasen:2014toa,Perego:2017wtu,Wollaeger:2017ahm,Kawaguchi:2018ptg}. Much more information about the property of the ejecta can be extracted if tighter constraint on the inclination angle is given. Indeed, for the recent BH-NS-like merger event GW190814~\cite{Abbott:2020khf}, the constraint on the ejecta mass is obtained by the upper limits to the EM signals, while the tightness of the constraint is shown to be dependent strongly on the viewing angle~\cite{Andreoni:2019qgh,Kawaguchi:2020osi} (by a factor of $\approx3$). \\

\noindent {\it - Future prospects}

The improvements of the source localization both in speed and accuracy are crucial and would be required as the first priority for kilonova follow-up observations. The detector sensitivity should be broadband because a low frequency band is important for quick localization (early warning) and a high frequency band is important for accurate localization. According to Tables~\ref{tab:BHNS-PE-errors} and~\ref{tab:BNS-PE-errors}, the median localized area is improved by up to a factor of $\approx 2$ by including KAGRA+ in the detector network (by upgrading bKAGRA to one of KAGRA+ configurations).  
This enables us to search the localized fields with deeper limiting magnitudes or with less latency. For example, the EM search deeper by $\approx1$ magnitude is in principle possible if the improvement in the localization area enabled us to spend 10 times longer exposure time. The follow-up by a 2-m class or larger class telescopes, such as Pan-STARRS~\citep{PanSTRR} and CTIO-Dark Energy Camera~\citep{DECam}, are crucial for the detection for such a situation~\citep{Nissanke:2012dj}. 
We also note that increase in the number of detectors with the stable operations will enhanced the chance for the multiple detectors to be operated at the same time, which is crucial for the source localization.

Tables~\ref{tab:BHNS-PE-errors} and~\ref{tab:BNS-PE-errors} show that the determination of the inclination angle is improved by up to $\approx 23\%$ by including KAGRA+ in the detector network (by upgrading bKAGRA to one of KAGRA+ configurations.  As mentioned above, such improvement will help us to provide the tighter constraint on ejecta properties and provide us a good opportunity to study the spatial distribution of ejecta. For example, numerical relativity simulations of a NS-binary merger (e.g.,~\cite{Radice:2016dwd,Sekiguchi:2015dma,Sekiguchi:2016bjd,Perego:2017wtu}) predict that the typical polar opening angle of the lanthanide-rich dynamical ejecta component ($\gtrsim30^\circ$) is comparable to or larger than the determination error of the inclination angle for $\Delta {\rm cos}\iota\approx 0.15$ (see Tables~\ref{tab:BHNS-PE-errors} and~\ref{tab:BNS-PE-errors}). Since the presence of the lanthanide-rich material in the line of sight  strongly affects the lightcurves~\cite{Kasen:2014toa}, the observations of kilonovae with the improved measurement of the inclination angle would enable us to examine the presence of such ejecta components.

\subsection{Short gamma-ray bursts}

\noindent {\it - Scientific objective}

The origin of sGRBs is a long-standing problem for more than 40 years
\cite{Nakar07,Berger14}.
A merger of BNS has been thought to be
the most promising candidate
\cite{Paczynski86,Goodman86,Eichler+89},
although an unequivocal evidence was missing.
Although it is widely accepted that
a GRB is produced by a relativistic jet
with a Lorentz factor larger than $\sim 100$,
there remain unresolved issues.
(1) The central engine is most likely
a spinning BH surrounded by an accretion disk,
while a millisecond magnetar remains a candidate.
(2) The jet formation mechanism is one of the central subjects in astrophysics.
It could be the Blandford-Znajek mechanism
or the neutrino annihilation mechanism.
For the former case, it is not known
how to realize the large-scale, poloidal configuration of the magnetic field.
(3) The jet composition is unknown.
It is likely dominated by baryon kinetic energy or magnetic energy.
(4) The emission mechanism is a puzzle.
The emission radius still has approximately four order-of-magnitude uncertainty.
The popular mechanism is synchrotron emission,
while there is no consensus how to realize
the observed spectral relations such as Amati and Yonetoku relations
\cite{Amati+02,Yonetoku+04}.
(5) The jet structure is unknown both in the radial and polar directions.
The long-lasting emission following the main prompt emission
such as extended emission ($\sim 10^2$ s)
and plateau emission ($\sim 10^4$ s)
suggest a long-lived central engine activity \cite{Kisaka+17}.

The detection of the GW event GW170817
\cite{GW170817PRL}
and the associated EM counterparts
\cite{mmaGW170817}
revolutionized the situation.
In particular, sGRB 170817A
detected two seconds ($\sim 1.7$ s) after GW170817
\cite{GW170817:GRB,GRB170817A_GBM,GRB170817A_INT}
and the following afterglows in radio to X-ray
\cite{Troja+17,Margutti+17,Haggard+17,Hallinan+17,Alexander+17,Lyman+18}
give the first direct clues to the origin of sGRBs.
However the situation is not so simple
because sGRB 170817A was very weak
with an isotropic-equivalent energy
$E_{\gamma,\rm iso} \sim 5.35 \times 10^{46}$ erg,
which is many orders of magnitude smaller than ordinary values.
On the other hand,
the afterglow observations, in particular
of superluminal motion in radio
\cite{Mooley+18b,Ghirlanda+18}
and the consistency between the spectral index
and the light curve slope after the luminosity peak
\cite{Troja+18b,Mooley+18c,Lamb+19},
strongly suggest that a relativistic jet is launched and
successfully breaks out the merger ejecta in this event
\cite{Nagakura+14}.
The jet power should be similar to those in the other normal sGRBs,
otherwise the jet cannot penetrate the merger ejecta \cite{Hamidani19}.
The most likely scenario is
that the jet is off-axis to our line-of-sight
and thereby faint in this event
\cite{IN18,GW170817:GRB,Granot+17},
although the emission mechanism is controversial
\cite{Kasliwal+17,Kisaka+18,Nakar+18,Nakar19}.
The afterglow observation,
in particular the slowly-rising light curve,
is not consistent with a top-hat jet \cite{Mooley+18a},
but for the first time strongly suggests a structured jet
\cite{Troja+18,Ruan+18,Margutti+18,D'Avanzo+18,Lazzati+18,Lyman+18,Troja+18b,Ghirlanda+18,Lamb+19}.
The angular structure is also important for solving the spectral puzzles of GRB~170817A \cite{Kisaka+18,IN19,Matsumoto19a,Matsumoto19b}.\\


\noindent {\it - Observations and measurements}

The GW signal is basically the same as that of NS mergers
with the dimensionless amplitude $h_c \propto f^{-1/6}$ and
$f_{\rm max} \sim$ a few kHz
(see Sec.~\ref{sec:BNS-evolution}).
The mass and mass ratio
are important for revealing the jet formation mechanism
since they determine the BH and disk masses after the merger.

Sky localization is essential to follow up EM counterparts
\cite{LVKScenario2018}.
In GW170817, the error region shrinks from 190 deg$^2$ to 31 deg$^2$
by adding Virgo data even though the SNR was $2.0$.
The survey with Subaru/Hyper Suprime-Cam
(one of the most powerful telescope in the world) covered 23.6 deg$^2$
and reached the 50\% completeness magnitude of 20.6 mag \cite{Tominaga+18}.
{At the design sensitivities of aLIGO, AdV, and KAGRA}, a typical source distance is further
(or the event closer than $40$ Mpc is rare
given the estimated event rate
$1 \times 10^{2}$ - $4\times 10^3$ Gpc$^{-3}$ yr$^{-1}$
\cite{GWTC1}).
Thus a localization error is desired to be less than $\sim 30$ deg$^{2}$ and it is achievable with the detector networks including KAGRA as shown in Tables~\ref{tab:BHNS-PE-errors} and~\ref{tab:BNS-PE-errors}.
Low latency should be also mandatory.
In GW170817, the alert with the localization error $\sim 31$ deg$^2$
was sent $\sim 5$ hours after the merger,
although the actual observations started
$\sim 10$ hours after the merger because of the interruption of the sea.
In {a couple of years,} at least minutes or even lower latency will be realized.
{Ground-based detectors} should try localization before the merger \cite{Cannon+12}
to allow other messengers (in particular with small field of view)
to observe the source at the merger time
as well as possible precursor emissions.

The inclination (or viewing angle) of the merging binary
is also a crucial parameter for sGRB studies.
It affects all the prompt, afterglow, and kilonova emissions.

The post-merger signals are also worth to search
for revealing the central engine of sGRBs \cite{Abbott+17}.
They include GW emission from
a hypermassive NS (or supramassive NS) and BH (see Sec.~\ref{sec:NS-remnant}).
There are many possibilities of GW emission mechanisms,
such as a BH ringdown,
NS quadrupolar $f$-mode,
magnetic-field-induced ellipticities,
unstable bar modes,
unstable $r$-modes and so on.
A possible range of GW duration is broad
from $\sim 10^{-3}$ s to $\sim 10^4$ s or even longer,
as implied by long-lasting activities of sGRBs.


A magnetar giant flare is also observed as a sGRB. Some GWs are expected from giant flares (see Sec.~\ref{sec:magnetar-flares})~\cite{Ioka+00}.
Currently there is no evidence of GWs associated with giant flares \cite{Abadie+10},
although the $f$-mode may not be the correct target \cite{Kashiyama+11}.
Possible candidates of giant flares include
GRB 070201 \cite{Abbott+07},
GRB 051103 \cite{Abadie+12}
and GRB 150906B \cite{Abbott+16}.\\

\noindent {\it - Future prospects}

As a summary, the frequency band is required to be broadband,
where the low-frequency band is important for quick localization (early warning)
and the high-frequency band is important for the merger time and the post merger signals.
In addition, inclination measurements are crucial. These are the same as those for a kilonova. See Sec.~\ref{sec:kilonovae}.
As a typcial distance to GW sources gets farther, only sGRBs remain as detectable EMnetic counterparts. Collaboration with X- and gamma-ray observatories such as SVOM, Einstein Probe, HiZ-GUNDAM and CTA will become more important.

\subsection{Long gamma-ray bursts}

\noindent {\it - Scientific objective}

Long GRBs, whose duration is longer than about 2~s, originate in the collapse of a massive star
\cite{piran04,kumar15}.
Prompt gamma-ray emissions and subsequent afterglows are emitted at the head of the relativistically moving,
narrowly collimated jet toward us.
The Lorentz factor of the jet is typically larger than $\sim10^2$.
Currently the jet launch mechanism as well as the radiation process is not yet fully understood.
The central engine of the relativistic jet is also unclear, but most likely massive accretion disk around a newly formed
BH.
It has been attempted to extract the information on the dynamics and/or structure of the jet from the observed 
EM emissions to tackle this problem.
However, the EM emission occurs far from the central engine, so that
previous EM observations had little direct information to answer the problem.\\

\noindent {\it - Observations and measurements}

In the collapsar model for long GRBs \cite{macfadyen99}, the highly spinnig core of a massive star collapses into a
rotating BH with an accretion disk, or sometimes a NS, producing a relativistic jet along the rotation axis.
Then, the rotating compact objects themselves emit GWs
\cite{stark1985,kobayashi03}. 
This is the cases of supernovae and hypernovae. So we yield to Sec.~\ref{sec:science-supernova} for more quantitative discussion. 
The accretion disk around the BH is massive ($\sim M_\odot$), so that  asymmetrical blobs arise via gravitational instabilities and may be responsible for GWs \cite{kobayashi03} as well as observed gamma-ray variability.
The characteristic amplitude of the GW emitted by the blob with a mass $m_b$
is estimated as
\begin{equation}
h_c(f)\sim 1\times 10^{-22} \left(\frac{M_c}{M_\odot} \right)^{5/6}\left(\frac{d}{100~{\rm Mpc}}\right)^{-1}\left(\frac{f}{f_{\rm max}}\right)^{-1/6}
\end{equation}
and the maximum frequency $f_{\rm max}\sim300(M_{\rm BH}/10M_\odot)^{-1}$~Hz, where
$d$ is the distance to the source,
$M_{\rm BH}$ is the BH mass, and the chirp mass
$M_c \sim 0.6M_\odot(m_b/0.1M_\odot)^{3/5}(M_{\rm BH}/10M_\odot)^{2/5}$.


In order to understand the jet launch mechanism, it should be clarified when and how the jet departs at the central engine. 
GW emission also takes place while the jet is accelerated.
Such gravitational radiation has a ``memory'', that is, the metric perturbation
does not return to its original value at the end of the jet acceleration \cite{braginsky87}.
Suppose that the mass $M$ is accelerated to a Lorentz factor $\gamma$.
The memory GW is antibeamed (nearly isotropic) except for the
on axis direction within the angle $\gamma^{-1}$ around the jet axis \cite{segalis01,sago04}.
The characteristic amplitude is given by
\begin{equation}
h_c(f)\sim1\times10^{-24}
\left(\frac{Mc^2}{10^{52}~{\rm erg}}\right)
\left(\frac{d}{100~{\rm Mpc}}\right)^{-1}
\left(\frac{f}{f_c}\right)^{-1}~~,
\label{eq:GWmemory_GRB_RY}
\end{equation}
where $f_c\sim T_{\rm dur}^{-1}\sim0.1(T_{\rm dur}/10~{\rm s})^{-1}$Hz,
and $T_{\rm dur}$ is the total duration of the burst.
When the acceleration takes over a time $\delta t$, the maximum frequency $f_{\rm max}$
is given by $f_{\rm max}\sim\delta t^{-1}\sim1~{\rm kHz}~(\delta t/10^{-3}\,{\rm s})^{-1}$.
Thus, expected amplitude looks too small to bee detected. 
However,  the amplitude may become much larger.
First, the energy source is not only limited to the rest mass energy of the jet matter $Mc^2$ but also other candidates
such as neutrino annihilation and magnetic field~\cite{suwa09}, which can potentially be, $E_s\sim10^{54}$~erg.
Second, the amplitude in the high-frequency range ($\sim f_{\rm max}$) is greatly enhanced if multiple ($\sim10^2$) short bursts
are taken into account \cite{sago04}.
Then, the signal is about two orders of magnitude larger than that estimated in Eq.~(\ref{eq:GWmemory_GRB_RY}):
\begin{equation}
h_c(f=100~{\rm Hz})\sim1\times10^{-23}\left(\frac{E_s}{10^{54}~{\rm erg}} \right)\left(\frac{d}{100~{\rm Mpc}} \right)^{-1}\;.
\end{equation}
Once such gravitational radiation is detected, we can extract important information on the time variability and active time
of the central engine as well as the jet structure.

Despite extensive searches for GWs from long GRBs,
so far the firm detection has not yet been obtained
\cite{Astone05_GRB,Abbott05_GRB,Aasi14_GRB,Abbott19_GRB}.\\



\noindent {\it - Future prospects}

For GWs from massive accretion disks or relativistic jet acceleration, observations with frequency less than 1~kHz is important. 
So the candidates of KAGRA+ (LF, 40kg, FDSQZ) have better sensitivities for the study of long GRBs.
The local rate density of usual high-luminosity long GRBs that are detectable by gamma-ray instruments
has been estimated as $\sim1$~Gpc$^{-3}$yr$^{-1}$ \cite{liang07_RY}.
This number is for events with GRB jets toward us to emit bright gamma-rays.
If the beaming factor of the jet, that is the fraction of solid angle of the gamma-ray emission to the full sky, is $f_b=1/200$, then
the intrinsic rate density is $\sim200$~Gpc$^{-3}$yr$^{-1}$. 
On the other hand, the GW emission is almost isotropic.
Hence the all-sky rate of the GW events within 100~Mpc is 
$\sim0.8$~yr$^{-1}$. 
%
Indeed, more frequent events in the local Universe are the low-luminosity long GRBs, the subclass of long GRBs \cite{liang07_RY}.
Their local rate density is two or three orders of magnitude higher than usual high-luminosity long GRBs,
so that the all sky rate of nearby ($<100$~Mpc) low-luminosity long GRBs is estimated as $\sim0.4$--4~yr$^{-1}$.
The beaming factor for this subclass is uncertain. However, if we roughly assume $f_b\sim 1/10$,
the unbeamed GW event rate is 4--40~yr$^{-1}$.

The event rates of GWs sounds promising for the current GW detectors even with bKAGRA at design sensitivity. However, the memory GW is more difficult to be detected out to 100 Mpc because the matched filtering technique for a chirp signal is not available and the memory signal is more like a burst signal. In this sense, observing with more detectors is crucial and KAGRA will be able to help veto fake signals and to enhance the network sensitivity.

\subsection{Fast radio bursts}
\label{sec:FRB}

\noindent {\it - Scientific objective}

Fast radio bursts (FRBs) are coherent radio transients with a duration of milliseconds. The most outstanding characteristic of FRBs is the dispersion measure (DM); they are as large as DM $\sim 1000\,{\rm pc}\,{\rm cm}^{-3}$, indicating that the sources are at cosmological distances. A FRB is originally discovered by Lorimer et al. in 2007 \cite{2007Sci...318..777L}. The number is significantly increased by wide-field-survey facilities like ASKAP and CHIME. See FRBCAT~\cite{FRBCAT} for an updated catalog. 

It is now known that a fraction (roughly $10 \%$) of the FRBs are repeating. 
The differences between repeating and non-repeating FRBs are still under debate. Host galaxies have been reported for a few FRBs; both for repeating and non-repeating ones. There is a diversity in the host galaxies; the host of FRB 121102 is a dwarf star forming galaxy while other host galaxies are medium-sized less-active galaxies. No multi-messenger counterpart has been reported. 

The origin of FRB is still not known, and the emission mechanism is rather uncertain. 
Given the similarities between FRB and some transient phenomena from Galactic young NSs, e.g., giant radio pulses from the Crab pulsar and magnetar giant flares, the young NS models have been most intensively investigated. See, e.g.~\cite{2017ApJ...839L...3K} and references therein. For a GW from a young NS, see Secs.~\ref{sec:magnetar-flares} and \ref{sec:science-stellar-oscillation}.
On the other hand, the most testable and interesting possibility for KAGRA and other ground-based interferometers would be the coalescing BNS model \cite{2013PASJ...65L..12T}; FRBs could be produced just before the merger, a fraction of the orbital energy being extracted by magnetic braking. Simultaneous detection of a GW and an FRB is the straightforward way to test this scenario.\\


\noindent {\it - Observations and measurements}

If a FRB is observed with a GW, the source will be determined. In general, arrival timing of the FRB relative to the GW is useful; the FRB will be emitted, say milliseconds after the the GWs. 
The timing of the FRB emission should be closely connected to the emission mechanism; information on the physical condition of the NS magnetosphere can be obtained. \\


\noindent {\it - Future prospects}

Ongoing and upcoming radio surveys could detect $O(1000)$ FRBs per year~(see e.g., \citep{2017ApJ...844..140C}), providing more opportunities to conduct multi-messenger observations. 
One can search the GW counterpart of known FRBs; in addition to Galactic young NSs, we can select relatively close extragalactic targets based on their DMs. Conversely, GW detectors could in principle give an alert to radio telescopes. Since the field of view of radio telescopes is relatively small (e.g., $\sim$0.1 deg for Parkes), a better sky localization with KAGRA is, as always, crucial. 


\section{Others}
\subsection{Cosmic strings}

\noindent {\it - Scientific objective}

Cosmic strings are one-dimensional cosmological defects that are predicted to form under phase transitions in the early Universe~\cite{Kibble76}. More recently, it has been recognized that cosmological models based on the string theory also result in string forming and growing to cosmic scales~\cite{Sarangi02,Jeannerot03}. A cosmic string is characterized by its tension, which is linked to the energy scale of the Universe when it is formed. Searching for cosmic strings will thus allow us to investigate the early time of the Universe that has never been probed before.\\

\noindent {\it - Observations and measurements}

A promising observational signature of cosmic strings is their gravitational radiation from loops~\cite{Damour00,Damour01,Damour05}. Loops can form through a process called intercommutation, in which two strings change partners upon encounter. An intercommutation of a loop by itself or of two strings on two points creates a loop, which oscillates and radiates their energy by GWs. Special points that are occasionally formed on loops, called cusps and kinks, are the important targets for GW detectors. Cusps are points on a loop that accompany large Lorentz boosts, and kinks are discontinuities on the tangent vectors of a loop~\cite{Vilenkin00}. Another interesting GW source is kinks on infinitely long cosmic strings~\cite{Matsui:2016xnp, Matsui:2019obe}, which are produced when two strings intersect. As the string network evolves with continuous intersections, a number of small kinks accumulate on infinite strings and their propagation and collisions generate GW bursts. Although the strain amplitude of each GW burst is small, a number of overlapped bursts due to the high event rate generate a GW background with a large amplitude, enough to be tested by the ground-based experiments. 
 
Search for cosmic strings have been conducted in the past, using data from LIGO and Virgo detectors~\cite{Abbott09,Aasi14,Abbott18}. These works have done two searches, a template-based matched-filtering search for individual burst signals (from cusps and kinks), and a search for the stochastic background generated by an overlapping population of these bursts. These searches have given constraints on the string parameters (e.g. tension) and models on the distribution of these loops in the Universe. The constraints depend on models of the probability distribution on the loop size. The searches using aLIGO's O1 data \cite{Abbott18} yielded a constraint of $G\mu < 5\times 10^{-8}$ for a simple loop model first developed in~\cite{Vilenkin00,Damour01}.

GW emission from these features are predicted to be linearly polarized, and have a power-law frequency spectrum of index $-4/3$ for cusps and $-5/3$ for kinks~\cite{Damour01}. The amplitude depends on the tension and size of the loop, as well as the distance to the loop. The spectrum also is considered to have a high-frequency cutoff due to the beamed nature of the signal. The high-frequency cutoff can take arbitrary values, but as bursts with higher frequency cutoff are more beamed \cite{Damour01}, bursts with lower high-frequency cutoffs can be seen more frequently than those with higher high-frequency cutoffs.\\

\noindent {\it - Future prospects}

Due to the power-law nature, the signals from cosmic string bursts generally have higher amplitudes in low frequencies. Hence the sensitivity to string signals will be most enhanced by improving the lower frequency band~\cite{Abbott18}. 
By defining the SNR ratio $r_{\alpha/\beta} \equiv r_{\alpha}/r_{\beta}$ for the configurations 
$\alpha$ and $\beta$ where $r_{\alpha}$ is defined in Eq.~(\ref{eq:SNR-ratio}) and a similar equation holds for the $\beta$ configuration, We can quantitatively compare performances of various configurations of possible KAGRA upgrades. For a cosmic string cusp, the amplitude is proportional to $f^{-4/3}$. If we take the high frequency cutoff at $f_{\rm high} = 30$~Hz, the SNR ratios from bKAGRA to KAGRA+ for a GW burst from a cosmic string cusp are LF 5.46, 40kg 1.54, FDSQZ 1.19, HF 0.39, Combined 2.82. Also significant sensitivity improvements in low frequencies are anticipated in the upgrades of aLIGO and AdV or the Einstein Telescope. Furthermore, because the string bursts are linearly polarized, using the information of polarization can be crucial for differentiating these signals from noise. This can be sufficiently done with a KAGRA detector in collaboration with the LIGO and Virgo detectors.

\subsection{Black-hole echoes}

\noindent {\it - Scientific objective}

The BH information loss paradox is a long standing problem. 
The problem looks complicated, but the rough idea can be captured in the following way.
Our assumptions are the followings:
\begin{enumerate}
\item Hawking evaporation can be described by a semi-classical picture. Namely, it 
can be described by the standard field theory in curved spacetime. 
This means that BHs radiate like a black body at the Hawking temperature $T_H=\kappa/2\pi$, 
where $\kappa$ is the surface gravity. In the case of a Schwarzschild BH, the surface gravity 
is given by $1/2r_g$ with $r_g$ being the horizon radius. This emitted radiation is perfectly 
entangled with the internal state of the BH. 
It would not be so difficult to convince yourself that this argument is likely. 
If we simply consider the evolution of initially pure quantum state, it remains so in 
the ordinary field theory. So, if the emitted radiation looks thermal, this thermal nature is 
obtained as a result of integrating out some unobserved degrees of freedom. 
In the case of a BH spacetime, these hidden degrees of freedom are inside the BH horizon. 

\item BH entropy is bounded by the area of the BH, {\em i.e.}, 
$S_{BH}<S_{\rm Bekenstein-Hawking}:=A/4G_N$, where 
$A$ is the area of the BH and $G_N$ is Newton's constant. Under the assumption 
that this bound is saturated, the BH thermodynamics holds. 
\end{enumerate}

These assumptions can be easily seen to be inconsistent. 
Let us consider a stationary process containing a BH emitting the Hawking radiation.  
To make it stationary, we keep throwing some matter continuously down into the BH.  
Then, in a mean time the total entropy of the emitted radiation exceeds 
$S_{\rm Bekenstein-Hawking}$. However, the internal state entangled 
with the emitted radiation should also have the same amount of entropy. 
Since the spacetime is stationary with an apparent horizon, no information 
can go out from the inside. This would contradict with the claim that the BH 
entropy is bounded by $S_{\rm Bekenstein-Hawking}$. 

To resolve the contradiction mentioned above, 
some of assumptions must be violated. An easy way is to abandon the 
assumption (2). However, many people believe that the BH entropy identified 
with its area is not 
just a theoretical illusion but has a deep physical meaning. 
As far as we wish to keep the 
assumption (2), we would be requested to abandon the assumption (1). 
One possibility is that the Hawking radiation is wrong. However, this is unlikely, although 
it is true that the explicit calculation of the Hawking radiation based on the field 
theory in curved spacetime is given only in weak coupling perturbative 
calculations. 
Then, at first glance radical but in a certain sense most conservative idea to resolve 
the contradiction will be to abandon the semi-classical picture. From the 
point of view of relativity, the horizon is not a special surface at all, 
unless we care about the global structure of the spacetime. 
Therefore it is difficult to imagine that 
something unusual happen particularly on the horizon. However, if we trace back the 
quanta emitted as the Hawking radiation, it infinitely blue-shifts in the near horizon limit. 
Hence, field theoretically the horizon might be somehow special, although the argument 
sounds a little acausal. 

From the discussion mentioned above, it would be well-motivated to doubt if the physics describing the region close to the BH horizon 
might be significantly modified from our familiar low energy physics. 
One simple but radical possibility is that the BH horizon is 
covered by a wall which reflects all the low energy excitations 
including GWs. Ordinary matter whose energy scale is much higher 
than the energy of the typical gravitons excited in the binary system 
will be absorbed by BHs,  
while the quanta whose energy is comparable or 
lower than $T_H$, i.e., whose wavelength is as long as the BH size, 
are selectively reflected by the wall. Interestingly, this critical energy scale 
also follows if we assume that the 
BH area is quantized by the Planck length 
squared~\cite{Bekenstein:1974jk,Mukhanov:1986me,Barcelo:2017lnx}. \\

\noindent {\it - Observations and measurements}

It is difficult to predict the exact waveform of the expected echoes, since the 
theoretical background is not so sound. 
What we can expect generically is the repetition of signals similar to the ringdown signal 
after merger with an equal time interval. 
Here, the uncertainties are in the boundary condition at the wall and the wall location. 
One possible ansatz about the properties of the wall 
is that the reflection rate at the wall is unity and the wall is 
placed at about a Planck distance from the horizon. Furthermore, 
we also need to make an assumption about the phase shift at the reflection. 
One simple assumption 
is that the phase shift is frequency independent or the frequency dependence is  
sufficiently weak. Under such an assumption, we could construct a template by solving 
the master equation for BH perturbation~\cite{Nakano}. 

There was a claim that some echo signal has been already detected~\cite{Abedi}.
However, very simple repetition of the same waveform was assumed in their analysis. 
Such an analysis is an interesting trial, but it is difficult to invent a theoretical model that 
supports that simple-minded assumption. In fact, our reanalysis indicates that 
the significance of the echo signal does not survive once we replace the 
template with a little more realistic one~\cite{Uchikata:2019frs}, i.e., the reflection rate at the angular momentum barrier calculated by using the BH perturbation theory is employed.
However, it is also true that the signal claimed in \cite{Abedi} does not disappear just by increasing the number of BH merger events including the events from LIGO/Virgo O2 observing run.   
\\
 
\noindent {\it - Future prospects}

Definitely, as the number of events increases, the nature of the signal that seem to be existing will be uncovered in near future.  
In order to discriminate the real signal coming from the sky 
from the unidentified noise of detectors correlated with 
GW events, it would be better to analyze the data from 
completely different detectors which have a similar sensitivity. 
In this sense it is quite interesting to analyse the data 
obtained by Virgo and KAGRA in the same manner.

\section{Conclusion}

In this article we have reviewed scientific cases available with the second-generation ground-based detectors including aLIGO, AdV, the baseline KAGRA (bKAGRA) and their future upgrades, A+, AdV+, and KAGRA+, respectively, and discuss KAGRA's scientific contributions to the global detector networks composed of the detectors above. For compact binaries composed of stellar-mass BHs and/or NSs, at least three detectors are necessary to localize sources and more detectors including KAGRA is preferable to obtain smaller sky areas to be searched. This is also true for sciences based on these binaries such as the tests of gravity (except for BH ringdown), the measurement of the Hubble constant, GW lensing, the measurement of EOS of a NS, and the multimessenger observations of short-gamma ray bursts and kilonovae. On the other hand, the discovery of new GW sources such as binary IMBHs, supernovae, magnetors, isolated pulsars, and low-mass X-ray binaries can be done with a single detector, though multiple detectors are preferable for events detected only by GWs. Therefore, for these sources, the higher duty cycle of a detector network is more important not to miss signals during detector down times. In summary, KAGRA will be able to contribute to broad science topics and play an important role in joint observations by the global detector networks.

\section*{Acknowledgment}
This work was supported by MEXT, JSPS Leading-edge Research Infrastructure Program, JSPS Grant-in-Aid for Specially Promoted Research 26000005, JSPS Grant-in-Aid for Scientific Research on Innovative Areas 2905: JP17H06358, JP17H06361 and JP17H06364, JSPS Core-to-Core Program A. Advanced Research Networks, JSPS Grant-in-Aid for Scientific Research (S) 17H06133, the joint research program of the Institute for Cosmic Ray Research, University of Tokyo, National Research Foundation (NRF) and Computing Infrastructure Project of KISTI-GSDC in Korea, Academia Sinica (AS), AS Grid Center (ASGC) and the Ministry of Science and Technology (MoST) in Taiwan under grants including AS-CDA-105-M06, Advanced Technology Center (ATC) of NAOJ, Mechanical Engineering Center of KEK, the LIGO project, and the Virgo project.

\newcommand{\ApJL}[3]{{{Astrophys. J. Lett.} {\bf #1}, #2 (#3).}}
\newcommand{\ApJ}[3]{{{Astrophys. J.} {\bf #1}, #2 (#3).}}
\newcommand{\MNRAS}[3]{{{Mon. Not. R. Astron. Soc.} {\bf #1}, #2 (#3).}}
\newcommand{\MNRASL}[3]{{{Mon. Not. R. Astron. Soc.} {\bf #1}, #2 (#3).}}
\newcommand{\arxiv}[1]{{arXiv:#1.}}
\newcommand{\JCAP}[4]{{{J. Cosmol. Astropart. Phys.} {\bf #1}, #4 (#2).}}
\newcommand{\CQG}[3]{{{Class. Quant. Grav.} {\bf #1}, #2 (#3).}}
\newcommand{\Nat}[3]{{{Nature} {\bf #1}, #2 (#3).}}
\newcommand{\GRG}[3]{{{Gen. Rel. Grav.} {\bf #1} #2 (#3).}}
\newcommand{\AnA}[3]{{{Astronomy \& Astrophysics} {\bf #1}, #2 (#3).}}
\newcommand{\ARAA}[3]{{{Ann. Rev. Astron. Astrophys.} {\bf #1}, #2 (#3).}}
\newcommand{\PRep}[3]{{{Physics Reports} {\bf #1}, #2 (#3).}}
\newcommand{\Sci}[3]{{{Science} {\bf #1}, #2 (#3).}}
\newcommand{\LRR}[3]{{{Liv. Rev. Rel.} {\bf #1}, #2 (#3).}}
\newcommand{\PTEP}[3]{{{Prog. Theor. Exp. Phys.} {\bf #1}, #2 (#1).}}
\newcommand{\JPCS}[3]{{{J. Phys. Conf. Ser.} {\bf #1} #2 (#3).}}
\newcommand{\NatCom}[3]{{{\it Nature Communications} {\bf #1}, #2 (#3).}}
\newcommand{\NatAst}[3]{{{\it Nature Astron.} {\bf #1}:#2 (#3).}}
\newcommand{\PASJ}[3]{{{\it PASJ} {\bf #1}, #2 (#3).}}
\newcommand{\PASA}[3]{{{\it PASA} {\bf #1} #2 (#3).}}
\newcommand{\MPLA}[3]{{{\it Modern Physics Letters A} {\bf #1}, #2 (#3).}}
\newcommand{\ARNPS}[3]{{{\it Annual Review of Nuclear and Particle Science} {\bf #1} #2 (#3).}}
\newcommand{\IJMPD}[3]{{{\it Int.\ J.\ Mod.\ Phys.\ D} {\bf #1}, #2 (#3).}}
\newcommand{\CRP}[3]{{{\it Comptes Rendus Physique} {\bf #1} #2 (#3).}}

\newcommand{\EPL}[4]{\href{https://doi.org/10.1209/epl/#4}{{\it Europhys. Lett.} {\bf #1}, #2 (#3).}}
\newcommand{\JOptB}[4]{\href{https://doi.org/10.1088/1464-4266/#4}{{\it J. Opt. B: Quantum Semiclass. Opt.}{\bf #1}, #2 (#3).}}
\newcommand{\LIGOdoc}[2]{\href{https://dcc.ligo.org/LIGO-#1}{{\it LIGO Document} {\bf #1} (#2).}}
\newcommand{\VIRdoc}[3]{\href{https://tds.virgo-gw.eu/?content=3&r=#3}{{\it Virgo document} {\bf VIR-#1} (#2).}}
\newcommand{\JGWdoc}[3]{\href{https://gwdoc.icrr.u-tokyo.ac.jp/cgi-bin/private/DocDB/ShowDocument?docid=#3}{{\it KAGRA document} {\bf JGW-#1} (#2)}}
\newcommand{\JETPL}[3]{{\it JETP Lett.} {\bf #1}, #2 (#3) [{\it Pisma Zh.\ Eksp. Teor. Fiz.}  {\bf #1}, #2 (#3).]}
\newcommand{\OME}[3]{\href{https://doi.org/10.1364/OME.#1.00#2}{{\it Opt. Mater. Express} {\bf #1}, #2 (#3)}}
\newcommand{\RPP}[4]{\href{https://doi.org/10.1088/1361-6633/#4}{{\it Rep. Prog. Phys.} {\bf #1}, #2 (#3).}}
\newcommand{\JOSAB}[4]{\href{https://doi.org/10.1364/JOSAB.#1.#4}{{\it J. Opt. Soc. Am. B} {\bf #1}, #2 (#3).}}
\newcommand{\AO}[4]{\href{https://doi.org/10.1364/AO.#1.#4}{{\it Appl. Opt.} {\bf #1}, #2 (#3).}}
\newcommand{\PSPIE}[4]{\href{https://doi.org/10.1117/#4}{{\it Proc. SPIE} {\bf #1}, #2 (#3).}}

\newcommand{\Authname}[2]{#2 #1} 
\newcommand{\LSC}{\Authname{Abbott}{B. P.} \etal (LIGO Scientific Collaboration)}
\newcommand{\LVC}{\Authname{Abbott}{B. P.} \etal (LIGO Scientific and Virgo Collaboration)}
\newcommand{\LVK}{\Authname{Abbott}{B. P.} \etal (LIGO Scientific, Virgo and KAGRA Collaboration)}


\begin{thebibliography}{999}
\bibitem{KAGRA:PTEP07}
T. Akutsu, {\it et al.} (KAGRA collaboration), in preparation as the series of article. {\it Overview of KAGRA: Future plans}.


\bibitem{aligo} 
J. Aasi \etal, (LIGO Scientific Collaboration, Advanced VIRGO),
\CQG{32}{074001}{2015}

\bibitem{AdV}
\Authname{Acernese}{F.} \etal,
\CQG{32}{024001}{2015}


\bibitem{GWTC1}
\LVC,
\PRX{9,031040,2019}

\bibitem{GW170817PRL} 
\LVC,
\PRL{119,161101,2017}

\bibitem{GW150914PRL} 
\LVC,
\PRL{116,061102,2016}

\bibitem{GW151226PRL} 
\LVC,
\PRL{116,241103,2016}

\bibitem{GW170104PRL} 
\LVC,
\PRL{118, 221101 ,2017}

\bibitem{GW170608ApJL} 
\LVC,
\ApJL{851}{L35}{2017}

\bibitem{GW170814PRL} 
\LVC, 
\PRL{119,141101,2017}

\bibitem{mmaGW170817}  
B. P. Abbott, {\it et al.} 
\ApJL{848}{L12}{2017}


\bibitem{A+}
\Authname{Miller}{J.} \etal, 
\PRD{91,062005,2015}

\bibitem{AdV+}
\Authname{Degallaix}{J.} for the Virgo Collaboration,
Advanced Virgo+ preliminary studies, VIR-0300A-18 (2018).

\bibitem{LVKScenario2018} 
B. P. Abbot \etal,
\LRR{21}{3}{2018}




\bibitem{KAGRA:NatureAstro}
T. Akutsu \etal, (KAGRA collaboration) 
Nat. Astron. \textbf{3}, 35 (2019).

\bibitem{kagraplus} 
\Authname{Michimura}{Y.} \etal,
\PRD{102,022008,2020}


















\bibitem{LIGO_ApJ}
\LVC,
\ApJL{818}{L22}{2016}

\bibitem{Belczynski_2004}
\Authname{Belczynski}{K.}, \Authname{Bulik}{T.} and \Authname{Rudak}{B.},
\ApJL{608}{L45}{2004}

\bibitem{Dominik_2012}
\Authname{Dominik}{M.}, \Authname{Belczynski}{K.}, \Authname{Fryer}{C.}, \Authname{Holz}{D.~E.}, \Authname{Berti}{E.}, \Authname{Bulik}{T.}, \Authname{Mandel}{I.} and \Authname{O'Shaughnessy}{R.},
\ApJ{759}{52}{2012}

\bibitem{Belczynski_2016}
\Authname{Belczynski}{K.}, \Authname{Holz}{D.~E.}, \Authname{Bulik}{T.} and \Authname{O'Shaughnessy}{R.},
\Nat{534}{512}{2016}

\bibitem{Mapeli_2016} 
\Authname{Mapelli}{M.},
\MNRAS{459}{3432}{2017}

\bibitem{Portegies Zwart_2000}
\Authname{Portegies Zwart}{S.~F.} and \Authname{McMillan}{S.~L.~W.},
\ApJL{528}{L17}{2000}

\bibitem{O'Leary_2009}
\Authname{O'Leary}{R.~M.}, \Authname{Kocsis}{B.} and \Authname{Loeb}{A.},
\MNRAS{395}{2127}{2009}

\bibitem{Chiba:2017rvs}
\Authname{Chiba}{T.} and \Authname{Yokoyama}{S.},
\PTEP{2017}{083E01}{pty087}

\bibitem{DeLuca:2019buf} 
\Authname{De Luca}{V.},  \Authname{Desjacques}{V.},  \Authname{Malhotra}{A.} and \Authname{Riotto}{A.},
\JCAP{05}{2019}{}{018}

\bibitem{Harada:2017fjm}
\Authname{Harada}{T.}, \Authname{Chul-Moon}{Y.}, \Authname{Kohri}{K.}, and \Authname{Nakao}{K.},
\PRD{96,083517,2017} Erratum: {\it ibid.} 99, 069904 (2019).

\bibitem{Rodriguez_2016a}
\Authname{Rodriguez}{C.~L.}, \Authname{Haster}{C.-J.}, \Authname{Chatterjee}{S.},
\Authname{Kalogera}{V.} and \Authname{Rasio}{F.~A.},
\ApJL{824}{L8}{2016}

\bibitem{Antonini_2016}
\Authname{Antonini}{F.} and \Authname{Rasio}{F. A.},
\ApJ{831}{187}{2016}

\bibitem{Stone_2017}
\Authname{Stone}{N.~C.}, \Authname{Metzger}{B.~D.} and \Authname{Haiman}{Z.},
\MNRAS{464}{946}{2017}

\bibitem{BaeKimLee}
Y.-B.~Bae, C.~Kim, H.~M.~Lee
\MNRAS{440}{2714}{2014}

\bibitem{Park_2017}
D.~Park, C.~Kim, H.~M.~Lee, Y.-B.~Bae, K.~Belczynski
\MNRAS{469}{4665}{2017}

\bibitem{Kinugawa_2014}
\Authname{Kinugawa}{T.}, \Authname{Inayoshi}{K.}, \Authname{Hotokezaka}{K.}, \Authname{Nakauchi}{D.} and \Authname{Nakamura}{T.},
\MNRAS{442}{2963}{2014}

\bibitem{Hartwig_2016}
\Authname{Hartwig}{T.}, \Authname{Volonteri}{M.}, \Authname{Bromm}{V.}, \Authname{Klessen}{R.~S.}, \Authname{Barausse}{E.}, \Authname{Magg}{M.} and \Authname{Stacy}{A.},
\MNRASL{460}{L74}{2016}

\bibitem{Inayoshi_2016}
\Authname{Inayoshi}{K.}, \Authname{Kashiyama}{K.}, \Authname{Visbal}{E.} and \Authname{Haiman}{Z.},
\MNRAS{461}{2722}{2016}

\bibitem{Inayoshi_2017} 
\Authname{Inayoshi}{K.}, \Authname{Hirai}{R.}, \Authname{Kinugawa}{T.} and \Authname{Hotokezaka}{K.},
\MNRAS{468}{5020}{2017}

\bibitem{Kushnir_2016}
\Authname{Kushnir}{D.}, \Authname{Zaldarriaga}{M.}, \Authname{Kollmeier}{J.~A.} and \Authname{Waldman}{R.},
\MNRAS{462}{844}{2016}
 
\bibitem{Rodriguez_2016b}
\Authname{Rodriguez}{C.~L.}, \Authname{Zevin}{M.}, \Authname{Pankow}{C.}, \Authname{Kalogera}{V.} and \Authname{Rasio}{F.~A.}
\ApJL{832}{L2}{2016}

\bibitem{Hotokezaka_Piran_2017}
\Authname{Hotokezaka}{K.} and \Authname{Piran}{T.},
\ApJ{842}{111}{2017}

\bibitem{Janka_2013}
\Authname{Janka}{T.}
\MNRAS{434}{1355}{2013}

\bibitem{Mandel_2016}
\Authname{Mandel}{I.}
\MNRAS{456}{578}{2016}

\bibitem{Phinney_2001}
\Authname{Phinney}{E.~S.},
\arxiv{astro-ph/0108028}

\bibitem{LIGO_background}
\LVC,
\PRL{116,131102,2016}


\bibitem{Carr_2016}
\Authname{Carr}{B.}, \Authname{K{\"u}hnel}{F.} and \Authname{Sandstad}{M.},
\PRL{94,083504,2016}

\bibitem{Sasaki_2016}
\Authname{Sasaki}{M.}, \Authname{Suyama}{T.}, \Authname{Tanaka}{T.} and \Authname{Yokoyama}{S.},
\PRL{117,061101,2016}

\bibitem{Sasaki:2018dmp}
\Authname{Sasaki}{M.}, \Authname{Suyama}{T.}, \Authname{Tanaka}{T.}, and \Authname{Yokoyama}{S.}, 
\CQG{35}{063001}{2018}



\bibitem{PostnovYungenson}
\Authname{Postnov}{K.~A.} and \Authname{Yungelson}{L.~R.}, 
\LRR{17}{3}{2014} 

\bibitem{Benacquista}
M. J. Benacquista, \LRR{5}{2}{2002}

\bibitem{Pekowsky_2013}
L.~Pekowsky, J.~Healy, D.~Shoemaker, P.~Laguna
\PRD{87,084008,2013}

\bibitem{LIGOScientific:2020stg}
\LVC,
\arxiv{2004.08342}

\bibitem{Bardeen:1972fi}
\Authname{Bardeen}{J.~M.} \etal,
\ApJ{178}{347}{1972}



\bibitem{Portegies_Zwart_2002}
\Authname{Portegies Zwart}{S.~F.} and \Authname{McMillan}{S.~L.~W.},
\ApJ{576}{899}{2002}

\bibitem{Freitag_2006}
\Authname{Freitag}{M.}, \Authname{G{\"u}rkan}{M.~A.} and \Authname{Rasio}{F.~A.},
\MNRAS{368}{141}{2006}

\bibitem{Amaro-Seoane_2010}
\Authname{Amaro-Seoane}{P.} and \Authname{Santamar{\'{\i}}a}{L.},
\ApJ{722}{1197}{2010}

\bibitem{Fragione_2018}
\Authname{Fragione}{G.}, \Authname{Ginsburg}{I.} and \Authname{Kocsis}{B.},
\ApJ{856}{92}{2018}

\bibitem{Shinkai_2017}
\Authname{Shinkai}{H.-a.}, \Authname{Kanda}{N.} and \Authname{Ebisuzaki}{T.},
\ApJ{835}{276}{2017}

\bibitem{Amaro-Seoane_2018}
\Authname{Amaro-Seoane}{P.},
\PRD{98,063018,2018}

\bibitem{Gurkan_2006}
\Authname{G{\"u}rkan}{M.~A.}, \Authname{Fregeau}{J.~M.} and \Authname{Rasio}{F.~A.},
\ApJL{640}{L39}{2006}

\bibitem{Hirano_2014}
\Authname{Hirano}{S.}, \Authname{Hosokawa}{T.}, \Authname{Yoshida}{N.}, \Authname{Umeda}{H.}, \Authname{Omukai}{K.}, \Authname{Chiaki}{G.} and \Authname{Yorke}{H.~W.},
\ApJ{780}{60}{2014}

\bibitem{Stacy_Bromm_2013}
\Authname{Stacy}{A.} and \Authname{Bromm}{V.},
\MNRAS{433}{1094}{2013}

\bibitem{Susa_2014}
\Authname{Susa}{H.}, \Authname{Hasegawa}{K.} and \Authname{Tominaga}{N.},
\ApJ{792}{32}{2014}

\bibitem{Inayoshi_Haiman_2014}
\Authname{Inayoshi}{K.}, \Authname{Haiman}{Z.},
\MNRAS{445}{1549}{2014}



\bibitem{Miller:2003sc}
\Authname{Miller}{M.~C.} \etal,
\IJMPD{13}{1}{2004}

\bibitem{Brown:2006pj}
\Authname{Brown}{D.~A.} \etal,
\PRL{99,201102,2007}

\bibitem{Berti:2004bd}
\Authname{Berti}{E.} \etal,
\PRD{71,084025,2005}

\bibitem{Isoyama:2018rjb}
\Authname{Isoyama}{S.} \etal,
\PTEP{2018}{073E01}{pty078}


\bibitem{Ori:2000zn}
\Authname{Ori}{A.} \etal,
\PRD{62,124022,2000}

\bibitem{Berti:2005ys}
\Authname{Berti}{E.} \etal,
\PRD{73,064030,2006}

\bibitem{Huerta:2010un}
\Authname{Huerta}{E.~A.} \etal,
\PRD{83,044020,2011}

\bibitem{Miller:2008fi}
\Authname{Miller}{M.~C.},
\CQG{26}{094031}{2009}




\bibitem{Tauris:2017}
T.~M.~Tauris \etal, \ApJL{846}{L170}{2017}

\bibitem{Chruslinska:2018}
\Authname{Chruslinska}{M.}, \Authname{Belczynski}{K.}, \Authname{Klencki1}{J.}, \Authname{Benacquista}{M.},
\MNRAS{474}{2937}{2018}

\bibitem{Pooley:2003}
D.~Pooley \etal, 
\ApJL{591}{L131}{2019}.

\bibitem{WeisbergNiceTaylor} 
J.~M.~Weisberg, D.~J.~Nice, J.~H.~Taylor,
\ApJ{722}{1030}{2010} 

\bibitem{Peters}
P.~C.~Peters, \PRB{136,1224,1964}

\bibitem{ozelFreire:2016}
\Authname{\"Ozel}{F.} and \Authname{Freire}{P.}, Annu. Rev. Astron. Astrophys, 54, 401 (2016).

\bibitem{Abbott:2020khf}
\LVC,
\ApJL{896}{L44}{2020}

\bibitem{PML2019}
\Authname{Pol}{N.}, \Authname{McLaughlin}{M.}, \Authname{Lorimer}{D.R.},
\ApJ{870}{71}{2019}

\bibitem{ATNFPulsarCatalogue}
\Authname{Manchester}{R. N.}, \Authname{Hobbs}{G. B.}, \Authname{Teoh}{A.} and \Authname{Hobbs}{M.},
\ApJ{129}{1993}{2005}; ATNF Pulsar Catalogue webpage, http://www.atnf.csiro.au/research/pulsar/psrcat/   

\bibitem{Abbott:2018wiz} 
\LVC,
  \PRX{9,011001,2019}



\bibitem{Lattimer:2015nhk} 
  J.~M.~Lattimer and M.~Prakash,
  Phys.\ Rept.\  {\bf 621}, 127 (2016).
  
\bibitem{Flanagan:2007ix} 
  E.~E.~Flanagan and T.~Hinderer,
  \PRD{77,021502,2008}

\bibitem{Hinderer:2007mb} 
  T.~Hinderer,
\ApJ{677}{1216}{2008}

\bibitem{Damour:2012yf} 
  T.~Damour, A.~Nagar and L.~Villain,
  Phys.\ Rev.\ D {\bf 85}, 123007 (2012)
    \PRD{85,123007,2012}

\bibitem{Wade:2014vqa} 
  L.~Wade, J.~D.~E.~Creighton, E.~Ochsner, B.~D.~Lackey, B.~F.~Farr, T.~B.~Littenberg and V.~Raymond,
  \PRD{89,103012,2014}

\bibitem{Hinderer:2009ca} 
  T.~Hinderer, B.~D.~Lackey, R.~N.~Lang and J.~S.~Read,
    \PRD{81,123016,2010}

\bibitem{Hotokezaka:2011dh} 
  K.~Hotokezaka, K.~Kyutoku, H.~Okawa, M.~Shibata and K.~Kiuchi,
    \PRD{83,124008,2011}

\bibitem{Buonanno:2009zt} 
  A.~Buonanno, B.~R.~Iyer, E.~Ochsner, Y.~Pan and B.~S.~Sathyaprakash,
    \PRD{80,084043,2009}
  
\bibitem{Blanchet:2013haa} 
  L.~Blanchet,
  Living Rev.\ Rel.\  {\bf 17}, 2 (2014).

\bibitem{Dietrich:2017aum} 
  T.~Dietrich, S.~Bernuzzi and W.~Tichy,
    \PRD{96,121501,2017}
    
\bibitem{Yagi:2016bkt}
  K.~Yagi and N.~Yunes,
  Phys.\ Rept.\  {\bf 681}, 1 (2017)

\bibitem{Abbott:2018exr}
  \LVC,
  \PRL{121,161101,2018}

\bibitem{Kastaun:2019bxo} 
  W.~Kastaun and F.~Ohme,
    \PRD{100,103023,2019}

\bibitem{LIGOScientific:2019eut} 
\LVC,
\CQG{37}{045006}{2020}

\bibitem{Mueller:1996pm} 
  H.~Mueller and B.~D.~Serot,
  Nucl.\ Phys.\ A {\bf 606}, 508 (1996).

\bibitem{De:2018uhw} 
  S.~De, D.~Finstad, J.~M.~Lattimer, D.~A.~Brown, E.~Berger and C.~M.~Biwer,
  \PRL{121,091102,2018} Erratum: {\it ibid}. 121, 259902 (2018).

\bibitem{Capano:2019eae}
  C.~D.~Capano {\it et al.},
\NatAst{4}{625}{2020}

\bibitem{Narikawa:2019xng} 
  T.~Narikawa, N.~Uchikata, K.~Kawaguchi, K.~Kiuchi, K.~Kyutoku, M.~Shibata and H.~Tagoshi,
\arxiv{1910.08971}

\bibitem{Kiuchi:2017pte} 
  K.~Kiuchi, K.~Kawaguchi, K.~Kyutoku, Y.~Sekiguchi, M.~Shibata and K.~Taniguchi,
    \PRD{96,084060,2017}

\bibitem{Kawaguchi:2018gvj} 
  K.~Kawaguchi, K.~Kiuchi, K.~Kyutoku, Y.~Sekiguchi, M.~Shibata and K.~Taniguchi,
    \PRD{97,044044,2018}
  
\bibitem{Dietrich:2019kaq} 
  T.~Dietrich, A.~Samajdar, S.~Khan, N.~K.~Johnson-McDaniel, R.~Dudi and W.~Tichy,
    \PRD{100,04403,2019}

\bibitem{Margalit:2017dij} 
  B.~Margalit and B.~D.~Metzger,
\ApJL{850}{L19}{2017}

\bibitem{Bauswein:2017vtn} 
  A.~Bauswein, O.~Just, H.~T.~Janka and N.~Stergioulas,
\ApJL{850}{L34}{2017}

\bibitem{Ruiz:2017due} 
  M.~Ruiz, S.~L.~Shapiro and A.~Tsokaros,
\PRD{97,021501(R),2018}

\bibitem{Annala:2017llu} 
  E.~Annala, T.~Gorda, A.~Kurkela and A.~Vuorinen,
    \PRL{120,172703,2018}

\bibitem{Zhou:2017pha} 
  E.~P.~Zhou, X.~Zhou and A.~Li,
    \PRD{97,083015,2018}

\bibitem{Fattoyev:2017jql} 
  F.~J.~Fattoyev, J.~Piekarewicz and C.~J.~Horowitz,
    \PRL{120,172702,2018}

\bibitem{Paschalidis:2017qmb} 
  V.~Paschalidis, K.~Yagi, D.~Alvarez-Castillo, D.~B.~Blaschke and A.~Sedrakian,
    \PRD{97,084038,2018}

\bibitem{Nandi:2017rhy} 
  R.~Nandi and P.~Char,
    \ApJ{857}{12}{2018}

\bibitem{Most:2018hfd}
  E.~R.~Most, L.~R.~Weih, L.~Rezzolla and J.~Schaffner-Bielich,
    \PRL{120,261103,2018}

\bibitem{Raithel:2018ncd}
  C.~Raithel, F.~\"Ozel and D.~Psaltis,
    \ApJL{857}{L23}{2018}

\bibitem{Landry:2018prl} 
  P.~Landry and R.~Essick,
    \PRD{99,084049,2019}

\bibitem{Baiotti:2019sew} 
L.~Baiotti,
Prog. Part. Nucl. Phys., {\bf{109}}, 103714 (2019).

\bibitem{Takami:2014tva} 
  K.~Takami, L.~Rezzolla and L.~Baiotti,
  \PRD{91,064001,2015}

\bibitem{Akmal:1998cf} 
  A.~Akmal, V.~R.~Pandharipande and D.~G.~Ravenhall,
  \PRC{58,1804,1998}

\bibitem{Douchin:2001sv}
  F.~Douchin and P.~Haensel,
  Astron. Astrophys. {\bf 380}, 151 (2001).

\bibitem{Narikawa:2018yzt} 
  T.~Narikawa, N.~Uchikata, K.~Kawaguchi, K.~Kiuchi, K.~Kyutoku, M.~Shibata and H.~Tagoshi,
  Phys.\ Rev.\ Research.\  {\bf 1}, 033055 (2019)



\bibitem{bai17} 
\Authname{Baiotti}{L.} and \Authname{Rezzolla}{L.}, RPPh, 80, 096901 (2017).

\bibitem{Duez2019}
\Authname{Duez}{M.~D.} and   \Authname{Zlochower}{Y.},
Rep. Prog. Phys., 82, 016902 (2019).

\bibitem{Paschalidis_and_Stergioulas_2017}
\Authname{Paschalidis}{V.} and \Authname{Stergioulas}{N.}, 
\LRR{20}{7}{2017}

\bibitem{Takami:2014zpa} 
\Authname{Takami}{K.}, \Authname{Rezzolla}{L.} and \Authname{Baiotti}{L.},
\PRL{113,091104,2014}

\bibitem{Maione2017}
\Authname{Maione}{F.}, \Authname{De Pietri}{R.}, \Authname{Feo}{A.} and   \Authname{L{\"o}ffler}{F.},
\PRD{96,063011,2017}

\bibitem{Breschi2019} 
\Authname{Breschi}{M.}, \Authname{Bernuzzi}{S.}, \Authname{Zappa}{F.}, \Authname{Agathos}{M.}, \Authname{Perego}{A.}, \Authname{Radice}{D.} and \Authname{Nagar}{A.},
\PRD{100,104029,2019}

\bibitem{Bauswein2011}
\Authname{Bauswein}{A.} and \Authname{Janka}{H.-T.},
\PRL{108,011101,2012}

\bibitem{Bose2017}
\Authname{Bose}{S.}, \Authname{Chakravarti}{K.}, \Authname{Rezzolla}{L.}, \Authname{Sathyaprakash}{B.~S.} and \Authname{Takami}{K.},
\PRL{120,031102,2018}

\bibitem{Chatziioannou2017}
\Authname{Chatziioannou}{K.}, \Authname{Clark}{J.~A.},
\Authname{Bauswein}{A.}, \Authname{Millhouse}{M.},   \Authname{Littenberg}{T.~B.}, and  \Authname{Cornish}{N.},
\PRD{96,124035,2017}

\bibitem{Torres-Rivas2019}
\Authname{Torres-Rivas}{A.}, \Authname{Chatziioannou}{K.}, \Authname{Bauswein}{A.} and \Authname{Clark}{J.~A.},
\PRD{99,044014,2019}

\bibitem{Weih2020} 
\Authname{Weih}{L.~R.}, \Authname{Hanauske}{M.} and \Authname{Rezzolla}{L.},
\PRL{124,171103,2020}

\bibitem{Bauswein2020} 
\Authname{Bauswein}{A.}, 
\Authname{Blacker}{S.},
\Authname{Vijayan}{V.},
\Authname{Stergioulas}{N.},
\Authname{Chatziioannou}{K.}
\Authname{Clark}{J.}
\Authname{Bastian}{N.}
\Authname{Blaschke}{D.~B.}
\Authname{Cierniak}{M.} and
\Authname{Fischer}{T.},
\arxiv{2004.00846}

\bibitem{bai08} 
\Authname{Baiotti}{L.}, \Authname{Giacomazzo}{B.} and \Authname{Rezzolla}{L.}, 
\PRD{78,084033,2008}

\bibitem{rez11}
\Authname{Rezzolla}{L.}, \Authname{Giacomazzo}{B.}, \Authname{Baiotti}{L.}, \Authname{Granot}{L.}, \Authname{Kouveliotou}{C.} and \Authname{Aloy}{M.A.}, 
\ApJL{732}{L6}{2011}

\bibitem{kiu15}
\Authname{Kiuchi}{K.},\Authname{Sekiguchi}{Y.}, \Authname{Kyutoku}{K.}, \Authname{Shibata}{M.}, \Authname{Taniguchi}{K.} and \Authname{Wada}{T.}, 
\PRD{92,064034,2015}

\bibitem{rad18} 
\Authname{Radice}{D.}, \Authname{Perego}{A.}, \Authname{Hotokezaka}{K.} \etal, \ApJ{869}{130}{2018}

\bibitem{ker63} 
R.~P.~Kerr, 
\PRL{11,237,1963}

\bibitem{van99}
\Authname{van Putten}{M.~H.~P.~M.}, 
\Sci{284}{115}{1999}

\bibitem{van19b} 
\Authname{van Putten}{M.~H.~P.~M.}, \Authname{Della Valle}{M.}, and \Authname{Levinson}{A.},
\ApJ{876}{L2}{2019}

\bibitem{dep19a}
\Authname{De Pietri}{R.}, \Authname{Drago}{A.}, \Authname{Feo}{A.}, \etal, 
\ApJ{881}{122}{2019}  

\bibitem{kli19}
\Authname{Klimo}{J.}, \Authname{Veselsky}{M.}, \Authname{Souliotis}{G.~A.}, \Authname{Bonasera}{A.}, Nucl. Phys. A, 992, 121640 (2019).

\bibitem{hae09} 
\Authname{Haensel}{P.}, \Authname{Zdunik}{J.~ L.}, \Authname{Bejger}{M.} \etal, A\&A, 502, 605 (2009).

\bibitem{con17} 
\Authname{Connaughton}{V.}, GCN, 21505 (2017).

\bibitem{sav17} 
\Authname{Savchenko}{V.} \etal, \ApJL{848}{L15}{2017}

\bibitem{moo18a} 
\Authname{Mooley}{K.~P.}, \Authname{Deller}{A.~T.}, \Authname{Gottlieb}{O.} \etal, \Nat{554}{207}{2018}

\bibitem{moo18b} 
\Authname{Mooley}{K.~P.}, \Authname{Deller}{A.~T.}, \Authname{Gottlieb}{O.} \etal, \Nat{561}{355}{2018}

\bibitem{cut02} 
\Authname{Cutler}{C.} and \Authname{Thorne}{K.~S.}, 
\arxiv{gr-qc/0204090}

\bibitem{hul75}
\Authname{Hulse}{R.~A.} and \Authname{Taylor}{J.~H.}, 
\ApJL{195}{L51}{1975}

\bibitem{wei78}
\Authname{Weiler}{K.~W.} and \Authname{Panagia}{N.},
A\&A, 70, 419 (1978).

\bibitem{GW170817:GRB}
\Authname{Abbott}{B.~P.} \etal,
\ApJL{848}{L13}{2017}

\bibitem{abb19a} 
\Authname{Abbott}{B.~P.} \etal, \ApJ{875}{160}{2019}


\bibitem{sun19} 
\Authname{Sun}{L.} and \Authname{Melatos}{A.}, 
\PRD{99,123003,2019}

\bibitem{gil19} 
R.~Gill, A.~Nathanail, and L.~Rezzolla, \ApJ{876}{139}{2019}

\bibitem{sma17} 
\Authname{Smartt}{S.~J.}, \Authname{Chen}{T.-W.}, \Authname{Jerkstrand}{A.} \etal, \Nat{551}{75}{2017}

\bibitem{pia17} 
\Authname{Pian}{E.}, \Authname{D'Avanzo}{P.}, \Authname{Benetti}{S.} \etal, \Nat{551}{67}{2017}

\bibitem{luc19} 
\Authname{Lucca}{M.} and \Authname{Sagunski}{L.}, 
JHEAp {\bf 27}, 33 (2020)

\bibitem{poo18} 
\Authname{Pooley}{D.}, \Authname{Kumar}{P.}, \Authname{Wheeler}{J.~C.}, \Authname{Grossan}{B.}, 
\ApJL{859}{L23}{2018}

\bibitem{van14} 
M.~H.~P.~M.~van~Putten, C.~Guidorzi, and F.~Frontera
\ApJ{786}{146}{2014}



\bibitem{van17} 
\Authname{van Putten}{M.~H.~P.~M.}, PTEP, 093F01 (2017).

\bibitem{van19a} 
\Authname{van Putten}{M.~H.~P.~M.} and \Authname{Della Valle}{M.}, \MNRAS{482}{L46}{2019}

\bibitem{GW170817:PMS}
\Authname{Abbott}{B. P.} \etal,
\ApJL{851}{L16}{2017}

\bibitem{van19c} 
\Authname{van Putten}{M.~H.~P.~M.}, \Authname{Levinson}{A.}, \Authname{Frontera}{F.}, \Authname{Guidorzi}{C.}, \Authname{Amati}{L.}, and \Authname{Della Valle}{M.}, EPJ Plus, 134, 547 (2019).

\bibitem{heo16}
\Authname{Heo}{J.-E.}, \Authname{Yoon}{S.},\Authname{Lee}{D.-S}, \etal,
New Astronomy {\bf 42}, 24 (2016).

\bibitem{and13} 
\Authname{Ando}{S.}, \Authname{Baret}{B.}, \Authname{Bartos}{I.}, \etal, Rev. Mod. Phys. 85 (2013).

\bibitem{aas14} 
\Authname{Aasi}{J.}, \Authname{Abbott}{B.P.}, \Authname{Abbott}{T.}, \etal,
\PRD{89,122004,2014}

\bibitem{ama18a}\Authname{Amati}{L.},\Authname{O'Brian}{P.},\Authname{G\"otz}{D.}, \etal, Adv. Space Sc., 62, 191 (2018).

\bibitem{stra18a}\Authname{Stratta}{G.},\Authname{Ciolfi}{R.},\Authname{Amati}{L.}, \etal, Adv. Space Sc., 62, 662 (2018).



\bibitem{Riles2017}
\Authname{Riles}{K.}, 
\MPLA{32}{1730035-685}{2017}

\bibitem{Ushomirsky2000}
\Authname{Ushomirsky}{G.} \etal,
\MNRAS{319}{902}{2000}

\bibitem{Bildsten}
\Authname{Bildsten}{L.},
\ApJL{501}{L89}{1998}

\bibitem{Mukherjee2018}
\Authname{Mukherjee}{A.} \etal,
\PRD{97,043016,2018}

\bibitem{Meadors2017}
\Authname{Meadors}{G. D.} \etal,
\PRD{95,042005,2017}

\bibitem{Abbott2017a}
\Authname{Abbott}{B. P.} \etal,
\ApJ{847}{47}{2017}

\bibitem{Abbott2017b}
\Authname{Abbott}{B. P.} \etal,
\PRD{95,122003,2017}

\bibitem{Abbott2019scox1}
\Authname{Abbott}{B. P.} \etal,
\PRD{100,122002,2019}

\bibitem{Kaluzienski1980}
\Authname{Kaluzienski}{L. J.} \etal,
\ApJ{241}{779}{1980}

\bibitem{Messenger2015}
\Authname{Messenger}{C.} \etal,
\PRD{92,023006,2015}

\bibitem{Meadors2018}
\Authname{Meadors}{G. D.} \etal,
\PRD{97,044017,2018}

\bibitem{Dreissigacker2019}
\Authname{Dreissigacker}{C.} \etal,
\PRD{100, 044009, 2019}

\bibitem{Smits_2011}
\Authname{Smits}{R.}, \Authname{Tingay}{S. J.}, \Authname{Wex}{N.}, \Authname{Kramer}{M.} and \Authname{Stappers}{B.},
\AnA{528}{A108}{2011}

\bibitem{Glampedakis_and_Gualtieri_2018}
\Authname{Glampedakis}{K.} and \Authname{Gualtieri}{L.}, 
Astrophys. Space Sci. Libr. {\bf{457}}, 673 (2018).

\bibitem{Sieniawska_and_Bejger_2019}
\Authname{Sieniawska}{M.} and \Authname{Bejger}{M.},
Universe {\bf{5}}, 217, (2019).

\bibitem{LVC_CW_SNR_APJ_2019}
\LVC, 
\ApJ{875}{122}{2019}

\bibitem{OnoEdaItoh}
\Authname{Ono}{K.}, \Authname{Eda}{K.}, and \Authname{Itoh}{Y.}, 
\PRD{91,084032,2015}

\bibitem{LVC_CW_O1CW_APJ_2017}
\LVC,
\ApJ{851}{71}{2017}

\bibitem{LVC_CW_O2CW_APJ_2019}
\LVC,
\ApJ{879}{10}{2019}

\bibitem{LVC_CW_NonTensorialCW_PRL_2018}
\LVC,
\PRL{120,031104,2018}

\bibitem{Johnson-McDaniel_2013a}
\Authname{Johnson-McDaniel}{N.~K.} and  \Authname{Owen}{B.~J.},
\PRD{88,044004,2013}

\bibitem{Johnson-McDaniel_2013b}
\Authname{Johnson-McDaniel}{N.~K.},
\PRD{88,044016,2013}

\bibitem{Pitkin_2011}
\Authname{Pitkin}{Matthew},
\MNRAS{415}{1849}{2011}



\bibitem{2005ApJ...628L..53I}
\Authname{Israel}{G.~L.} \etal,
\ApJL{628}{L53}{2005}

\bibitem{2001MNRAS.327..639I}
\Authname{Ioka}{K.},
\MNRAS{327}{639}{2001}

\bibitem{2011PhRvD..83j4014C}
\Authname{Corsi}{A.} \etal,
\PRD{83,104014,2011}

\bibitem{2011MNRAS.418..659L}
\Authname{Levin }{Y.} \etal,
\MNRAS{418}{659}{2011}

\bibitem{2012PhRvD..85b4030Z}
\Authname{Zink}{B.} \etal,
\PRD{85,1002030,2012}

\bibitem{Abbott:2007zzb}
\LSC
\PRD{76,062003,2007}

\bibitem{2019ApJ...874..16}
\LSC
\ApJ{874}{163}{2019}



\bibitem{KS99}
\Authname{Kokkotas}{K.} \etal,
\LRR{2}{2}{1999}

\bibitem{AK96}
\Authname{Andersson}{N.} \etal,
\PRL{77,4134,1996}

\bibitem{AK98}
\Authname{Andersson}{N.} \etal,
\MNRAS{299}{1059}{1998}

\bibitem{F87}
\Authname{Finn}{L. S.}
\MNRAS{227}{265}{1987}

\bibitem{STM01}
\Authname{Sotani}{H.} \etal,
\PRD{65,024010,2001}

\bibitem{MPBGF03}
\Authname{Miniutti}{G.} \etal,
\MNRAS{338}{389}{2003}

\bibitem{SYMT11} 
\Authname{Sotani}{H.} \etal,
\PRD{83,024014,2011}

\bibitem{RG92}
\Authname{Reisenegger}{R.} \etal,
\ApJ{395}{240}{1992}

\bibitem{FGP07}
\Authname{Ferrari}{V.} \etal,
Class. Quant. Grav. \textbf{24}, 5093 (2007).

\bibitem{PAH16}
\Authname{Passamonti}{A.} \etal,
\MNRAS{455}{1489}{2016}

\bibitem{A98}
\Authname{Andersson}{N.}
\ApJ{502}{708}{1998}

\bibitem{FM98}
\Authname{Friedman}{J. L.} \etal,
\ApJ{502}{714}{1998}

\bibitem{AFMSTW03}
\Authname{Arras}{P.} \etal,
\ApJ{591}{1129}{2003}

\bibitem{GGKZ11}
\Authname{Gaertig}{E.} \etal,
\PRL{107,101102,2011}

\bibitem{ST16}
\Authname{Sotani}{H.} \etal,
\PRD{94,044043,2016}

\bibitem{SKTK17}
\Authname{Sotani}{H.} \etal,
\PRD{96,063005,2017}

\bibitem{MRBV18}
\Authname{Morozova}{V.} \etal,
\ApJ{861}{10}{2018}

\bibitem{SKTK19}
\Authname{Sotani}{H.} \etal,
\PRD{99,123024,2019}

\bibitem{Sotani19}
\Authname{Sotani}{H.}, \Authname{Sumiyoshi}{K.}
\PRD{100,083008,2019}

\bibitem{TCPF18}
\Authname{Torres-Forn\'{e}}{A.} \etal,
\MNRAS{474}{5272}{2018}

\bibitem{TCPOF19}
\Authname{Torres-Forn\'{e}}{A.} \etal,
\MNRAS{482}{3967}{2019}



\bibitem{casA}
\Authname{Grefenstette}{B.~W.}, \Authname{Reynolds}{S.~P.} and \Authname{Harrison}{F.~A.} \etal,
\ApJ{802}{15}{2015}

\bibitem{tanaka17}
\Authname{Tanaka}{M.}, \Authname{Maeda}{K.}, \Authname{Mazzali}{P.~A.}, \Authname{Kawabata}{K.~S.} and \Authname{Nomoto}{K.}
\ApJ{837}{105}{2017}

\bibitem{bethe}
\Authname{Bethe}{H.~A.},
\RMP{62,801,1990}

\bibitem{Kotake13}
\Authname{Kotake}{K.},
\CRP{14}{318}{2013}

\bibitem{bernhard16}
\Authname{M{\"u}ller}{B.},
\PASA{33}{e048}{2016} 

\bibitem{bernhard13}
\Authname{M{\"u}ller}{B.}, \Authname{Janka}{H.-T.} and \Authname{Marek}{A.},
\ApJ{766}{43}{2013}

\bibitem{janka16}
\Authname{Janka}{H.-T.}, \Authname{Melson}{T.} and \Authname{Summa}{A.}, 
\ARNPS{66}{341}{2016}

\bibitem{burrows13}
\Authname{Burrows}{A.},
\RMP{85,245,2013}

\bibitem{Kotake12}
\Authname{Kotake}{K.}, \Authname{Sumiyoshi}{K.}, \Authname{Yamada}{S.}, \etal,
 \PTEP{2012}{01A301}{pts009}

\bibitem{radice18}
\Authname{Radice}{D.}, \Authname{Morozova}{V.}, \Authname{Burrows}{A.}, \Authname{Vartanyan}{D.} and \Authname{Nagakura}{H.},
\ApJL{876}{L9}{2019}{}

\bibitem{andresen18}
\Authname{Andresen}{H.}, \Authname{M{\"u}ller}{E.}, \Authname{Janka}{H.-T.}, \etal,
\MNRAS{486}{2238}{2019}

\bibitem{andresen17}
\Authname{Andresen}{H.}, \Authname{M{\"u}ller}{B.}, \Authname{M{\"u}ller}{E.} and \Authname{Janka}{H.-T.},
\MNRAS{468}{2032}{2017}{stx618}

\bibitem{kuroda16}
\Authname{Kuroda}{T.}, \Authname{Kotake}{K.} and \Authname{Takiwaki}{T.},
\ApJL{829}{L14}{2016}

\bibitem{astone18}
\Authname{Astone}{P.}, \Authname{Cerd{\'a}-Dur{\'a}n}{P.} and \Authname{Di Palma}{I.} \etal,
\PRD{98,122002,2018} 

\bibitem{thierry15}
\Authname{Foglizzo}{T.}, \Authname{Kazeroni}{R.} and \Authname{Guilet}{J.} \etal,
\PASA{32}{e009}{2015}

\bibitem{ott_rev}
\Authname{Ott}{C.~D.},
\CQG{26}{063001}{2009} 

\bibitem{hayama15}
\Authname{Hayama}{K.}, \Authname{Kuroda}{T.}, \Authname{Kotake}{K.} and \Authname{Takiwaki}{T.},
\PRD{92,122001,2015} 

\bibitem{powell19}
\Authname{Suvorova}{S.}, \Authname{Powell}{J.} and \Authname{Melatos}{A.},
\PRD{99,123012,2019}

\bibitem{gossan16}
\Authname{Gossan}{S.~E.}, \Authname{Sutton}{P.},
\Authname{Stuver}{A.} \etal,
\PRD{93,042002,2016}

\bibitem{Hayama16}
\Authname{Hayama}{K.}, \Authname{Kuroda}{T.}, \Authname{Nakamura}{K.} and \Authname{Yamada}{S.}, 
\PRL{116,151102,2016} 

\bibitem{Hayama18}
\Authname{Hayama}{K.}, \Authname{Kuroda}{T.}, \Authname{Kotake}{K.} and \Authname{Takiwaki}{T.},
\MNRASL{477}{L96}{2018}{sly055}

\bibitem{forne19}
\Authname{Torres-Forn{\'e}}{A.}, \Authname{Cerd{\'a}-Dur{\'a}n}{P.}, \Authname{Passamonti}{A.}, \Authname{Obergaulinger}{M.} and \Authname{Font}{J.~A.},
\MNRAS{482}{3967}{2019}

\bibitem{morozova18}
\Authname{Morozova}{V.}, \Authname{Radice}{D.}, \Authname{Burrows}{A.} and \Authname{Vartanyan}{D.},
\ApJ{861}{10}{2018}

\bibitem{sotani17}
\Authname{Sotani}{H.}, \Authname{Kuroda}{T.}, \Authname{Takiwaki}{T.} and \Authname{Kotake}{K.},
\PRD{96,063005,2017} 

\bibitem{kagra17}
\Authname{Akutsu}{T.}, \Authname{Ando}{M.} and \Authname{Araki}{S.}, \etal,
\PTEP{2018}{013F01}{ptx180}

\bibitem{woos06}
\Authname{Woosley}{S.~E.} and \Authname{Heger}{A.},
\ApJ{637}{914}{2006}

\bibitem{Hayama15}
\Authname{Hayama}{K.}, \Authname{Kuroda}{T.}, \Authname{Kotake}{K.} and \Authname{Takiwaki}{T.}
\PRD{92,122001,2015}



\bibitem{inflation1}
\Authname{Starobinsky}{A.~A.},
\PLB{91,99,1980}

\bibitem{inflation2}
\Authname{Sato}{K.},
\MNRAS{195}{467}{1981}{}

\bibitem{inflation3}
\Authname{Kazanas}{D.} 
\ApJL{241}{L59}{1980}

\bibitem{inflation4}
\Authname{Guth}{A.~H.},
\PRD{23,347,1981}

\bibitem{scalar1}
\Authname{Mukhanov}{V.~F.} and \Authname{Chibisov}{G.~V.},
\JETPL{33}{532}{1981}

\bibitem{scalar2}
\Authname{Linde}{A.~D.},
\PLB{116,335,1982}

\bibitem{scalar3}
\Authname{Hawking}{S.~W.} 
\PLB{115,295,1982}

\bibitem{scalar4}
\Authname{Starobinsky}{A.~A.},
\PLB{117,175,1982}

\bibitem{scalar5}
\Authname{Guth}{A.~H.} and \Authname{Pi}{S.~Y.},
\PRL{49,1110,1982}

\bibitem{tensor1}
\Authname{Starobinsky}{A.~A.},
\JETPL{30}{682}{1979}

\bibitem{tensor2}
\Authname{Rubakov}{V.~A.},  \Authname{Sazhin}{M.~V.} and \Authname{Veryaskin}{A.~V.},
\PLB{115,189,1982}

\bibitem{tensor3}
\Authname{Abbott}{L.~F.} and \Authname{Wise}{M.~B.},
\NPB{244,541,1984}

\bibitem{LIGOScientific:2019vic} 
\Authname{Abbott}{B. P.} \etal (LIGO/Virgo collaborations),
\PRD{100,061101,2009}

\bibitem{Kuroyanagi:2008ye} 
\Authname{Kuroyanagi}{S.}, \Authname{Chiba}{T.} and \Authname{Sugiyama}{N.},
\PRD{79,103501,2009}

\bibitem{Kuroyanagi:2014qaa} 
\Authname{Kuroyanagi}{S.},  \Authname{Tsujikawa}{S.},  \Authname{Chiba}{T.} and \Authname{Sugiyama}{N.},
\PRD{90,063513,2014}

\bibitem{Brandenberger:2006xi} 
\Authname{Brandenberger}{R.~H.}, \Authname{Nayeri}{A.}, \Authname{Patil}{S.~P.} and \Authname{Vafa}{C.},
\PRL{98,231302,2007}

\bibitem{Piao:2004tq} 
\Authname{Piao}{Y.~-S.} and \Authname{Zhang}{Y.~-Z.},
\PRD{70,063513,2004}
 
\bibitem{Baldi:2005gk} 
\Authname{Baldi}{M.}, \Authname{Finelli}{F.} and \Authname{Matarrese}{S.},
\PRD{72,083504,2005}
 
\bibitem{Kobayashi:2010cm} 
\Authname{Kobayashi}{T.}, \Authname{Yamaguchi}{M.} and \Authname{Yokoyama}{J.~'i.},
\PRL{105,231302,2010}

\bibitem{Calcagni:2004as} 
\Authname{Calcagni}{G.} and \Authname{Tsujikawa}{S.},
\PRD{70,103514,2004}
 
\bibitem{Calcagni:2013lya} 
\Authname{Calcagni}{G.},  \Authname{Kuroyanagi}{S.},  \Authname{Ohashi}{J.} and \Authname{Tsujikawa}{S.},
\JCAP{1403}{2014}{}{025}

\bibitem{Cook:2011hg} 
\Authname{Cook}{J.~L.} and \Authname{Sorbo}{L.},
\PRD{85,023534,2012} Erratum: {\it ibid}. 86, 069901 (2012).

\bibitem{Mukohyama:2014gba} 
\Authname{Mukohyama}{S.}, \Authname{Namba}{R.}, \Authname{Peloso}{M.} and \Authname{Shiu}{G.},
\JCAP{08}{2014}{}{036}
  
\bibitem{Mohanty:2014kwa} 
\Authname{Mohanty}{S.} and \Authname{Nautiyal}{A.},
\arxiv{1404.2222}
  
\bibitem{Biagetti:2013kwa} 
\Authname{Biagetti}{M.}, \Authname{Fasiello}{M.} and \Authname{Riotto}{A.},
\PRD{88,103518,2013}
  
\bibitem{Fujita:2018ehq} 
\Authname{Fujita}{T.}, \Authname{Kuroyanagi}{S.}, \Authname{Mizuno}{S.} and \Authname{Mukohyama}{S.},
\PLB{789,215,2019}

\bibitem{Peebles:1998qn} 
\Authname{Peebles}{P.~J.~E.} and \Authname{Vilenkin}{A.},
Quintessential inflation.
\PRD{59,063505,1999}

\bibitem{Giovannini:1998bp} 
\Authname{Giovannini}{M.},
\PRD{58,083504,1998}
  
\bibitem{Giovannini:1999bh} 
\Authname{Giovannini}{M.},
\PRD{60,123511,1999}

\bibitem{Giovannini:1999qj}
\Authname{Giovannini}{M.},
Class. Quant. Grav. \textbf{16}, 2905 (1999).
     
\bibitem{Tashiro:2003qp} 
\Authname{Tashiro}{H.}, \Authname{Chiba}{T.} and \Authname{Sasaki}{M.},
Class. Quant. Grav. \textbf{21}, 1761 (2004).

\bibitem{Giovannini:2008tm}
\Authname{Giovannini}{M.},
Class. Quant. Grav. \textbf{26}, 045004 (2009).

\bibitem{Figueroa:2018twl} 
\Authname{Figueroa}{D.~G.} and \Authname{Tanin}{E.~H.},
\JCAP{10}{2019}{}{050}
  
\bibitem{Ahmad:2019jbm} 
\Authname{Ahmad}{S.}, \Authname{Felice}{A.~De}, \Authname{Jaman}{N.}, \Authname{Kuroyanagi}{S.} and \Authname{Sami}{M.},
\PRD{100,103525,2019}

\bibitem{Allen:1997ad} 
\Authname{Allen}{B.}, \Authname{Romano}{J.D.},
\PRD{59,102001,1999}



\bibitem{Grojean:2006bp}
\Authname{Grojean}{C.} \etal,
\PRD{75,043507,2007}

\bibitem{Caprini:2015zlo}
\Authname{Caprini}{C.} \etal,
\JCAP{04}{2016}{}{001}

\bibitem{Kakizaki:2015wua}
\Authname{Kakizaki}{M.} \etal,
\PRD{92,115007,2015}

\bibitem{Hashino:2018wee}
\Authname{Hashino}{K.} \etal,
\PRD{99,075011,2019}
    
\bibitem{Dev:2016feu}
\Authname{Dev}{P. S. B} \etal,
\PRD{93,104001,2016}



\bibitem{Kuroyanagi:2018csn}
\Authname{Kuroyanagi}{S.}, \Authname{Chiba}{T.} and  \Authname{Takahashi}{T.},
\JCAP{11}{2018}{}{038}

\bibitem{Allen:1996gp} 
\Authname{Allen}{B.} and \Authname{Ottewill}{A.~C.},
\PRD{56,545,1997}

\bibitem{TheLIGOScientific:2016xzw} 
\LVC,
\PRL{118,121102,2017}

\bibitem{Nishizawa:2009bf} 
\Authname{Nishizawa}{A.},  \Authname{Taruya}{A.},  \Authname{Hayama}{K.},  \Authname{Kawamura}{S.} and \Authname{Sakagami}{M.~a.},
\PRD{79,082002,2009}

\bibitem{Abbott:2018utx} 
\LVC,
\PRL{120,201102,2018}

\bibitem{Drasco:2002yd} 
\Authname{Drasco}{S.} and \Authname{Flanagan}{E.~E.},
\PRD{67,082003,2003}

\bibitem{Thrane:2013kb} 
\Authname{Thrane}{E.}, 
\PRD{87,043009,2013}

\bibitem{Martellini:2014xia} 
\Authname{Martellini}{L.} and \Authname{Regimbau}{T.},
\PRD{89,124009,2014}

\bibitem{Brito:2017wnc}
\Authname{Brito}{R.} {\it et al.},
\PRL{119,131101,2017}

\bibitem{Bose:2005fm} 
\Authname{Bose}{S.},
\PRD{71,082001,2005}

\bibitem{Figueroa:2018xtu} 
\Authname{Figueroa}{D.~G.}, \Authname{Megias}{E.}, \Authname{Nardini}{G.}, \Authname{Pieroni}{M.}, \Authname{Quiros}{M.}, \Authname{Ricciardone}{A.} and \Authname{Tasinato}{G.},
PoS GRASS {\bf 2018}, 036 (2018)



\bibitem{LiEtAl:2012a}
\Authname{Li}{T.~G.~F.}, \Authname{Del Pozzo}{W.}, \Authname{Vitale}{S.}, \Authname{Van Den Broeck}{C.}, \Authname{Agathos}{M.}, \Authname{Veitch}{J.}, \Authname{Grover}{K.}, \Authname{Sidery}{T.}, \Authname{Sturani}{R.} and \Authname{Vecchio}{A.},
\PRD{85,082003,2012}

\bibitem{LiEtAl:2012b}
\Authname{Li}{T.~G.~F.}, \Authname{Del Pozzo}{W.}, \Authname{Vitale}{S.}, \Authname{Van Den Broeck}{C.}, \Authname{Agathos}{M.}, \Authname{Veitch}{J.}, \Authname{Grover}{K.}, \Authname{Sidery}{T.}, \Authname{Sturani}{R.} and \Authname{Vecchio}{A.},
\JPCS{363}{012028}{2012}

\bibitem{Agathos:2013upa}
\Authname{Agathos}{M.~}, \Authname{Del Pozzo}{W.~}, \Authname{Li}{T.~G.~F.~}, \Authname{Van Den Broeck}{C.~}, \Authname{Veitch}{J.~} and \Authname{Vitale}{S.~},
\PRD{89,082001,2014}

\bibitem{Cornish:2011ys}
\Authname{Cornish}{N.~}, \Authname{Sampson}{L.~}, \Authname{Yunes}{N.~} and \Authname{Pretorius}{F.~},
\PRD{84,062003,2011}

\bibitem{Sampson:2013jpa}
\Authname{Sampson}{L.~}, \Authname{Cornish}{N.~} and \Authname{Yunes}{N.~},
\PRD{89,064037,2014}

\bibitem{Meidam:2017dgf}
\Authname{Meidam}{J.~} \etal,
\PRD{97,044033,2018}

\bibitem{Ghosh:2016qgn}
\Authname{Ghosh}{A.~} \etal,
\PRD{94,021101(R),2016}

\bibitem{Ghosh:2017gfp}
\Authname{Ghosh}{A.~} \etal,
\CQG{35}{014002}{2018}

\bibitem{TheLIGOScientific:2016src}
\LVC,
\PRL{116,221101,2016}

\bibitem{TheLIGOScientific:2016pea}
\LVC,
\PRX{6,041015,2016}

\bibitem{Abbott:2018lct}
\LVC,
\PRL{123,011102,2019}

\bibitem{LIGOScientific:2019fpa}
\LVC,
\PRD{100,104036,2019}



\bibitem{Khoury:2003aq} 
  \Authname{Khoury}{J.} and \Authname{Weltman}{A.},
  \PRD{93,171104,2004}

\bibitem{Vainshtein:1972sx}
\Authname{Vainshtein}{A.~I.},
\PLB{39,393,1972}  

\bibitem{YunesPretorius2009}
\Authname{Yunes}{N.} and \Authname{Pretorius}{F.},
\PRD{80,122003,2009}
 
 \bibitem{Yunes:2009bv}
 \Authname{Yunes}{N.}, \Authname{Pretorius}{F.} and \Authname{Spergel}{D.},
  \PRD{81,064018,2010}

\bibitem{Yagi:2011yu}
  \Authname{Yagi}{K.}, \Authname{Tanahashi}{N.} and \Authname{Tanaka}{T.},
  \PRD{83,084036,2011}
 
\bibitem{Will:1989sk} 
  \Authname{Will}{C.~M.} and \Authname{Zaglauer}{H.~W.},
  \ApJ{346}{366}{1989}
  
\bibitem{Yagi:2011xp} 
  \Authname{Yagi}{K.}, \Authname{Stein}{L.~C.}, \Authname{Yunes}{N.} and \Authname{Tanaka}{T.},
  \PRD{85,064022,2012} Erratum:  {\it ibid}. 93, 029902 (2016).

\bibitem{Will:1997bb} 
  \Authname{Will}{C.~M.},
  \PRD{57,2061,1998}

\bibitem{Hassan:2011zd} 
 \Authname{Hassan}{S.~F.} and \Authname{Rosen}{R.~A.},
  \JHEP{1202,126,2012}
 
\bibitem{Yagi:2012vf} 
  \Authname{Yagi}{K.}, \Authname{Yunes}{N.} and \Authname{Tanaka}{T.},
  \PRL{109,251105,2012}
  Erratum: {\it ibid}. 116, 169902 (2016).

\bibitem{Ramazanoglu2016}
\Authname{Ramazanoglu}{F.~M.} and \Authname{Pretorius}{F.},  
\PRD{93,064005,2016}



\bibitem{Lombriser:2015sxa}
\Authname{Lombriser}{L.} and \Authname{Taylor}{A.},
\JCAP{03}{2016}{}{031}

\bibitem{Saltas:2014dha}
\Authname{Saltas}{I. D.}, \Authname{Sawicki}{I.}, \Authname{Amendola}{L.}, and \Authname{Kunz}{M.},
\PRL{113,191101,2014}

\bibitem{Nishizawa:2018srh}
\Authname{Nishizawa}{A.} and \Authname{Kobayashi}{T.},
\PRD{98,124018,2018}

\bibitem{Zhao:2019xmm}
\Authname{Zhao}{W.}, 
\Authname{Zhu}{T.},
\Authname{Qiao}{J.}, and 
\Authname{Wang}{A.},
\PRD{101,024002,2020}

\bibitem{Blas:2016qmn}
\Authname{Blas}{D.},
\Authname{Ivanov}{M.~M.},
\Authname{Sawicki}{I.}, and 
\Authname{Sibiryakov}{S.},
JETP Lett. 103, 624 (2016).

\bibitem{Cornish:2017jml}
\Authname{Cornish}{N.},
\Authname{Blas}{D.}, and 
\Authname{Nardini}{G.},
\PRL{119,161102,2017}

\bibitem{Nishizawa2014PRD}
\Authname{Nishizawa}{A.} and \Authname{Nakamura}{T.},
\PRD{90,044048,2014}

\bibitem{Baker:2017hug}
\Authname{Baker}{T.}, \Authname{Bellini}{E.}, \Authname{Ferreira}{P.G.}, \Authname{Lagos}{M.}, \Authname{Noller}{J.}, and \Authname{Sawicki}{I.},
\PRL{119,251301,2017}

\bibitem{Creminelli:2017sry}
\Authname{Creminelli}{P.} and \Authname{Vernizzi}{F.},
\PRL{119,251302,2017}

\bibitem{Sakstein:2017xjx}
\Authname{Sakstein}{J.} and \Authname{Jain}{B.},
\PRL{119,251303,2017}

\bibitem{Ezquiaga:2017ekz}
\Authname{Ezquiaga}{J.M.} and \Authname{Zumalacarregui}{M.},
\PRL{119,251304,2017}

\bibitem{Bernus:2019rgl}
\Authname{Bernus}{L.}, \Authname{Minazzoli}{O.}, \Authname{Fienga}{A.}, \Authname{Gastineau}{M.}, \Authname{Laskar}{J.}, \Authname{Deram}{P.},
\PRL{123,161103,2019}

\bibitem{Arai:2017hxj}
\Authname{Arai}{S.} and \Authname{Nishizawa}{A.},
\PRD{97,104038,2018}

\bibitem{Nishizawa:2019rra}
\Authname{Nishizawa}{A.} and \Authname{Arai}{S.},
\PRD{99,104038,2019}

\bibitem{Nishizawa:2017nef}
\Authname{Nishizawa}{A.},
\PRD{97,104037,2018}

\bibitem{Audley:2017drz}
\Authname{Audley}{H.} \etal,
\arxiv{1702.00786}

\bibitem{Sato:2017dkf}
\Authname{Sato}{S.} \etal,
J. Phys. Conf. Ser. 840, 012010 (2017).

\bibitem{Yagi:2009zm}
\Authname{Yagi}{K.} and \Authname{Tanaka}{T.},
\PRD{81,064008,2010}

\bibitem{Yagi:2009zz}
\Authname{Yagi}{K.} and \Authname{Tanaka}{T.},
Prog. Theor. Phys. 123, 1069 (2010).

\bibitem{Crisostomi:2017ugk}
\Authname{Crisostomi}{M.}, \Authname{Noui}{K.}, \Authname{Charmousis}{C.}, and \Authname{Langlois}{D.},
\PRD{97,044034,2018}






\bibitem{Eardley1973}
\Authname{Eardley}{D.} \etal,
\PRL{30,884,1973}

\bibitem{Will2005}
\Authname{Will}{C. M.},
\LRR{9}{3}{2006}

\bibitem{Takeda2018}
\Authname{Takeda}{H.} \etal,
\PRD{98,022008,2018}

\bibitem{Hagihara:2018azu}
\Authname{Hagihara}{Y.}, \Authname{Era}{N.}, \Authname{Iikawa}{D.}, \Authname{Asada}{H.},
\PRD{98,064035,2018}

\bibitem{Hagihara:2019ihn}
\Authname{Hagihara}{Y.}, \Authname{Era}{N.}, \Authname{Iikawa}{D.}, \Authname{Nishizawa}{A.}, \Authname{Asada}{H.},
\PRD{100,064010,2019}

\bibitem{Hagihara:2019rny}
\Authname{Hagihara}{Y.}, \Authname{Era}{N.}, \Authname{Iikawa}{D.}, \Authname{Takeda}{N.}, \Authname{Asada}{H.},
\PRD{101,041501,2020}

\bibitem{Chatziioannou:2012rf}
\Authname{Chatziioannou}{K.}, \Authname{Yunes}{N.}, \Authname{Cornish}{N.},
\PRD{86,022004,2012}



\bibitem{Dreyer:2003bv}
\Authname{Dreyer}{O.}, \Authname{Kelly}{B.~J.}, \Authname{Krishnan}{B.}, \Authname{Finn}{L.~S.}, \Authname{Garrison}{D.} and \Authname{Lopez-Aleman}{R.},
\CQG{21}{787}{2004}

\bibitem{Berti:2007zu} 
\Authname{Berti}{E.}, \Authname{Cardoso}{J.}, \Authname{Cardoso}{V.} and \Authname{Cavaglia}{M.},
\PRD{76,104044,2007}

\bibitem{Sakai:2017ckm} 
\Authname{Sakai}{K.}, \Authname{Oohara}{K.~I.}, \Authname{Nakano}{H.}, \Authname{Kaneyama}{M.} and \Authname{Takahashi}{H.},
\PRD{96,044047,2017}
  
\bibitem{Carullo:2018sfu} 
\Authname{Carullo}{G.} \etal, 
\PRD{98,104020,2018}
  
\bibitem{arXiv:1902.07527} 
\Authname{Carullo}{G.}, \Authname{Pozzo}{W.~Del} and \Authname{Veitch}{J.},
\PRD{99,123029,2019} Erratum: {\it ibid}. 100, 089903 (2019).

\bibitem{Berti:2018vdi} 
\Authname{Berti}{E.}, \Authname{Yagi}{K.}, \Authname{Yang}{H.} and \Authname{Yunes}{N.},
\GRG{50}{49}{2018}

\bibitem{Mohanty:1997eu} 
\Authname{Mohanty}{S.~D.}, 
\PRD{57,630,1998}

\bibitem{Baibhav:2017jhs} 
\Authname{Baibhav}{V.}, \Authname{Berti}{E.}, \Authname{Cardoso}{V.} and \Authname{Khanna}{G.},
\PRD{97,044048,2018}

\bibitem{Giesler:2019uxc} 
\Authname{Giesler}{M.}, \Authname{Isi}{M.}, \Authname{Scheel}{M.~A.} and \Authname{Teukolsky}{S.~A.},
Phys. Rev. X, {\bf 9}, 041060 (2019).

\bibitem{Isi:2019aib} 
\Authname{Isi}{M.}, \Authname{Giesler}{M.}, \Authname{Farr}{W.~M.}, \Authname{Scheel}{M.~A.} and \Authname{Teukolsky}{S.~A.},
\PRL{123,111102,2019}

\bibitem{Bhagwat:2019dtm} 
\Authname{Bhagwat}{S.}, \Authname{Forteza}{X.~J.}, \Authname{Pani}{P.} and \Authname{Ferrari}{V.},
\PRD{101,044033,2020}

\bibitem{Ota:2019bzl} 
\Authname{Ota}{I.} and \Authname{Chirenti}{C.},
\PRD{101,104005,2020}

\bibitem{Baibhav:2018rfk} 
\Authname{Baibhav}{V.} and \Authname{Berti}{E.}, 
\PRD{99,024005,2019}

\bibitem{Berti:2016lat} 
\Authname{Berti}{E.}, \Authname{Sesana}{A.}, \Authname{Barausse}{E.}, \Authname{Cardoso}{V.} and \Authname{Belczynski}{K.},
\PRL{117,101102,2016}

\bibitem{Healy:2017psd} 
\Authname{Healy}{J.}, \Authname{Lousto}{C.~O.}, \Authname{Zlochower}{Y.} and \Authname{Campanelli}{M.},
\CQG{34}{224001}{2017}

\bibitem{RITcatalog}
\url{https://ccrg.rit.edu/content/data/rit-waveform-catalog} 

\bibitem{Uchikata_prep}
\Authname{Uchikata}{N.} \etal, in preparation.

  

\bibitem{Schutz:1986Nature}
\Authname{Schutz}{B.~F.}, 
\Nat{323}{310}{1986}

\bibitem{Holz:2005ApJ}
\Authname{Holz}{D.~E.} and \Authname{Hughes}{S.~A.},
\ApJ{629}{15}{2005}

\bibitem{Planck2018cosmology}
\Authname{Aghanim}{N.} \etal, (Planck Collaboration),
\arxiv{1807.06209}

\bibitem{Schoneberg:2019wmt}
\Authname{Schoneberg}{N.}, \Authname{Lesgourgues}{J.}, \Authname{Hooper}{D.~C.},  
\JCAP{10}{2019}{}{029}

\bibitem{Cuceu:2019for}
\Authname{Cuceu}{A.}, \Authname{Farr}{J.}, \Authname{Lemos}{P.}, \Authname{Font-Ribera}{A.}
\JCAP{10}{2019}{}{044}

\bibitem{Riess:2018byc}
\Authname{Riess}{A.~G.} \etal,
\ApJ{861}{126}{2018}

\bibitem{Riess:2019cxk}
\Authname{Riess}{A.~G.} \etal,
\ApJ{876}{85}{2019}

\bibitem{Collett:2019hrr}
\Authname{Collett}{T.}, \Authname{Montanari}{F.}, \Authname{Rasanen}{S.}
\PRL{123,231101,2019}

\bibitem{Wong:2019kwg}
\Authname{Wong}{K.~C.} \etal,
\arxiv{1907.04869}

\bibitem{Chen:2019ejq}
\Authname{Chen}{G.~C.~-F.} \etal,
\MNRAS{490}{1743}{2019}

\bibitem{Freedman:2019jwv}
\Authname{Freedman}{W.~L.} \etal,
\ApJ{882}{34}{2019}

\bibitem{Cutler:2009qv}
\Authname{Cutler}{C.} and \Authname{Holz}{D.~E.},
\PRD{80,104009,2009}


\bibitem{Nishizawa:2016ood}
\Authname{Nishizawa}{A.},
\PRD{96,101303,2017}

\bibitem{MacLeod:2008PRD}
\Authname{MacLeod}{C.~L.} and \Authname{Hogan}{C.~J.},
\PRD{77,043512,2008}

\bibitem{Abbott:2017xzu} 
\Authname{Abbott}{B.~P.} \etal, (LIGO Scientific, Virgo, 1M2H, Dark Energy Camera GW-E, DES, DLT40, Las Cumbres Observatory, VINROUGE, MASTER Collaborations), 
\Nat{551}{85}{2017}

\bibitem{Soares-Santos:2019irc}
\Authname{Soares-Santos}{M.} \etal, (DES, LIGO Scientific, Virgo Collaborations),
\arxiv{1901.01540}

\bibitem{Hotokezaka:2018dfi} 
\Authname{Hotokezaka}{K.} \etal, 
\NatAst{3}{940}{2019}

\bibitem{Vitale:2018wlg}
\Authname{Vitale}{S.} and \Authname{Chen}{H.~-Y.},
\PRL{121,021303,2018}

\bibitem{Seto:2017swx}
\Authname{Seto}{N.} and \Authname{Kyutoku}{K.},
\MNRAS{475}{4133}{2018}

\bibitem{Chen:2017rfc} 
\Authname{Chen}{H.~-Y.}, \Authname{Fishbach}{M.}, and \Authname{Holz}{D.~E.}, 
\Nat{562}{545}{2018}

\bibitem{Gupte:2018pht}
\Authname{Gupte}{N.} and \Authname{Bartos}{I.}, 
\arxiv{1808.06238}

\bibitem{Howell:2018nhu}
\Authname{Howell}{E.~J.}, \Authname{Ackley}{K.}, \Authname{Rowlinson}{A.} and \Authname{Coward}{D.}, 
\MNRAS{485}{1435}{2019}

\bibitem{Mogushi:2018ufy}
\Authname{Mogushi}{K.}, \Authname{Cavaglia}{M.} and \Authname{Siellez}{K.}, 
\arxiv{1811.08542}

\bibitem{Gehrels:2015ApJ}
\Authname{Gehrels}{N.} \etal,
\ApJ{820}{136}{2016}



\bibitem{Turner:1984ch}
\Authname{Turner}{E.~L.}, \Authname{Ostriker}{J.~P.} and \Authname{Gott}{J.~R.},
\ApJ{284}{1}{1984}

\bibitem{Treu:2010uj}
\Authname{Treu}{T.},
\ARAA{48}{87}{2010}

\bibitem{Broadhurst:2018saj}
\Authname{Broadhurst}{T.}, \Authname{Diego}{J.~M.} and \Authname{Smoot}{G.},
\arxiv{1802.05273}

\bibitem{Broadhurst:2019ijv}
\Authname{Broadhurst}{T.}, \Authname{Diego}{J.~M.} and \Authname{Smoot}{G.},
\arxiv{1901.03190}

\bibitem{Christian:2018vsi}
\Authname{Christian}{P.}, \Authname{Vitale}{S.} and \Authname{Loeb}{A.},
\PRD{98,103022,2018}

\bibitem{Ng:2017yiu}
\Authname{Ng}{K.~K.~Y.} \etal, 
\PRD{97,023012,2018}

\bibitem{Smith:2017mqu}
\Authname{Smith}{G.~P.} \etal, 
\MNRAS{475}{3823}{2018}

\bibitem{Li:2018prc}
\Authname{Li}{S.-S.} \etal, 
\MNRAS{476}{2220}{2018}

\bibitem{Finn:1995ah}
\Authname{Finn}{L.~S.},
\PRD{53,2878,1996}

\bibitem{Wang:1996as}
\Authname{Wang}{Y.}, \Authname{Stebbins}{A.} and \Authname{Turner}{E.~L.},
\PRL{77,2875,1996}

\bibitem{Nakamura:1997sw}
\Authname{Nakamura}{T.~T.},
\PRL{80,1138,1998}

\bibitem{Jung:2017flg}
\Authname{Jung}{S.} and \Authname{Shin}{C.~S.},
\PRL{122,041103,2019}

\bibitem{Lai:2018rto}
\Authname{Lai}{K.-H.} \etal,
\PRD{98,083005,2018}

\bibitem{Takahashi:2003ix}
\Authname{Takahashi}{R.} and \Authname{Nakamura}{T.},
\ApJ{595}{1039}{2003}

\bibitem{Collett:2016dey}
\Authname{Collett}{T.~E.} and \Authname{Bacon}{D.},
\PRL{118,091101,2017}

\bibitem{Fan:2016swi}
\Authname{Fan}{X.-L.} \etal, 
\PRL{118,091102,2017}



\bibitem{DECam}
\Authname{Bernstein}{J.~P.} \etal,
\ApJ{753}{152}{2012}

\bibitem{PanSTRR}
\url{http://pan-starrs.ifa.hawaii.edu} 

\bibitem{LSST}
\url{https://www.lsstcorporation.org} 

\bibitem{Nissanke:2012dj}
S.,~Nissanke, M.~Kasliwal, and A.~Georgieva,
\ApJ{767}{124}{2013}

\bibitem{Rosswog:1998hy}
S.~Rosswog, M.~Liebendoerfer, F.~K. Thielemann, M.~B. Davies, W.~Benz, and T.~Piran,
Astron. Astrophys. {\bf{341}}, 499 (1999).

\bibitem{Hotokezaka:2012ze}
K.~Hotokezaka, K.~Kiuchi, K.~Kyutoku, H.~Okawa, Y.~Sekiguchi, M.~Shibata, and K.~Taniguchi,
\PRD{87,024001,2013}

\bibitem{Lattimer:1974slx}
J.~M.~Lattimer and D.~N.~Schramm,
\ApJL{192}{L145}{1974}

\bibitem{Eichler:1989ve}
\Authname{Eichler}{D.}, \Authname{Livio}{M.}, \Authname{Piran}{T.}, \Authname{Schramm}{D.~N.},
\Nat{340}{126}{1989}

\bibitem{Korobkin:2012uy}
O.~Korobkin, S.~Rosswog, A.~Arcones, and C.~Winteler,
\MNRAS{426}{1940}{2012}

\bibitem{Wanajo:2014wha}
S.~Wanajo, Y.~Sekiguchi, N.~Nishimura, K.~Kiuchi, K.~Kyutoku, and M.~Shibata,
\ApJL{789}{L39}{2014}

\bibitem{Li:1998bw}
L.~-X.~Li and B.~Paczynski,
\ApJL{507}{L59}{1998}

\bibitem{Kulkarni:2005jw}
S.~R.~Kulkarni,
\arxiv{astro-ph/0510256}

\bibitem{Metzger:2010sy}
B.~D.~Metzger, G.~Martinez-Pinedo, S.~Darbha, E.~Quataert, A.~Arcones, D.~Kasen, R.~Thomas, P.~Nugent, I.~V. Panov, and N.~T.~Zinner,
\MNRAS{406}{2650}{2010}

\bibitem{Kasen:2013xka}
D.~Kasen, N.~R.~Badnell, and J.~Barnes,
\ApJ{774}{25}{2013}

\bibitem{Tanaka:2013ana}
M.~Tanaka and K.~Hotokezaka,
\ApJ{775}{113}{2013}

\bibitem{Kasen:2014toa}
D.~Kasen, R.~Fern{\'a}ndez, and B.~Metzger,
\MNRAS{450}{1777}{2015}

\bibitem{Barnes:2016umi}
J.~Barnes, D.~Kasen, M.-R.~Wu, and G.~Mart{\'i}nez-Pinedo,
\ApJ{829}{110}{2016}

\bibitem{Wollaeger:2017ahm}
R.~T.~Wollaeger \etal,
\MNRAS{478}{3298}{2018}

\bibitem{Tanaka:2017qxj}
M.~Tanaka et~al.
Publ. Astron. Soc. Jap. 69, psx12 (2017).

\bibitem{Tanaka:2017lxb}
M.~Tanaka et~al,
\ApJ{852}{109}{2018}

\bibitem{Kasliwal:2017ngb}
M.~M.~Kasliwal \etal,
\Sci{358}{1559}{2017}

\bibitem{Cowperthwaite:2017dyu}
P.~S. Cowperthwaite et~al.
\ApJL{848}{L17}{2017}

\bibitem{Kasen:2017sxr}
D.~Kasen, B.~Metzger, J.~Barnes, E.~Quataert, and E.~Ramirez-Ruiz,
\Nat{551}{80}{2017}

\bibitem{Villar:2017wcc}
V.~Ashley Villar \etal,
\ApJL{851}{L21}{2017}

\bibitem{Perego:2017wtu}
A.~Perego, D.~Radice, and S.~Bernuzzi,
\ApJL{850}{L37}{2017}

\bibitem{Tanvir:2017pws}
N.~R.~Tanvir et~al,
\ApJL{848}{L27}{2017}

\bibitem{Kawaguchi:2018ptg}
K.~Kawaguchi, M.~Shibata, and M.~Tanaka,
\ApJL{865}{L21}{2018}

\bibitem{GBM:2017lvd}
B.~P.~Abbott \etal,
\ApJL{848}{L12}{2017}

\bibitem{Rezzolla:2017aly}
L.~Rezzolla, E.~R.~Most, and L.~R.~Weih,
\ApJL{852}{L25}{2018}

\bibitem{Shibata:2017xdx}
M.~Shibata, S.~Fujibayashi, K.~Hotokezaka, K.~Kiuchi, K.~Kyutoku, Y.~Sekiguchi, and M.~Tanaka,
\PRD{96,123012,2017}

\bibitem{Kiuchi:2019lls}
K.~Kiuchi, K.~Kyutoku, M.~Shibata, and K.~Taniguchi,
\ApJL{876}{L31}{2019}

\bibitem{Foucart:2018rjc}
F.~Foucart, T.~Hinderer, and S.~Nissanke,
\PRD{98,081501,2018}

\bibitem{Kawaguchi:2016ana}
K.~Kawaguchi, K.~Kyutoku, M.~Shibata, and M.~Tanaka,
\ApJ{825}{52}{2016}

\bibitem{Gottlieb:2017mqv}
O.~Gottlieb, E.~Nakar, and T.~Piran,
\MNRAS{473}{576}{2018}

\bibitem{Arcavi:2018mzm}
I.~Arcavi,
\ApJL{855}{L23}{2018}

\bibitem{Matsumoto:2018gzy}
T.~Matsumoto,
\MNRAS{481}{1008}{2018}

\bibitem[{{The LIGO Scientific Collaboration and the Virgo
  Collaboration}(2019)}]{gcn25320}
{The LIGO Scientific Collaboration and the Virgo Collaboration}. 2019, GCN
  Circular, 25320

\bibitem{Andreoni:2019qgh}
\Authname{Andreoni}{I.} \etal,
\ApJ{890}{131}{2020}

\bibitem{Kawaguchi:2020osi}
\Authname{Kawaguchi}{K.}, \Authname{Shibata}{M.}, \Authname{Tanaka}{M.},
\ApJ{893}{153}{2020}

\bibitem{Radice:2016dwd}
\Authname{Radice}{D.}, \Authname{Galeazzi}{F.}, \Authname{Lippuner}{J.},\Authname{Roberts}{L.~F.}, \Authname{Ott}{C.~D.}, \Authname{Rezzolla}{L.}
\MNRAS{460}{3255}{2016}

\bibitem{Sekiguchi:2015dma}
\Authname{Sekiguchi}{Y.}, \Authname{Kiuchi}{K.}, \Authname{Kyutoku}{K.}, \Authname{Shibata}{M.}
\PRD{91,064059,2015}

\bibitem{Sekiguchi:2016bjd}
\Authname{Sekiguchi}{Y.}, \Authname{Kiuchi}{K.}, \Authname{Kyutoku}{K.}, \Authname{Shibata}{M.},  \Authname{Taniguchi}{K.}
\PRD{93,124046,2016}

\bibitem{Cannon+12}
\Authname{Cannon}{K.} \etal,
\ApJ{748}{136}{2012}

\bibitem{Kocevski+18}
\Authname{Kocevski}{D.} \etal,
\ApJ{862}{152}{2018}

\bibitem{Nakar07}
\Authname{Nakar}{E},
\PRep{442}{166}{2007}

\bibitem{Berger14} 
\Authname{Berger}{E.},
\ARNPS{52}{43}{2014}

\bibitem{Paczynski86}
\Authname{Paczy{\'n}ski}{B.},
\ApJL{308}{L43}{1986}

\bibitem{Goodman86}
\Authname{Goodman}{J.},
\ApJL{308}{L47}{1986}

\bibitem{Eichler+89}
\Authname{Eichier}{D.}, \Authname{Livio}{M.}, \Authname{Piran}{T.} and \Authname{Schramm}{D.~N.},
\Nat{340}{126}{1989}

\bibitem{Amati+02} 
\Authname{Amati}{L.} \etal,
\AnA{390}{81}{2002}

\bibitem{Yonetoku+04} 
\Authname{Yonetoku}{D.}, \Authname{Murakami}{T.}, \Authname{Nakamura}{T.}, \Authname{Yamazaki}{R.}, \Authname{Inoue}{A.~K.} and \Authname{Ioka}{K.}, 
\ApJ{609}{935}{2004}

\bibitem{Kisaka+17} 
\Authname{Kisaka}{S.}, \Authname{Ioka}{K.}, \Authname{Sakamoto}{T.},
\ApJ{846}{142}{2017}

\bibitem{GRB170817A_GBM} 
\Authname{Goldstein}{A.} \etal,
\ApJL{848}{L14}{2017}

\bibitem{GRB170817A_INT} 
\Authname{Savchenko}{V.} \etal,
\ApJL{848}{L15}{2017}

\bibitem{Troja+17} 
\Authname{Troja}{E.} \etal,
\Nat{551}{71}{2017}

\bibitem{Margutti+17} 
\Authname{Margutti}{R.} \etal,
\ApJL{848}{L20}{2017}

\bibitem{Haggard+17} 
\Authname{Haggard}{D.}, \Authname{Nynka}{M.}, \Authname{Ruan}{J.~J.}, \Authname{Kalogera}{V.}, \Authname{Cenko}{S.~B.}, \Authname{Evans}{P.} and \Authname{Kennea}{J.~A.},
\ApJL{848}{L25}{2017}

\bibitem{Hallinan+17} 
\Authname{Hallinan}{G.} \etal,
\Sci{358}{1579}{2017}

\bibitem{Alexander+17} 
\Authname{Alexander}{K.~D.} \etal,
\ApJL{848}{L21}{2017}

\bibitem{Lyman+18} 
\Authname{Lyman}{J.~D.} \etal,
\NatAst{2}{751}{2018}

\bibitem{Mooley+18b}
\Authname{Mooley}{K.~P.} \etal,
\Nat{561}{355}{2018}

\bibitem{Ghirlanda+18} 
\Authname{Ghirlanda}{G.} \etal,
\Sci{363}{968}{2019}

\bibitem{Troja+18b}
\Authname{Troja}{E.}, \Authname{van Eerten}{H.}, \Authname{Ryan}{G.}, \Authname{Ricci}{R.}, \Authname{Burgess}{J.~M.}, \Authname{Wieringa}{M.}, \Authname{Piro}{L.}, \Authname{Cenko}{S.~B.} and \Authname{Sakamoto}{T.},
\MNRAS{489}{1919}{2019}

\bibitem{Mooley+18c} 
\Authname{Mooley}{K.~P.} \etal,
\ApJL{868}{L11}{2018}

\bibitem{Lamb+19}
\Authname{Lamb}{G.~P.} \etal,
\ApJL{870}{L15}{2019}

\bibitem{Nagakura+14} 
\Authname{Nagakura}{H.}, \Authname{Hotokezaka}{K.}, \Authname{Sekiguchi}{Y.}  \Authname{Shibata}{M.} and \Authname{Ioka}{K.},
\ApJL{784}{L28}{2014}

\bibitem{Hamidani19}
\Authname{Hamidani}{H.}, \Authname{Kiuchi}{K.}, \Authname{Ioka}{K.}, 
\MNRAS{491}{3192}{2020}

\bibitem{IN18} 
\Authname{Ioka}{K.} and \Authname{Nakamura}{T.},
\PTEP{2018}{043E02}

\bibitem{Granot+17} 
\Authname{Granot}{J.}, \Authname{Guetta}{D.} and \Authname{Gill}{R.},
\ApJL{850}{L24}{2017}

\bibitem{Kasliwal+17} 
\Authname{Kasliwal}{M.~M.} \etal,
\Sci{358}{1559}{2017}

\bibitem{Kisaka+18} 
\Authname{Kisaka}{S.}, \Authname{Ioka}{K.}, \Authname{Kashiyama}{K.} and \Authname{Nakamura}{T.},
\ApJ{867}{39}{2018}

\bibitem{Nakar+18}
\Authname{Nakar}{E.}, \Authname{Gottlieb}{O.}, \Authname{Piran}{T.}, \Authname{Kasliwal}{M.~M.} and \Authname{Hallinan}{G.},
\ApJ{867}{18}{2018}

\bibitem{Troja+18} 
\Authname{Troja}{E.} \etal,
\MNRASL{478}{L18}{2018}

\bibitem{Nakar19}
\Authname{Nakar}{E.}, 
\arxiv{1912.05659}

\bibitem{Mooley+18a}
\Authname{Mooley}{K.~P.} \etal,
\Nat{554}{207}{2018}

\bibitem{Ruan+18} 
\Authname{Ruan}{J.~J.},  \Authname{Nynka}{M.}, \Authname{Haggard}{D.}, \Authname{Kalogera}{V.} and \Authname{Evans}{P.},
\ApJL{853}{L4}{2018}

\bibitem{Margutti+18} 
\Authname{Margutti}{R.} \etal,
\ApJL{856}{L18}{2018}

\bibitem{D'Avanzo+18} 
\Authname{D'Avanzo}{P.} \etal,
\AnA{613}{L1}{2018}

\bibitem{Lazzati+18} 
\Authname{Lazzati}{D.}, \Authname{Perna}{R.}, \Authname{Morsony}{B.~J.}, \Authname{Lpez-Cmara}{D.}, \Authname{Cantiello}{M.}, \Authname{Ciolfi}{R.}, \Authname{Giacomazzo}{B.} and \Authname{Workman}{J.~C.},
\PRL{120,241103,2018}

\bibitem{IN19}
\Authname{Ioka}{K.} and \Authname{Nakamura}{T.}, \MNRAS{487}{4884}{2019}

\bibitem{Matsumoto19a}
\Authname{Matsumoto}{T.}, \Authname{Nakar}{E.}, \Authname{Piran}{T.},
\MNRAS{483}{1247}{2019}

\bibitem{Matsumoto19b}
\Authname{Matsumoto}{T.}, \Authname{Nakar}{E.}, \Authname{Piran}{T.},
\MNRAS{486}{1563}{2019}

\bibitem{Tominaga+18} 
\Authname{Tominaga}{N.} \etal,
\PASJ{70}{28}{2018}

\bibitem{Usman+18} 
\Authname{Usman}{S.~A.}, \Authname{Mills}{J.~C.} and \Authname{Fairhurst}{S.},
\ApJ{877}{82}{2019}

\bibitem{Abbott+17} 
\LVC,
\ApJL{851}{L16}{2017}

\bibitem{Gehrels+06} 
\Authname{Gehrels}{N.} \etal,
\Nat{444}{1044}{2006}

\bibitem{GalYam+06} 
\Authname{Gal-Yam}{A.} \etal,
\Nat{444}{1053}{2006}

\bibitem{Ioka+00} 
\Authname{Ioka}{K.},
\MNRAS{327}{639}{2001}

\bibitem{Abadie+10} 
\Authname{Abadie}{J.} \etal (LIGO Scientific and Virgo Collaborations),
\ApJ{734}{L35}{2011}

\bibitem{Kashiyama+11} 
\Authname{Kashiyama}{K.}, \Authname{Ioka}{K.},
\PRD{83,081302,2011}

\bibitem{Abbott+07} 
\Authname{Abbott}{B.} \etal (LIGO Scientific Collaboration),
\ApJ{681}{1419}{2008}

\bibitem{Abadie+12} 
\Authname{Abadie}{J.} \etal (LIGO Scientific Collaboration),
\ApJ{755}{2}{2012}

\bibitem{Abbott+16} 
\Authname{Abbott}{B.~P.} \etal (LIGO Scientific and Virgo and IPN Collaborations),
\ApJ{841}{89}{2017}



\bibitem{piran04}
\Authname{Piran}{T.},
\RMP{76,1143,2004}

\bibitem{kumar15}
\Authname{Kumar}{P.} and \Authname{Zhang}{B.},
\PRep{561}{1}{2015}

\bibitem{macfadyen99}
\Authname{MacFadyen}{A.~I.} and \Authname{Woosley}{S.~E.},
\ApJ{524}{262}{1999}

\bibitem{stark1985}
\Authname{Stark}{R.~F.} and \Authname{Piran}{T.}, 
\PRL{55,891,1985}

\bibitem{kobayashi03}
\Authname{Kobayashi}{S.} and \Authname{M{\'e}sz{\'a}ros}{P.},
\ApJ{589}{861}{2003}

\bibitem{braginsky87}
\Authname{Braginsky}{V.~B.} and \Authname{Thorne}{K.~S.},
\Nat{327}{123}{1987}

\bibitem{segalis01}
\Authname{Segalis}{E.~B.} and \Authname{Ori}{A.},
\PRD{64,064018,2001}

\bibitem{sago04}
\Authname{Sago}{N.} \etal, 
\PRD{70,104012,2004}

\bibitem{suwa09}
\Authname{Suwa}{Y.} and \Authname{Murase}{K.},
\PRD{80,123008,2009}

\bibitem{Astone05_GRB}
\Authname{Astone}{P.} \etal,
\PRD{71,042001,2005}

\bibitem{Abbott05_GRB}
\Authname{Abbott}{B.} \etal,
\PRD{72,042002,2005}

\bibitem{Aasi14_GRB}
\Authname{Aasi}{J.} \etal,
\PRL{113,011102,2014}

\bibitem{Abbott19_GRB}
\LVC,
\ApJ{886}{75}{2019}

\bibitem{fynbo06}
\Authname{Fynbo}{J. P. U.} \etal,
\Nat{444}{1047}{2006}

\bibitem{dellavalle06}
\Authname{Della Valle}{M.} \etal,
\Nat{444}{1050}{2006}

\bibitem{galyam06}
\Authname{Gal-Yam}{A.} \etal,
\Nat{444}{1053}{2006}

\bibitem{gehrels06}
\Authname{Gehrels}{N.} \etal,
\Nat{444}{1044}{2006}

\bibitem{yang15}
\Authname{Yang}{B.} \etal,
\NatCom{6}{7323}{2015}

\bibitem{liang07_RY}
E. Liang, B. Zhang, F. Virgili, \etal,
\ApJ{662}{1111}{2007}

\bibitem{ioka01_ApJL}
K. Ioka, K. and T. Nakamura, 
\ApJ{554}{L163}{2001}

\bibitem{yamazaki02_ApJL}
R. Yamazaki, K. Ioka, T. Nakamura,
\ApJ{571}{L31}{2002}



\bibitem{2007Sci...318..777L} 
\Authname{Lorimer}{D.~R.}, \Authname{Bailes}{M.}, \Authname{McLaughlin}{M.~A.}, \Authname{Narkevic}{D.~J.} and \Authname{Crawford}{F.},
\Sci{318}{777}{2007}

\bibitem{FRBCAT}
FRBCAT, http://frbcat.org/

\bibitem{2017Natur.541...58C} 
\Authname{Chatterjee}{S.} \etal, 
\Nat{541}{58}{2017}

\bibitem{2017ApJ...834L...8M} 
\Authname{Marcote}{B.} \etal,
\ApJL{834}{L8}{2017}

\bibitem{2019Natur.566..235C} 
CHIME/FRB Collaboration,
\Nat{566}{235}{2019}

\bibitem{2013PASJ...65L..12T} 
\Authname{Totani}{T.},
\PASJ{65}{L12}{2013}

\bibitem{2017ApJ...839L...3K} 
\Authname{Kashiyama}{K.} and \Authname{Murase}{K.},
\ApJL{839}{L3}{2017}

\bibitem{2016ApJ...818...94K} 
\Authname{Kashiyama}{K.}, \Authname{Murase}{K.}, \Authname{Bartos}{I.}, \Authname{Kiuchi}{K.} and \Authname{Margutti}{R.},
\ApJ{818}{94}{2016}

\bibitem{2017ApJ...844..140C} 
\Authname{Chawla}{P.} \etal, 
\ApJ{844}{140}{2017}




\bibitem{Kibble76}
\Authname{Kibble}{T. W. B.},
J. Phys. A \textbf{9}, 1387 (1976).

\bibitem{Sarangi02} 
\Authname{Sarangi}{S.} and \Authname{Tye}{S. H. H.}, 
\PLB{536,185,2002}

\bibitem{Jeannerot03}
\Authname{Jeannerot}{R.} \etal,
\PRD{68,103514,2003}

\bibitem{Damour00}
\Authname{Damour}{T.} and \Authname{Vilenkin}{A.},
\PRL{85,3761,2000}

\bibitem{Damour01}
\Authname{Damour}{T.} and \Authname{Vilenkin}{A.},
\PRD{64,064008,2001}

\bibitem{Damour05}
\Authname{Damour}{T.} and \Authname{Vilenkin}{A.}, 
\PRD{71,063510,2005}

\bibitem{Vilenkin00}
\Authname{Vilenkin}{A.} and \Authname{Shellard}{E.},
Cosmic strings and other Topological Defects, 
Cambridge University Press, 2000.

\bibitem{Matsui:2016xnp} 
\Authname{Matsui}{Y.}, \Authname{Horiguchi}{K.}, \Authname{Nitta}{D.} and \Authname{Kuroyanagi}{S.},
\JCAP{11}{2016}{}{005}
 
\bibitem{Matsui:2019obe} 
\Authname{Matsui}{Y.} and \Authname{Kuroyanagi}{S.},
\PRD{100,123515,2019}

\bibitem{Abbott09}
\LSC,
\PRD{80,062002,2009}

\bibitem{Aasi14}
\LVC,
\PRL{112,131101,2014}

\bibitem{Abbott18}
\LVC,
\PRD{97,102002,2018}



\bibitem{Bekenstein:1974jk} 
\Authname{Bekenstein}{J.D.},
  Lett.\ Nuovo Cim.\  {\bf 11}, 467 (1974).

\bibitem{Mukhanov:1986me} 
\Authname{Mukhanov}{V.F.},
\JETPL{44}{63}{1986}

\bibitem{Barcelo:2017lnx} 
\Authname{Barceló}{C.}, \Authname{Carballo-Rubio}{R.} and \Authname{Garay}{L.J.},
\JHEP{1705,054,2017}

\bibitem{Nakano}
\Authname{Nakano}{H.}, \Authname{Sago}{N.}, \Authname{Tagoshi}{H.} and \Authname{Tanaka}{T.},
\PTEP{2017}{071E01}{ptx093}

\bibitem{Abedi}
\Authname{Abedi}{J.}, \Authname{Dykaar}{H.} and \Authname{Afshordi}{N.},
\PRD{96,082004,2017}

\bibitem{Uchikata:2019frs} 
 \Authname{Uchikata}{N.}, \Authname{Nakano}{H.}, 
 \Authname{Narikawa}{T.}, \Authname{Sago}{N.}, 
 \Authname{Tagoshi}{H.} and \Authname{Tanaka}{T.},
\PRD{100,062006,2019}

\end{thebibliography}
\end{document}